\documentclass[12pt]{article}
\renewcommand{\baselinestretch}{1.6}

\usepackage{amsfonts}
\usepackage{amsgen,amsmath,amstext,amsbsy,amsopn,amssymb,soul,subfigure}

\usepackage[pdftex]{graphicx}
\usepackage{epstopdf}
\usepackage{natbib} 
\usepackage[dvipsnames,table,dvipsnames*, svgnames*]{xcolor}
\usepackage{comment}
\usepackage{blkarray}
\usepackage{algorithm}
\usepackage{algpseudocode}
\usepackage{booktabs}
  
\definecolor{blue}{RGB}{050,050,250}
\definecolor{green}{RGB}{000,150,100}
\definecolor{purple}{RGB}{220,040,250}

\def\red{\color{red}}

\newtheorem{Theorem}{Theorem}

\newtheorem{Lemma}{Lemma}
\newtheorem{Remark}{Remark}
\newtheorem{Corollary}{Corollary}

\newtheorem{Proposition}{Proposition}

\newcommand{\be}{\begin{equation}}
\newcommand{\ee}{\end{equation}}

\newcommand{\V}{{\rm V}}

\newcommand{\rank}{{\rm rank}}

\newcommand{\diag}{{\rm diag}}
\newcommand{\argmin}{\mathop{\rm arg\min}}

\def\mybox#1{\vskip1mm \begin{center} \bf \red
        \hspace{.0\textwidth}\vbox{\hrule\hbox{\vrule\kern6pt
\parbox{.95\textwidth}{\kern6pt#1\vskip6pt}\kern6pt\vrule}\hrule}
        \end{center} \vskip-5mm}
 
\begin{document}

\title{Structured Matrix Completion with Applications to Genomic Data Integration$^1$}

\author{Tianxi Cai, ~  T. Tony Cai, ~and~  Anru Zhang}

\date{}

\maketitle

\footnotetext[1]{Tianxi Cai is Professor of Biostatistics, Department of Biostatistics,  Harvard University, Boston, MA (E-mail: tcai@hsph.harvard.edu); T. Tony Cai is Dorothy Silberberg Professor of Statistics, Department of Statistics, The Wharton School, University of Pennsylvania, Philadelphia, PA (E-mail: tcai@wharton.upenn.edu); Anru Zhang is a Ph.D. student, Department of Statistics, The Wharton School, University of Pennsylvania, Philadelphia, PA (E-mail: anrzhang@wharton.upenn.edu). The research of Tianxi Cai was supported in part by  NIH Grants R01 GM079330 and U54 H6007963; the research of Tony Cai and Anru Zhang  was supported in part by NSF Grants DMS-1208982  and DMS-1403708, and NIH Grant R01 CA127334. }


\maketitle

\vspace{-.4in}

\begin{abstract}

Matrix completion has attracted significant recent attention in many fields including statistics, applied mathematics and electrical engineering.  
Current literature on matrix completion focuses primarily on independent sampling models under which the individual observed entries are  sampled independently. Motivated by applications in genomic data integration, we propose a new framework of structured matrix completion (SMC) to treat structured missingness by design. Specifically, our proposed method aims at efficient matrix recovery when a subset of the rows and columns of an approximately low-rank matrix are observed.  We provide theoretical justification for the proposed SMC method and derive lower bound for the estimation errors, which together establish the optimal rate of recovery over certain classes of approximately low-rank matrices.  Simulation studies show that the method performs well in finite sample under a variety of configurations. The method is applied to integrate several ovarian cancer genomic studies with different extent of genomic measurements, which enables us to construct more accurate prediction rules for ovarian cancer survival.
\end{abstract}


\section{Introduction}
\label{sec.intro}

Motivated by an array of applications, matrix completion has attracted significant recent attention in different fields including statistics, applied mathematics and electrical engineering.  The central goal of matrix completion is to recover a high-dimensional low-rank matrix based on a subset of its entries.  Applications include recommender systems \citep{koren2009matrix}, genomics \citep{chi2013genotype}, multi-task learning \citep{argyriou2008convex},  sensor localization \citep{biswas2006semidefinite, singer2010uniqueness},   and computer vision \citep{chen2004recovering, tomasi1992shape}, among many others. 
 
Matrix completion has been well studied under  the uniform sampling model, where observed entries are assumed to be sampled uniformly at random.  The best known approach is perhaps the constrained nuclear norm minimization (NNM), which has been shown to yield near-optimal results when the sampling distribution of the observed entries is uniform \citep{candes2009exact, candes2010power, gross2011recovering, recht2011simpler,candes2011tight}. For  estimating approximately low-rank matrices from uniformly sampled noisy observations, several penalized or constrained NNM estimators, which are based on  the same principle as the well-known Lasso and Dantzig selector for sparse signal recovery, were proposed and analyzed  \citep{keshavan2010matrix, mazumder, koltchinskii2011neumann, koltchinskii2011nuclear, rohde2011estimation}. In many applications, the entries are sampled independently but not uniformly.  In such a setting, \cite{salakhutdinov2010collaborative} showed that the standard NNM methods do not perform well, and proposed a weighted NNM method, which depends on the true sampling distribution. In the case of unknown sampling distribution, \cite{foygel2011learning} introduced an empirically-weighted NNM method. \cite{cai2013matrix} studied a max-norm constrained minimization method for the recovery of a low-rank matrix  based on the noisy observations under the non-uniform sampling model. It was shown that the max-norm constrained least squares estimator is rate-optimal under the Frobenius norm loss and  yields a more stable approximate recovery guarantee with respect to the sampling distributions.

The focus of matrix completion has so far been on the recovery of a low-rank matrix based on independently sampled entries.  Motivated by applications in genomic data integration, we introduce in this paper a new framework of matrix completion called \emph{structured matrix completion} (SMC), where a subset of the rows and a subset of the columns of an approximately low-rank matrix are observed and the goal is to reconstruct the whole matrix based on the observed rows and columns. We first discuss the genomic data integration problem before introducing the SMC model.

\subsection{Genomic Data Integration}

When analyzing genome-wide studies (GWS) of association, expression profiling or methylation,  ensuring adequate power of the analysis is one of 
the most crucial goals due to the high dimensionality of the genomic markers under consideration. Because of cost constraints, GWS typically
have small to moderate sample sizes and hence limited power. One approach to increase the power is to integrate information from multiple GWS 
of the same phenotype. However, some practical complications may hamper the feasibility of such integrative analysis.  
Different GWS often involve different platforms with distinct genomic coverage. 
For example, whole genome next generation sequencing (NGS) studies would provide mutation information on all loci while older technologies 
for genome-wide association studies (GWAS) would only provide information on a small subset of loci. 
In some settings, certain studies may provide a wider range of genomic data than others. For example, one study may provide extensive genomic measurements
including gene expression, miRNA and DNA methylation while other studies may only measure gene expression. 

To perform integrative analysis of studies with different extent of genomic measurements, the naive complete observation only approach may suffer from low power.
For the GWAS setting with a small fraction of loci missing, many imputation methods have been proposed in 
recent years to improve the power of the studies. Examples of useful methods include
haplotype reconstruction, $k$-nearest neighbor, regression and singular value decomposition methods 
\citep{scheet2006fast,li2006mach,browning2009unified,troyanskaya2001missing,kim2005missing,wang2006missing}. 
Many of the haplotype phasing methods
are considered to be highly effective in recovering missing 
genotype information \citep{yu2007methods}. These methods, while useful, are often computationally intensive. In addition, when one study has a much denser coverage than the other, the fraction of missingness could be high and an exceedingly large number of observation would need  to be imputed. It is unclear whether it is 
statistically or computationally feasible to extend these methods to such settings. 
Moreover, haplotype based methods cannot be extended to incorporate other types of genomic data such as gene expression and miRNA data. 


When integrating multiple studies with different extent of genomic measurements, the observed data can be viewed as complete rows and columns of a large matrix $A$ and the 
missing components can be arranged as a submatrix of $A$. As such, the missingness in $A$ is structured by design.
In this paper, we propose a novel SMC method for imputing the missing submatrix of $A$. 
As shown in Section \ref{application.sec}, by imputing the missing miRNA measurements and constructing prediction rules based on the imputed data, it is possible to significantly improve the prediction performance.


\subsection{Structured Matrix Completion Model}

Motivated by the applications mentioned above, this paper considers SMC where a subset of  rows and columns are observed. Specifically, 
we observe  $m_1< p_1$ rows and $m_2<p_2$ columns  of a matrix $A\in \mathbb{R}^{p_1\times p_2}$ and the goal is to recover the whole matrix. Since the singular values are invariant under row/column permutations, it can be assumed without loss of generality that we observe the first $m_1$ rows and $m_2$ columns of $A$ which can be written in a block form:
\begin{equation}\label{eq:A_block}
A = \begin{blockarray}{ccc}
m_2 & p_2 - m_2 & \\
\begin{block}{[cc]c}
A_{11} & A_{12} & m_1 \\
A_{21} & {\color{gray} A_{22}} & p_1-m_1\\
\end{block}
\end{blockarray}
\end{equation}
where  $A_{11}$, $A_{12}$, and $A_{21}$ are observed and the goal is to recover the missing block $A_{22}$. 
See Figure \ref{fig:A_Z_illustration}(a) in Section \ref{sec.procedure} for a graphical display of the data.
%
Clearly there is no way to recover  $A_{22}$ if $A$ is an arbitrary matrix. However, in many applications such as genomic data integration discussed earlier, $A$ is approximately low-rank, which makes it possible to recover  $A_{22}$ with accuracy. 
In this paper, we introduce a method based on the singular value decomposition (SVD) for the recovery of $A_{22}$ when $A$ is approximately low-rank. 

It is important to note that the observations here are much more ``structured" comparing to the previous settings of matrix completion. As the observed entries are in full rows or full columns, the existing methods based on NNM are not suitable. As mentioned earlier, constrained NNM methods have been widely used in matrix completion problems based on independently observed entries. 
However, for the problem considered in the present paper, these methods do not utilize the structure of the observations and do not guarantee precise recovery even for exactly low-rank matrix $A$ (See Remark \ref{rm:nuclear_norm_fail} in Section \ref{sec.procedure}). Numerical results in Section \ref{simulation.sec}  show that NNM methods do not perform well in SMC.

In this paper we propose a new SMC method that can be easily implemented by a fast algorithm which only involves basic matrix operations and the SVD. The main idea of our recovery procedure is based on the Schur Complement. In the ideal case when $A$ is exactly low rank, the Schur complement of the missing block, $A_{22} - A_{21}A_{11}^\dagger A_{12}$, is zero and thus $A_{21}A_{11}^\dagger A_{12}$ can be used to recover $A_{22}$ exactly. When $A$ is approximately low rank, $A_{21}A_{11}^\dagger A_{12}$ cannot be used directly to estimate $A_{22}$. For  this case, we transform the observed blocks using SVD; remove some unimportant rows and columns based on thresholding rules; and subsequently apply a similar procedure to recover $A_{22}$. 

Both its theoretical and numerical  properties are studied. It is shown that the estimator recovers low-rank matrices accurately and is robust against small perturbations. A lower bound result shows that the estimator is rate optimal for a class of approximately low-rank matrices. 
Although it is required for the theoretical analysis that there is a significant gap between the singular values of the true low-rank matrix and those of the perturbation, simulation results indicate that this gap is not really necessary in practice and the estimator recovers $A$ accurately whenever the singular values of $A$ decay sufficiently fast.
 
\subsection{Organization of the Paper}

The rest of the paper is organized as follows. In Section  \ref{sec.procedure}, we introduce in detail the proposed SMC methods when $A$ is exactly or  approximately low-rank. The theoretical properties of the estimators are analyzed in Section \ref{analysis.sec}. Both upper and lower bounds for the recovery accuracy under the Schatten-$q$ norm loss are established. Simulation results are shown in Section \ref{simulation.sec} to investigate the numerical performance of the proposed methods. A real data application to genomic data integration is given in Section \ref{application.sec}. Section \ref{discussion.sec} discusses a few practical issues related to real data applications. For reasons of space, the proofs of the main results and additional simulation results are given in the supplement \citep{supplement}. 
Some key technical tools used in the proofs of the main theorems are also developed and proved in the supplement. 

\section{Structured Matrix Completion: Methodology}
\label{sec.procedure}

In this section, we propose procedures to recover the submatrix $A_{22}$ based on the observed blocks $A_{11}$,  $A_{12}$,  and $A_{21}$. We begin with basic notation and definitions that will be used in the rest of the paper. 
 
For a matrix $U$, we use $U_{[\Omega_1, \Omega_2]}$ to represent its sub-matrix with row indices $\Omega_1$ and column indices $\Omega_2$. We also use the Matlab syntax to represent index sets. Specifically for integers $a\leq b$, ``$a:b$" represents $\{a, a+1,\cdots, b\}$; and ``:" alone represents the entire index set. Therefore, $U_{[:, 1:r]}$ stands for the first $r$ columns of $U$ while $U_{[(m_1+1):p_1, :]}$ stands for the $\{m_1+1, ..., p_1\}^{th}$ rows of $U$. 
For the matrix $A$ given in \eqref{eq:A_block}, we use the notation $A_{\bullet 1}$ and $A_{1 \bullet}$ to denote $[A_{11}^{\intercal},A_{21}^{\intercal}]^{\intercal}$ and $[A_{11},A_{12}]$, respectively.
For a matrix $B\in \mathbb{R}^{m\times n}$, let $B =U\Sigma V^{\intercal}= \sum_i \sigma_i(B)u_iv_i^{\intercal}$ be the SVD, where $\Sigma=\diag\{\sigma_1(B), \sigma_2(B), ...\}$
with $\sigma_1(B) \geq \sigma_2(B) \geq \cdots \geq 0$ being the singular values of $B$ in decreasing order.  The smallest singular value $\sigma_{\min(m, n)}$, which will be denoted by $\sigma_{\min}(B)$, plays an important role in our analysis. 
We also define $B_{\max(r)} = \sum_{i=1}^r \sigma_i(B)u_iv_i^{\intercal}$ and $B_{-\max(r)} = B - B_{\max(r)} = \sum_{i\geq r+1}\sigma_i(B)u_iv_i^{\intercal}$. For $1\leq q\leq \infty$,  the Schatten-$q$ norm $\|B\|_q$ is defined to be the vector $q$-norm of the singular values of $B$, i.e. $\|B\|_q = \left(\sum_i \sigma_i^q(B)\right)^{1/q}$. Three special cases are of particular interest: when $q = 1$, $\|B\|_1 = \sum_i \sigma_i(B)$ is the nuclear (or trace) norm of $B$ and will be denoted as $\|B\|_*$; when $q = 2$, $\|B\|_2 = \sqrt{\sum_{i, j}B_{ij}^2}$ is the Frobenius norm of $B$ and will be denoted as $\|B\|_F$; when $q = \infty$, $\|B\|_\infty = \sigma_1(B)$ is the spectral norm of $B$ that we simply denote as $\|B\|$. 
 For any matrix $U\in \mathbb{R}^{p\times n}$, we use $P_{U}\equiv U\left(U^{\intercal}U\right)^{\dagger}U^{\intercal}\in\mathbb{R}^{p\times p}$ to denote the projection operator onto the column space of $U$. Throughout, we assume that $A$ is {\em approximately rank $r$} in that for some integer $0 < r \leq \min(m_1, m_2)$, there is a significant gap between  $\sigma_r(A)$ and $\sigma_{r+1}(A)$ and the tail $\|A_{-\max(r)}\|_q=\left(\sum_{k\ge r+1} \sigma_{k}^q(A)\right)^{1/q}$ is small.
The gap assumption enables us to provide a theoretical upper bound on the accuracy of the estimator, while it is not necessary in practice (see Section \ref{simulation.sec} for more details).

\subsection{Exact Low-rank Matrix Recovery}
\label{exact_low_rank.sec}

We begin with the relatively easy case where $A$ is exactly of rank $r$. In this case, a simple analysis indicates that $A$ can be perfectly recovered as shown in the following proposition.

\begin{Proposition}\label{th:noiseless}
Suppose $A$ is of rank $r$, 
the SVD of $A_{11}$ is $A_{11} = U\Sigma V^{\intercal}$, where $U\in\mathbb{R}^{p_1 \times r}, \Sigma\in\mathbb{R}^{r\times r},$ and $V\in\mathbb{R}^{p_2\times r}$. If
$$\rank([A_{11} \; A_{12}] ) = \rank\left(\begin{bmatrix}
A_{11}\\
A_{21}
\end{bmatrix}
\right) = \rank(A) = r,$$
then $\rank(A_{11}) = r$ and $A_{22}$ is exactly given by
\begin{equation}\label{eq:A_22_exact_low_rank}
 A_{22} = A_{21}(A_{11})^\dagger A_{12} = A_{21}V(\Sigma)^{-1}U^{\intercal}A_{12}.
\end{equation}
\end{Proposition}
\begin{Remark}\label{rm:nuclear_norm_fail}
{\rm
Under the same conditions as Proposition \ref{th:noiseless}, the NNM
\begin{equation}\label{eq:nuclear_norm_minimization}
\hat A_{22} = \argmin_{B} \left\|\begin{bmatrix}
A_{11} & A_{12}\\
A_{21} & B
\end{bmatrix}\right\|_\ast
\end{equation} 
fails to guarantee the exact recovery of $A_{22}$. Consider the case where $A$ is a $p_1\times p_2$ matrix with all entries being 1. Suppose we observe arbitrary $m_1$ rows and $m_2$ columns, the NNM would yield $\hat A_{22}\in \mathbb{R}^{(p_1 - m_1)\times (p_2- m_2)}$ with all entries being $\left(1\wedge \sqrt{\frac{m_1m_2}{(p_1 - m_1)(p_2-m_2)}}\right)$ (See Lemma \ref{lm:nuclear_norm_1} in the Supplement). Hence when $m_1m_2 < (p_1 - m_1)(p_2 - m_2)$, i.e., when the size of the observed blocks are much smaller than that of $A$, the NNM fails to recover exactly  the missing block $A_{22}$. See also the numerical comparison in Section \ref{simulation.sec}. The NNM \eqref{eq:nuclear_norm_minimization} also fails to recover $A_{22}$ with high probability in a random matrix setting where  $A=B_1B_2^T$ with $B_1\in \mathbb{R}^{p_1\times r}$ and $B_2\in \mathbb{R}^{p_2\times r}$ being i.i.d. standard Gaussian matrices. See  Lemma \ref{lm:nuclear_norm_fail} in the Supplement for further details.
In addition to \eqref{eq:nuclear_norm_minimization}, other variations of NNM have been proposed in the literature, including {\em penalized} NNM \citep{Toh,mazumder},
\begin{equation}\label{eq:PNNM} 
\hat A^{PN} = \argmin_Z \left\{\frac{1}{2}\sum_{(i_k, j_k)\in \Omega}( Z_{i_k, j_k} - A_{i_k, j_k})^2 + t\|Z\|_\ast \right\};
\end{equation}
and {\em constrained} NNM with relaxation \citep{Cai_Candes_Shen},
\begin{equation}\label{eq:CNNM}
\hat A^{CN} = \argmin_{Z} \left\{\|Z\|_\ast: \; |Z_{i_k, j_k} - A_{i_k, j_k}|\leq t  \mbox{ for } (i_k, j_k)\in \Omega \right\},
\end{equation}
where $\Omega = \{(i_k, j_k): A_{i_k,j_k} \mbox{ observed}, 1 \le i_k \le p_1, 1\le j_k \le p_2\}$ and $t$ is the tunning parameter. However, these NNM methods may not be suitable for SMC especially when only a small number of rows and columns are observed. In particular, when $m_1 \ll p_1, m_2\ll p_2$, $A$ is well spread in each block $A_{11}, A_{12}, A_{21}, A_{22}$, we have $\|[A_{11} ~ A_{12}]\|_\ast \ll\|A\|_\ast$, $[A_{12}]_\ast \ll \|A\|_\ast$. Thus,
$$\left\|\begin{bmatrix}
A_{11} & A_{12}\\
A_{21} & 0
\end{bmatrix}\right\|_\ast \leq \left\|\begin{bmatrix}
A_{11}\\
A_{21}
\end{bmatrix}\right\|_\ast + \left\|\begin{bmatrix}
A_{12}
\end{bmatrix}\right\|_\ast \ll \left\|\begin{bmatrix}
A_{11} & A_{12}\\
A_{21} & A_{22}
\end{bmatrix}\right\|_\ast.  $$
In the other words, imputing $A_{22}$ with all zero yields a much smaller nuclear norm than imputing with the true $A_{22}$ and hence NNM methods would generally  fail to recover $A_{22}$ under such settings.}

\end{Remark}

Proposition \ref{th:noiseless} shows that, when $A$ is exactly low-rank,  $A_{22}$ can be recovered precisely by $A_{21}(A_{11})^\dagger A_{12}$. Unfortunately, this result heavily relies on the exactly low-rank assumption that cannot be directly used for approximately low-rank matrices. In fact, even with a small perturbation to $A$, the inverse of $A_{11}$ makes the formula $A_{21}(A_{11})^\dagger A_{12}$ unstable, which may lead to the failure of recovery. In practice, $A$ is often not exactly low rank but approximately low rank. Thus for the rest of the paper, we focus on the latter setting.

\subsection{Approximate Low-rank Matrix Recovery}
\label{approx_low_rank.sec}

 Let $A = U\Sigma V^{\intercal}$ be the SVD of an approximately low rank matrix $A$
and partition  $U\in \mathbb{R}^{p_1\times p_1}, V \in \mathbb{R}^{p_2 \times p_2}$ and $\Sigma\in \mathbb{R}^{p_1 \times p_2}$ into blocks as
\begin{equation}\label{eq:block}
U = \begin{blockarray}{ccc}
 r & p_1-r & \\
\begin{block}{[cc]c}
  U_{11} & U_{12} & m_1 \\
  U_{21} & U_{22} & p_1 - m_1\\
\end{block}
\end{blockarray},\ V = \begin{blockarray}{ccc}
 r & p_2-r & \\
\begin{block}{[cc]c}
  V_{11} & V_{12} & m_2 \\
  V_{21} & V_{22} & p_2 - m_2\\
\end{block}
\end{blockarray}, \ \Sigma = 
\begin{blockarray}{ccc}
 r & p_2-r & \\
\begin{block}{[cc]c}
  \Sigma_{1} & 0 & r \\
  0 & \Sigma_{2} & p_1-r\\
\end{block}
\end{blockarray}
\end{equation}
Then $A$ can be decomposed as $A = A_{\max(r)} + A_{-\max(r)}$ where $A_{\max(r)}$  is of rank $r$ with the largest $r$ singular values of $A$ and $A_{-\max(r)}$ is general but with small singular values. Then
\begin{equation}\label{eq:A_1, A_2}
A_{\max(r)} = U_{\bullet  1}\Sigma_1 V_{\bullet  1}^{\intercal} = \begin{blockarray}{ccc}
 m_2 & p_2-m_2 & \\
\begin{block}{[cc]c}
  U_{11}\Sigma_{1}V_{11}^{\intercal} &   U_{11}\Sigma_{1}V_{21}^{\intercal} & m_1 \\
    U_{21}\Sigma_{1}V_{11}^{\intercal} &   U_{21}\Sigma_{1}V_{21}^{\intercal} & p_1-m_1\\
\end{block}
\end{blockarray}
, \quad \mbox{and} \quad A_{-\max(r)} = U_{\bullet 2}\Sigma_2 V_{\bullet 2}^{\intercal} .
\end{equation}
Here and in the sequel, we use the notation $U_{\bullet k}$ and $U_{k \bullet}$ to denote $[U_{1k}^{\intercal},U_{2k}^{\intercal}]^{\intercal}$ and $[U_{k1},U_{k2}]$, respectively.
Thus, $A_{\max(r)}$ can be viewed as a rank-$r$ approximation to $A$ and obviously
$$U_{21}\Sigma_{1}V_{21}^{\intercal} = \{U_{21}\Sigma_{1}V_{11}^{\intercal}\} \{U_{11}\Sigma_{1}\V_{11}^{\intercal}\}^{-1} \{U_{11}\Sigma_{1}V_{21}^{\intercal}\}.$$
We will use the observed $A_{11}$, $A_{12}$ and $A_{21}$ to obtain estimates of $U_{\bullet 1}$, $V_{\bullet 1}$ and $\Sigma_{1}$
and subsequently recover $A_{22}$ using an estimated $U_{21}\Sigma_{1}V_{21}^{\intercal}$.

When $r$ is known, i.e., we know where the gap is located in the singular values of $A$, a simple procedure can be implemented to estimate $A_{22}$ as described in Algorithm 1 below by estimating $U_{\bullet 1}$ and $V_{\bullet 1}$ 
using the principal components of $A_{\bullet 1}$ and $A_{1\bullet}$.
\begin{algorithm}[H]
\caption{Algorithm for Structured Matrix Completion with a given $r$}
\begin{algorithmic}[1]
\State Input: $A_{11}\in \mathbb{R}^{m_1\times m_2}, A_{12}\in\mathbb{R}^{(p_1-m_1)\times m_2}, A_{21}\in \mathbb{R}^{m_1\times (p_2-m_2)}$.
\State Calculate the SVD of $A_{\bullet 1}$ and $A_{1\bullet}$ to obtain $A_{\bullet 1} = U^{(1)}\Sigma^{(1)} V^{(1)\intercal}, \ A_{1 \bullet} = U^{(2)}\Sigma^{(2)}V^{(2)\intercal}$.
\State Suppose $M, N$ are orthonormal basis of $U_{11}, V_{11}$. We estimate the column space of $U_{11}$ and $V_{11}$ by $\hat M = U^{(2)}_{[:, 1:r]}, \hat N = V^{(1)}_{[:, 1:r]}. $
\State Finally we estimate $A_{22}$ as
\begin{equation}\label{eq:hat A}
\hat A_{22} = A_{21}\hat N (\hat M^{\intercal} A_{11} \hat N)^{-1} \hat M^{\intercal} A_{12}.
\end{equation}
\end{algorithmic}\label{al:with_r}
\end{algorithm}

\begin{figure}[htbp]
\begin{center}
\subfigure[heatmap of block-wise $A$]{\includegraphics[height = 2in, width=2.6in]{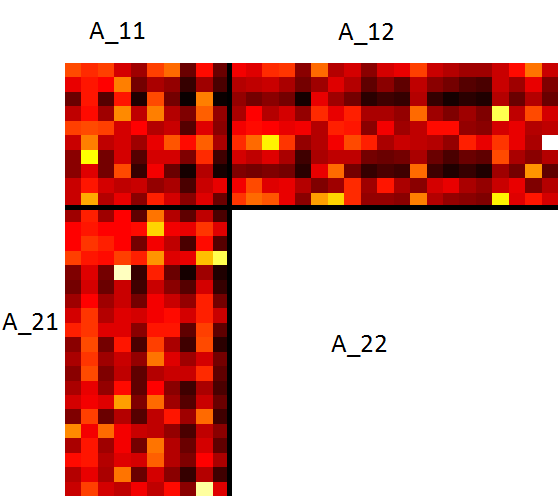}} 
\subfigure[heatmap of block-wise $Z$ after rotation]{\includegraphics[height = 2in, width=2.6in]{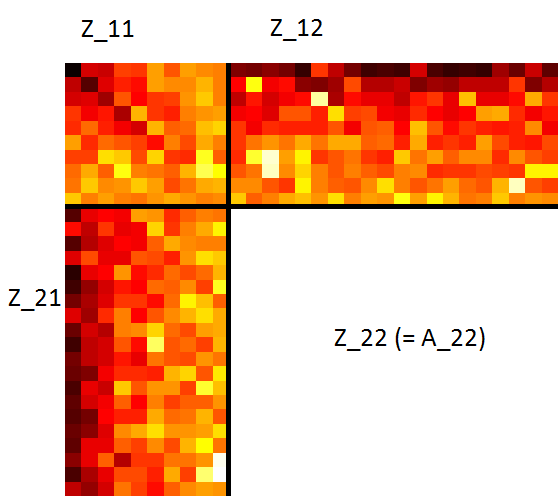}}
\caption{ Illustrative example with $A\in \mathbb{R}^{30\times 30}$, $m_1 = m_2 = 10$. (A darker block corresponds to larger magnitude.)}\vspace{-.2in}
\label{fig:A_Z_illustration}
\end{center}
\end{figure}

\begin{figure}[htbp]
\begin{center}
\subfigure[Intermediate step when $\hat r = 9$]{\includegraphics[height = 2in, width=2.7in]{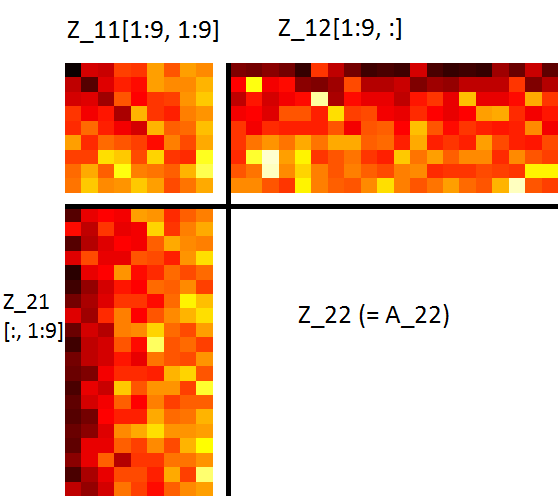}} 
\subfigure[Identify the position to truncate at $\hat r = 4$]{\includegraphics[height = 2in, width=2.7in]{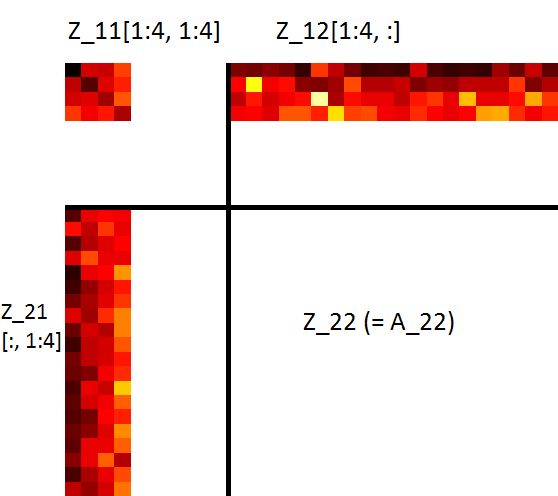}}
\caption{Searching for the appropriate position to truncate from $\hat r = 10$ to 1.}\vspace{-.2in}
\label{fig:thresholding}
\end{center}
\end{figure}

However, Algorithm 1 has several major limitations. First, it relies on a given $r$ which is typically unknown in practice. Second, the algorithm need to calculate the matrix divisions, which may cause serious precision issues when the matrix is near-singular or the rank $r$ is mis-specified. To overcome these difficulties, we propose another Algorithm which essentially first estimates $r$ with $\hat{r}$ and then apply Algorithm 1 to recover $A_{22}$.
Before introducing the algorithm of recovery without knowing $r$, it is helpful to illustrate the idea with heat maps in Figures \ref{fig:A_Z_illustration} and \ref{fig:thresholding}. 
%

Our procedure has three steps.
\begin{enumerate}
\item First, we move the significant factors of $A_{\bullet1}$ and $A_{1\bullet}$ to the front by rotating the columns of $A_{\bullet 1}$ and the rows of $A_{1 \bullet}$ based on the SVD,
$$A_{\bullet 1} = U^{(1)}\Sigma^{(1)} V^{(1)\intercal}, \quad  A_{1 \bullet} = U^{(2)}\Sigma^{(2)}V^{(2)\intercal}. $$
After the transformation, we have $Z_{11}, Z_{12}, Z_{21}$,
$$Z_{11} = U^{(2)\intercal} A_{11}V^{(1)},\quad  Z_{12} = U^{(2)\intercal} A_{12} ,\quad  Z_{21}= A_{21}V^{(1)}, \quad Z_{22} = A_{22}.$$
Clearly $A$ and $Z$ have the same singular values since the transformation is orthogonal. As shown in Figure \ref{fig:A_Z_illustration}(b), the amplitudes of the columns of $Z_{\bullet 1}=[Z_{11}^{\intercal}, Z_{21}^{\intercal}]^{\intercal}$ and the rows of $Z_{1\bullet}=[Z_{11},Z_{12}]$ are decaying. 

\item When $A$ is exactly of rank $r$,  the $\{r+1, \cdots, m_1\}^{th}$ rows and $\{r+1, \cdots, m_2\}^{th}$  columns of $Z$ are zero. Due to the small perturbation term $A_{-\max(r)}$, the back columns of $Z_{\bullet 1}$ and rows of $Z_{1\bullet}$ are small but non-zero. In order to recover $A_{\max(r)}$, the best rank $r$ approximation to $A$, a natural idea is to first delete these back rows of $Z_{1\bullet}$ and columns of $Z_{\bullet1}$, i.e. the $\{r+1, \cdots, m_1\}^{th}$ rows and $\{r+1, \cdots, m_2\}^{th}$ columns of $Z$. 

However, since $r$ is unknown, it is unclear how many back rows and columns should be removed. It will be helpful to have an estimate for $r$, $\hat{r}$, and then use $Z_{21,[:, 1:\hat r]}$,  $Z_{11,[1:\hat r, 1:\hat r]}$ and $Z_{12[1:\hat r, :]}$ to recover $A_{22}$. It will be shown that a good choice of $\hat r$ would satisfy that $Z_{11, [1:\hat r, 1:\hat r]}$ is non-singular and $\|Z_{21,[1:\hat r, 1:\hat r]}Z_{11,[1:\hat r, 1:\hat r]}^{-1}\| \leq T_R$, where $T_R$ is some constant to be specified later. Our final estimator for $r$ would be the largest $\hat{r}$ that satisfies this condition, which can be identified 
recursively from $\min(m_1,m_2)$ to 1 (See Figure \ref{fig:thresholding}). 

\item Finally, similar to \eqref{eq:A_22_exact_low_rank}, $A_{22}$ can be estimated by 
\begin{equation}\label{eq:hat_A_22_illustration}
\hat A_{22} = Z_{21,[:, 1:\hat r]}Z_{11, [1:\hat r, 1:\hat r]}^{-1}Z_{12, [1:\hat r, :]},
\end{equation}
\end{enumerate}

The method we propose can be summarized as the following algorithm. 
\begin{algorithm}[H]
\caption{Algorithm of Structured Matrix Completion with unknown $r$}
\begin{algorithmic}[1]
\State Input: $A_{11}\in\mathbb{R}^{m_1\times m_2}, A_{12}^{m_1\times (p_2-m_2)}, A_{21}^{(p_1 - m_1)\times m_2}$. Thresholding level: $T_R $, (or $T_C$).
\State Calculate the SVD $A_{\bullet 1} = U^{(1)} \Sigma^{(1)} V^{(1)\intercal}$, $A_{1 \bullet} = U^{(2)}\Sigma^{(2)} V^{(2)\intercal}$.
\State Calculate $Z_{11}\in\mathbb{R}^{m_1\times m_2}, Z_{12}\in\mathbb{R}^{m_1\times (p_2 - m_2)}, Z_{21}\in\mathbb{R}^{(p_1 - m_1)\times m_2}$
$$Z_{11} = U^{(2)\intercal} A_{11}V^{(1)},\quad  Z_{12} = U^{(2)\intercal} A_{12} ,\quad  Z_{21}= A_{21}V^{(1)}.$$
\For {s = $\min(m_1, m_2)$ : -1: 1} \quad (Use iteration to find $\hat r$)
\State Calculate $D_{R, s}\in \mathbb{R}^{(p_1 - m_1)\times s}$ (or $D_{C, s}\in\mathbb{R}^{s\times (p_2 - m_2)}$) by solving linear equation system,
$$D_{R, s} = Z_{21, [:, 1:s]} Z_{11, [1:s, 1:s]}^{-1} \quad (\text{or} \quad D_{C, s} = Z_{11, [1:s, 1:s]}^{-1}Z_{12, [1:s, :]}) $$
\If {$Z_{11, [1:s, 1:s]}$ is not singular and $\|D_{R, s}\| \leq T_R$ ( or $\|D_{C, s}\| \leq T_C$)}
\State $\hat r = s$; {\bf break} from the loop;
\EndIf
\EndFor
\If {($\hat r$ is not valued)} $\hat r = 0$. \EndIf
\State Finally we calculate the estimate as
$$\hat A_{22} 
 = Z_{21,[:, 1:\hat r]}Z_{11, [1:\hat r, 1:\hat r]}^{-1}Z_{12, [1:\hat r, :]}$$
\end{algorithmic}\label{al:without_r}
\end{algorithm}

It can also be seen from Algorithm \ref{al:without_r} that the estimator $\hat r$ is constructed based on either the row thresholding rule $\|D_{R, s}\| \leq T_R$ or  the column thresholding rule $\|D_{C, s}\|\leq T_C$. Discussions on the choice between $D_{R,s}$ and $D_{C,s}$ are given in the next section. Let us focus for now  on the row thresholding based on $D_{R, s} = Z_{21, [:, 1:s]}Z_{11, [1:s, 1:s]}^{-1}$.
It is important to note that $Z_{21[:,1:r]}$ and $Z_{11,[1:r,1:r]}$ approximate $U_{21}\Sigma_1$ and $\Sigma_1$, respectively. 
The idea behind the proposed $\hat r$ is that when $s > r$, $Z_{21[:,1:s]}$ and $Z_{11,[1:s,1:s]}$ are nearly singular and hence $D_{R,s}$
may either be deemed singular or with unbounded norm. 
When $s = r$, $Z_{11, [1:s, 1:s]}$ is non-singular with $\|D_{R, s}\|$ bounded 
by some constant, as we show in Theorem \ref{th:main_without_r}. Thus, we  estimate $\hat{r}$ as the largest $r$ such that 
$Z_{11, [1:s, 1:s]}$ is non-singular with $\|D_{R, s}\| < T_R$.

\section{Theoretical Analysis}
\label{analysis.sec}

In this section, we investigate the theoretical properties of the algorithms introduced in Section \ref{sec.procedure}. Upper bounds for the estimation errors of Algorithms 1 and 2 are presented in Theorems \ref{th:hat A} and \ref{th:main_without_r}, respectively, and the lower-bound results are given in Theorem \ref{th:lower_bound}. These  bounds together establish the optimal rate of recovery over certain classes of approximately low-rank matrices. The choices of tuning parameters $T_R$ and $T_C$ are discussed in Corollaries \ref{cr:random_column_row} and \ref{cr:randomUV}.

\begin{Theorem}\label{th:hat A}
Suppose $\hat A$ is given by the procedure of Algorithm 1. Assume 
\begin{equation}\label{eq:assumption}
\sigma_{r+1}(A) \le {1\over 2} \sigma_r(A)\cdot\sigma_{\min}(U_{11})\cdot \sigma_{\min} (V_{11}),
\end{equation}  Then 
for any $1\leq q\leq \infty$, 
\begin{equation}\label{ineq:th1_bound}
\left\|\hat A_{22} - A_{22}\right\|_q \leq 3\|A_{-\max(r)}\|_q \left(1+\frac{1}{\sigma_{\min}(U_{11})}\right)\left(1+\frac{1}{\sigma_{\min}(V_{11})}\right) 
\end{equation}
\end{Theorem}

\begin{Remark}{\rm
It is helpful to explain intuitively why Condition \eqref{eq:assumption} is needed. When $A$ is approximately low-rank,  the dominant low-rank component of $A$, $A_{\max(r)}$, serves as a good approximation to $A$, while the residual $A_{-\max(r)}$ is ``small". The goal is to  recover $A_{\max(r)}$ well.
Among the three observed blocks, $A_{11}$ is the most important  and it is necessary to have $A_{\max(r)}$ dominating $A_{-\max(r)}$ in $A_{11}$. Note that $A_{11} = A_{\max(r), [1:m_1, 1:m_2]} + A_{-\max(r), [1:m_1, 1:m_2]}$, 
$$\sigma_r(A_{\max(r), [1:m_1, 1:m_2]}) = \sigma_r(U_{11}\Sigma_1V_{11}^{\intercal}) \geq \sigma_{\min}(U_{11})\sigma_r(A)\sigma_{\min}(V_{11}),$$ 
$$\|A_{-\max(r), [1:m_1, 1:m_2]}\| = \|U_{12}\Sigma_2V_{12}^{\intercal}\| \leq \sigma_{r+1}(A).$$ 
We thus require Condition \eqref{eq:assumption} in Theorem \ref{th:hat A} for the theoretical analysis.
}
\end{Remark} 

Theorem \ref{th:hat A} gives an upper bound for the estimation accuracy of Algorithm 1 under the assumption that there is a significant gap between $\sigma_r(A)$ and $\sigma_{r+1}(A)$ for some known $r$. It is noteworthy that there are possibly multiple values of $r$ that satisfy Condition  \eqref{eq:assumption}. In such a case, the bound \eqref{ineq:th1_bound} applies to all such $r$ and the largest $r$ yields the strongest result. 

We now turn to Algorithm \ref{al:without_r}, where the knowledge of $r$ is not assumed. Theorem \ref{th:main_without_r} below shows that for properly chosen $T_R$ or $T_C$, Algorithm 2 can lead to accurate recovery of $A_{22}$. 
\begin{Theorem}\label{th:main_without_r}
Assume that there exists $r \in [1, \min(m_1, m_2)]$ such that
\begin{equation}\label{ineq:assumption_theorem2}
\sigma_{r+1}(A) \leq {1\over 4} \sigma_r(A)\cdot \sigma_{\min}(U_{11})\sigma_{\min}(V_{11}).
\end{equation}
Let $T_R$ and $T_C$ be two constants satisfying
\[
T_R \geq \frac{1.36}{\sigma_{\min}(U_{11})}+0.35 \quad \mbox{and}\quad T_C\geq \frac{1.36}{\sigma_{\min}(V_{11})} + 0.35.
\]
Then for $1\leq q\leq \infty$, $\hat A_{22}$ given by Algorithm 2 satisfies
\begin{align}\label{eq:th_without_r}
& \left\|\hat A_{22} - A_{22}\right\|_q \leq 6.5T_R\left(\frac{1}{\sigma_{\min}(V_{11})} + 1\right) \|A_{-\max(r)}\|_q \\
\mbox{or}\quad & \left\|\hat A_{22} - A_{22}\right\|_q \leq 6.5T_C \left(\frac{1}{\sigma_{\min}(U_{11})} + 1\right)\|A_{-\max(r)}\|_q \nonumber
\end{align}
when $\hat{r}$ is estimated based on the thresholding rule $\|D_{R, s}\|\leq T_R$ or $\|D_{C, s}\| \leq T_C$, respectively.
\end{Theorem}

Besides $\sigma_r(A)$ and $\sigma_{r+1}(A)$, Theorems \ref{th:hat A} and \ref{th:main_without_r} involve $\sigma_{\min}(U_{11})$ and $\sigma_{\min}(V_{11})$, 
two important quantities that reflect how much the low-rank matrix $A_{\max(r)} = U_{\bullet 1}\Sigma_1V_{\bullet 1}^{\intercal}$ is concentrated on the first $m_1$ rows and $m_2$ columns. We should note that $\sigma_{\min}(U_{11})$ and $\sigma_{\min}(V_{11})$ depend on the singular vectors of $A$ and $\sigma_r(A)$ and $\sigma_{r+1}(A)$ are the singular values of $A$. The lower bound in Theorem \ref{th:lower_bound} below indicates that $\sigma_{\min}(U_{11})$, $\sigma_{\min}(V_{11})$, and the singular values of $A$  together quantify the difficulty of the problem:
recovery of $A_{22}$ gets harder as $\sigma_{\min}(U_{11})$ and $\sigma_{\min}(V_{11})$ become smaller  or the $\{r+1, \cdots, \min(p_1, p_2)\}^{th}$ singular values become larger. Define the class of approximately rank-$r$ matrices $\mathcal{F}_r(M_1, M_2)$ by
\begin{equation}
\mathcal{F}_r(M_1, M_2) = \left\{A\in \mathbb{R}^{p_1\times p_2}:\begin{array}{ll}
 \sigma_{\min}(U_{11}) \geq M_1, \sigma_{\min}(V_{11}) \geq M_2,\\ \sigma_{r+1}(A) \le {1\over 2}\sigma_r(A)\sigma_{\min}(U_{11})\sigma_{\min}(V_{11})
 \end{array} \right\}.
\end{equation}

\begin{Theorem}[Lower Bound]\label{th:lower_bound}
Suppose $r \leq \min(m_1, m_2, p_1-m_1, p_2 -m_2)$ and $0<M_1, M_2 < 1$, then for all $1\leq q\leq \infty$,
\begin{equation}
\inf_{\hat A_{22}}\sup_{A\in \mathcal{F}_{r}(M_1, M_2)} \frac{\|\hat A_{22} - A_{22}\|_q}{\|A_{-\max(r)}\|_q} \geq \frac{1}{4}\left(\frac{1}{M_1}+1\right)\left(\frac{1}{M_2}+1\right).
\end{equation}
\end{Theorem}
\begin{Remark}
{\rm
Theorems \ref{th:hat A}, \ref{th:main_without_r} and \ref{th:lower_bound} together immediately yield the optimal rate of recovery over the class $\mathcal{F}_r(M_1 M_2)$,
\begin{equation}
\inf _{\hat A_{22}}\sup_{A\in \mathcal{F}_r(M_1, M_2)} \frac{\|\hat A_{22}-A_{22}\|_q}{\|A_{-\max(r)}\|_q} \asymp \left(\frac{1}{M_1} + 1\right)\left(\frac{1}{M_2}+1\right)
\ \mbox{for $0\leq M_1, M_2 <1$, $1\leq q\leq \infty$.}
\end{equation}
}
\end{Remark}

Since $U_{11}$ and $V_{11}$ are determined by the SVD of  $A$ and   $\sigma_{\min}(U_{11}) $ and $\sigma_{\min}(V_{11})$ are unknown based only on  $A_{11}, A_{12}, $ and $A_{21}$, it is thus not straightforward to choose the tuning parameters $T_R$ and $T_C$ in a principled way. Theorem \ref{th:main_without_r} also does not provide information on the choice between row and column thresholding.
Such a choice generally depends on the problem setting. 
We consider below two settings where either the row/columns of $A$ are randomly sampled or $A$ is itself a random low-rank matrix. In such settings, when $A$ is approximately rank $r$ and at least $O(r\log r)$ number of rows and columns are observed, Algorithm \ref{al:without_r} gives accurate recovery of $A$ with fully specified tuning parameter.
We first consider in Corollary \ref{cr:random_column_row} a fixed matrix $A$ with the observed $m_1$ rows and $m_2$ columns selected uniformly randomly. 

\begin{Corollary}[Random Rows/Columns]\label{cr:random_column_row}
Let $A = U\Sigma V^{\intercal}$ be the SVD of $A\in\mathbb{R}^{p_1\times p_2}$. Set
\begin{equation}
W_r^{(1)} = \frac{p_1}{r}\max_{1\leq i\leq p_1}\sum_{j=1}^r U_{ij}^2 \quad\mbox{\rm and}\quad  W_r^{(2)} = \frac{p_2}{r}\max_{1\leq i\leq p_2} \sum_{j=1}^r V_{ij}^2.
\end{equation} 
Let $\Omega_1\subset\{1,\cdots, p_1\}$ and $\Omega_2\subset\{1, \cdots, p_2\}$ be  respectively the index set of the observed $m_1$ rows and $m_2$ columns.
Then $A$ can be decomposed as
\begin{equation}\label{eq:coro_A_decompose}
A_{11} = A_{[\Omega_1, \Omega_2]}, \; A_{21} = A_{[\Omega_1^c, \Omega_2]}, \; A_{12} = A_{[\Omega_1, \Omega_2^c]}, \; A_{22} = A_{[\Omega_1^c, \Omega_2^c]}. 
\end{equation}
\begin{enumerate}
\item Let $\Omega_1$ and $\Omega_2$ be independently and uniformly selected from $\{1, \cdots, p_1\}$ and $\{1, \cdots, p_2\}$ with or without replacement, respectively. Suppose there exists $r \leq \min(m_1, m_2)$ such that 
$$\sigma_{r+1}(A)\leq \frac{1}{6}\sigma_r(A)\sqrt{\frac{m_1m_2}{p_1p_2}}.$$
and the number of rows and number of columns we observed satisfy
$$m_1 \geq 12.5rW_r^{(1)}(\log (r) + c), \quad m_2 \geq 12.5rW_r^{(2)}(\log (r)+c), \quad \mbox{for some constant $c > 1$.}$$
Algorithm 2 with either column thresholding with the break condition $\|D_{R,s}\|\leq T_R$ where $T_R = 2\sqrt{\frac{p_1}{m_1}}$ or row thresholding with the break condition $\|D_{C,s}\|\leq T_C$ where $T_C = 2\sqrt{\frac{p_2}{m_2}}$ satisfies, for all $1\leq q\leq \infty$, 
$$\|\hat A_{22} - A_{22}\|_q \leq 29\|A_{-\max(r)}\|_q \sqrt{\frac{p_1p_2}{m_1m_2}} \quad \mbox{with probability $\ge 1-4\exp(-c)$.}$$

\item If $\Omega_1$ is uniformly randomly selected from $\{1,\cdots, p_1\}$ with or without replacement ($\Omega_2$ is not necessarily random), and there exists $r \leq m_2$ such that 
$$\sigma_{r+1}(A) \leq {1\over 5} \sigma_r(A)\sigma_{\min}(V_{11})\sqrt{\frac{m_1}{p_1}}$$
and the number of observed rows satisfies
\begin{equation}\label{eq:th_number_row}
m_1 \geq 12.5 rW_r^{(1)} \left(\log (r) + c\right) \quad \mbox{for some constant $c > 1$,}
\end{equation}
then Algorithm 2 with the break condition $\|D_{R, s}\| \leq T_R$ where $T_R\geq 2\sqrt{\frac{p_1}{m_1}}$ satisfies, for all $1\leq q\leq \infty$,
$$\left\|\hat A_{22} - A_{22}\right\|_q \leq 6.5\|A_{-\max(r)}\|_qT_R\left(\frac{1}{\sigma_{\min}(V_{11})} + 1\right) \quad \mbox{with probability $\ge 1-2\exp(-c)$.}$$
%
\item Similarly, if  $\Omega_2$ is uniformly randomly selected from $\{1,\cdots, p_2\}$ with or without replacement ($\Omega_1$ is not necessarily random) and there exists $r \leq m_2$ such that 
$$\sigma_{r+1}(A) \leq {1\over 5}\sigma_r(A)\sigma_{\min}(U_{11})\sqrt{\frac{m_2}{p_2}},$$
and the number of observed columns satisfies
\begin{equation}\label{eq:th_number_col}
m_2 \geq 12.5 rW_r^{(2)} \left(\log (r) + c\right) \quad \mbox{for some constant $c > 1$,}
\end{equation}
then Algorithm 2 with the break condition $\|D_{C, s}\| \leq T_C$ where $T_C\geq 2\sqrt{\frac{p_2}{m_2}}$ satisfies, for all $1\leq q\leq \infty$,
$$\left\|\hat A_{22} - A_{22}\right\|_q \leq 6.5\|A_{-\max(r)}\|_qT_C\left(\frac{1}{\sigma_{\min}(U_{11})} + 1\right) \quad \mbox{with probability $\ge 1-2\exp(-c)$.}$$

\end{enumerate}
\end{Corollary}

\begin{Remark}
{\rm
The quantities $W^{(1)}_r$ and $W^{(2)}_r$ in Corollary \ref{cr:random_column_row} measure the variation of amplitude of each row or each column of $A_{\max(r)}$. When $W^{(1)}_r$ and $W^{(2)}_r$ become larger, a small number of rows and columns in $A_{\max(r)}$ would have larger amplitude than others, while these rows and columns would be missed with large probability in the sampling of $\Omega$, which means the problem would become harder. Hence, more observations for the matrix with larger $W^{(1)}_r$ and $W^{(2)}_r$ are needed as shown in \eqref{eq:th_number_row}.
}
\end{Remark}

We now consider  the case where $A$ is a random matrix.
\begin{Corollary}[Random Matrix]\label{cr:randomUV}
Suppose $A\in\mathbb{R}^{p_1\times p_2}$ is a random matrix generated by $A =  U\Sigma V^{\intercal}$, where the singular values $\Sigma$ and singular space $V$ are fixed, and $U$ has orthonormal columns that are randomly sampled based on the Haar measure. Suppose we observe the first $m_1$ rows and first $m_2$ columns of $A$. Assume there exists $r < \frac{1}{2}\min(m_1, m_2)$ such that 
$$\sigma_{r+1}(A) \leq \frac{1}{5}\sigma_r(A)\sigma_{\min}(V_{11})\sqrt{\frac{m_1}{p_1}}.$$
Then there exist uniform constants $c, \delta>0$ such that if $m_1 \geq cr$, $\hat A_{22}$ is given by Algorithm \ref{al:without_r} with the break condition $\|D_{R, s}\|\leq T_R$, where $T_R\geq 2\sqrt{\frac{p_1}{m_1}}$, we have for all $1\leq q\leq \infty$,
$$\left\|\hat A_{22} - A_{22}\right\|_q \leq 6.5\|A_{-\max(r)}\|_q T_R\left(\frac{1}{\sigma_{\min}(V_{11})} + 1\right) \quad \mbox{with probability at least $1 - e^{-\delta m_1}$.}$$
Parallel results hold for the case when $U$ is fixed and $V$ has orthonormal columns that are randomly sampled based on the Haar measure, and we observe the first $m_1$ rows and first $m_2$ columns of $A$. Assume there exists $r<\frac{1}{2}\min (m_1, m_2)$ such that
$$\sigma_{r+1}(A)\leq \frac{1}{5}\sigma_r(A)\sigma_{\min}(U_{11})\sqrt{\frac{m_2}{p_2}}. $$ 
Then there exist unifrom constants $c,\delta>0$ such that if $m_2\geq cr$, $\hat A_{22}$ is given by Algorithm 2 with column thresholding with the break condition $\|D_{C,s}\|\leq T_C$, where $T_C \geq 2\sqrt{\frac{p_2}{m_2}}$, we have for all $1\leq q\leq \infty$,
$$\left\|\hat A_{22} - A_{22}\right\|_q \leq 6.5\|A_{-\max(r)}\|_q T_C\left(\frac{1}{\sigma_{\min}(U_{11})} + 1\right)  \quad \mbox{with probability at least $1-e^{-\delta m_2}$.}$$
\end{Corollary}

\section{Simulation}
\label{simulation.sec}

 In this section, we show results from extensive simulation studies that examine the numerical performance of  Algorithm \ref{al:without_r} on randomly generated matrices for various values of $p_1$, $p_2$, $m_1$ and $m_2$.  
 We first consider settings where a gap between some adjacent singular values exists, as required by our theoretical analysis. 
 Then we investigate settings where the singular 
 values decay smoothly with no significant gap between adjacent singular values.
 The results show that the proposed procedure performs well even when there is no significant gap, as long as the singular values decay at a reasonable rate.

 We also examine how sensitive the proposed estimators are to the choice of the threshold 
 and the choice between row and column thresholding. In addition, we compare the performance of the SMC method with that of the NNM method. 
 Finally, we  consider a setting similar to the real data application discussed in the next section. 
Results shown below are based on 200-500 replications for each configuration. Additional simulation results on the effect of $m_1$, $m_2$ and ratio $p_1/m_1$ are provided in the supplement. 
Throughout, we generate the random matrix A from $A = U\Sigma V$, where  the singular values of the diagonal matrix $\Sigma$ are 
chosen accordingly for different settings. The singular spaces $U$ and $V$ are drawn randomly from the Haar measure. Specifically, we generate i.i.d. standard Gaussian matrix $\tilde U \in \mathbb{R}^{p_1\times \min (p_1, p_2)}$ and $\tilde V\in \mathbb{R}^{p_2\times \min(p_1, p_2)}$, then apply the QR decomposition to $\tilde{U}$ and $\tilde{V}$ and assign $U$ and $V$ with the $Q$ part of the result. 

\begin{figure}[htbp]
\begin{center}
\includegraphics[width=.45\linewidth]{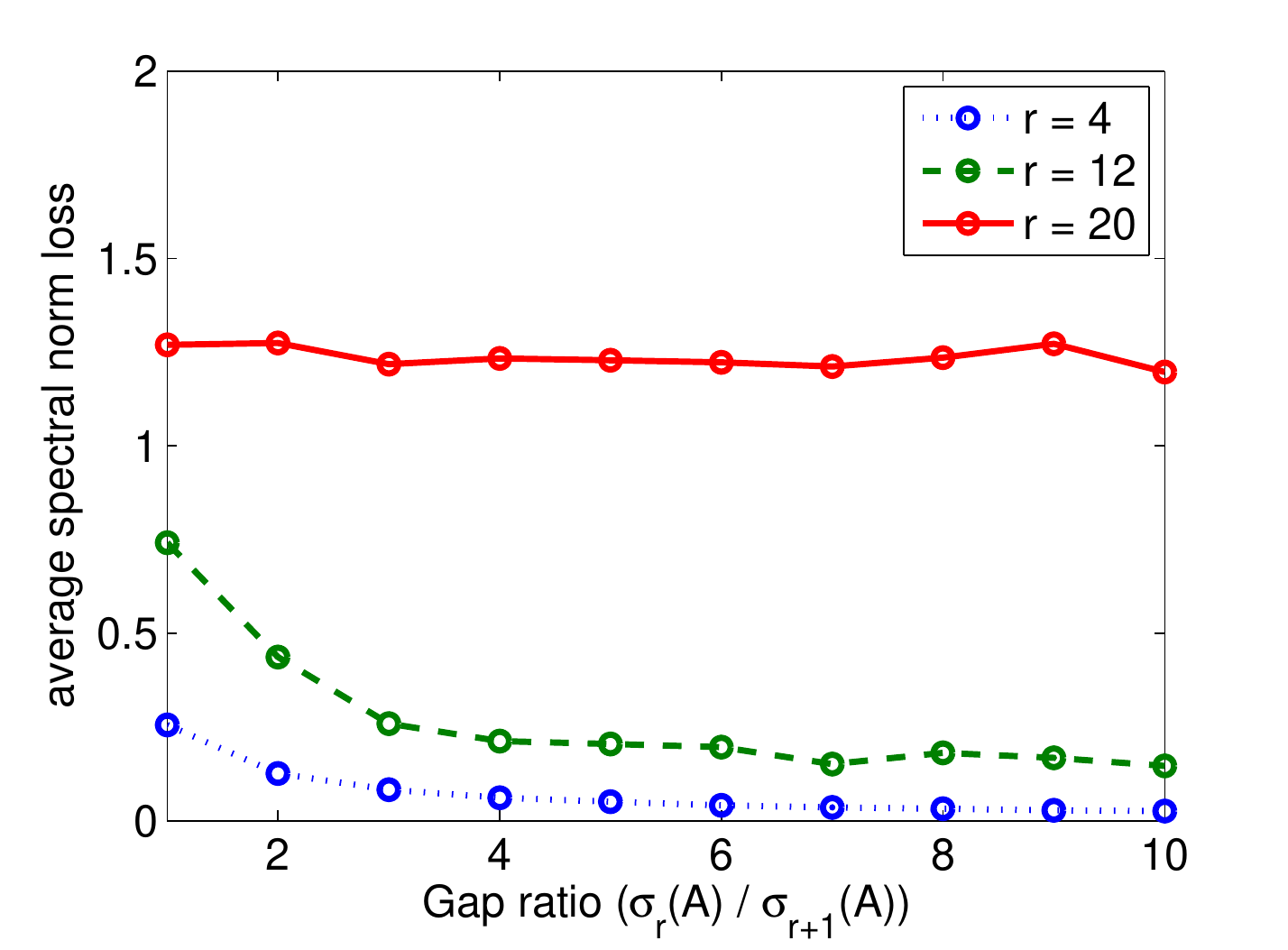}
\includegraphics[width=.45\linewidth]{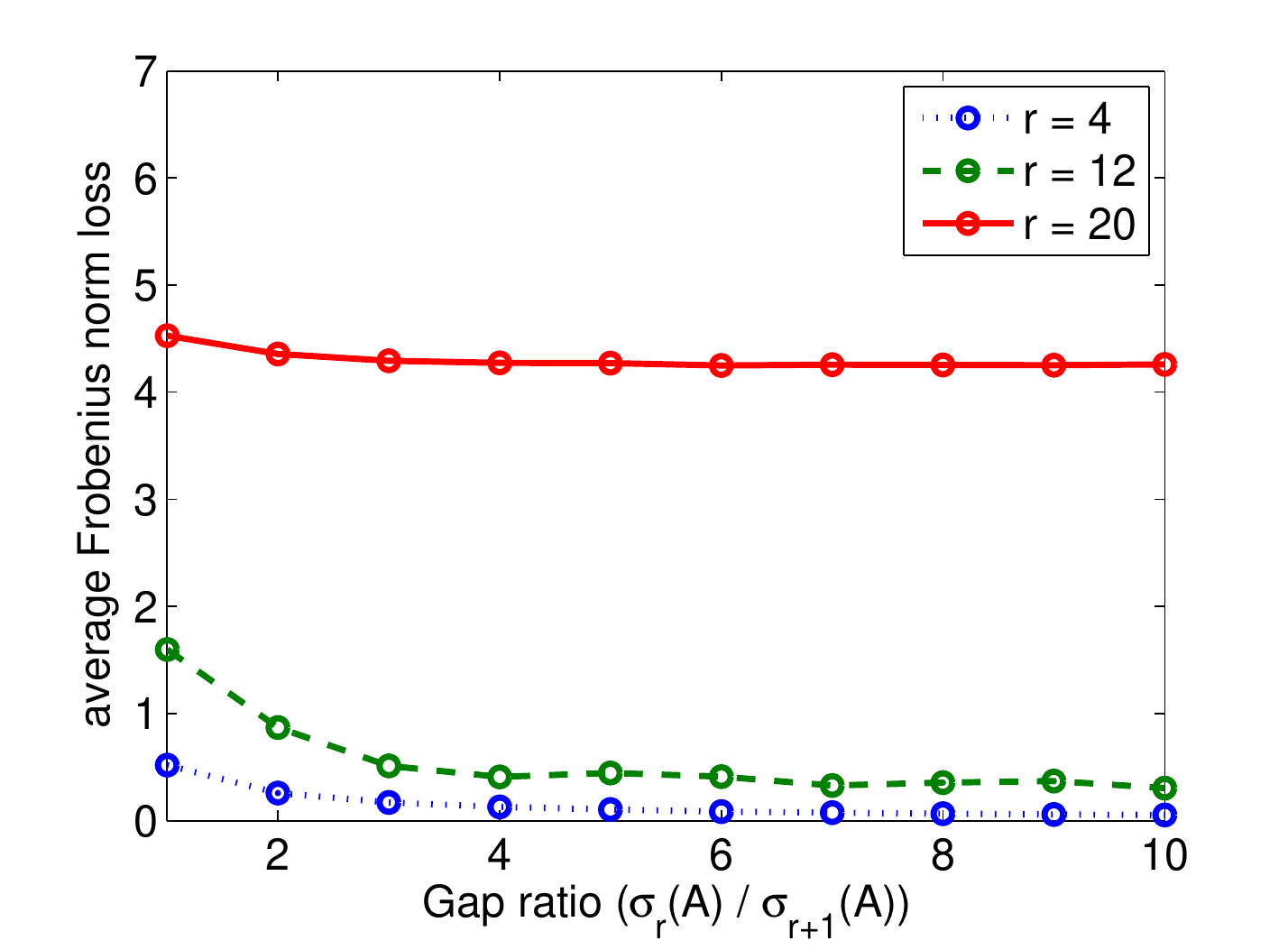}
\caption{Spectral norm loss (left panel) and Frobenius norm loss (right panel) when there is a gap between $\sigma_r(A)$ and $\sigma_{r+1}(A)$. The singular value values of $A$ are given by \eqref{eq:singular_value_gap}, $p_1 = p_2 = 1000$, and $ m_1 = m_2 = 50$.}\vspace{-.2in}
\label{fig:gap}
\end{center}
\end{figure}

We first consider the performance of Algorithm \ref{al:without_r} when a significant gap between the $r^{th}$ and $(r+1)^{th}$ singular values of $A$. We fixed $p_1 = p_2 = 1000, m_1 = m_2 = 50$ and choose the singular values as
\begin{equation}\label{eq:singular_value_gap}
\{\underbrace{1, \cdots, 1}_{r}, ~ g^{-1}1^{-1}, ~ g^{-1}2^{-1}, ~ \cdots\}, \quad g = 1, 2, \cdots, 10, \quad r = 4, 12 \text{ and } 20.
\end{equation}
Here $r$ is the rank of the major low-rank part $A_{\max(r)}$, $g = \frac{\sigma_r(A)}{\sigma_{r+1}(A)}$ is the gap ratio between the $r^{th}$ and $(r+1)^{th}$ singular values of $A$. 
The average loss of $\hat A_{22}$ from Algorithm 2 with the row thresholding and $T_R = 2\sqrt{p_1/m_1}$ under both the spectral norm and Frobenius norm losses are given in Figure \ref{fig:gap}.
The results suggest that our algorithm performs better when $r$ gets smaller and gap ratio $g = \sigma_r(A)/\sigma_{r+1}(A)$ gets larger. Moreover, even when $g = 1$, namely there is no significant gap between any adjacent singular values, our algorithm still works well for small $r$. As will be seen in the following simulation studies,  this is generally the case as long as the singular values of $A$  decay sufficiently fast.

\begin{figure}[htbp]
\begin{center}
\begin{tabular}{cc}
\includegraphics[width =3.2in,height=2.0in]{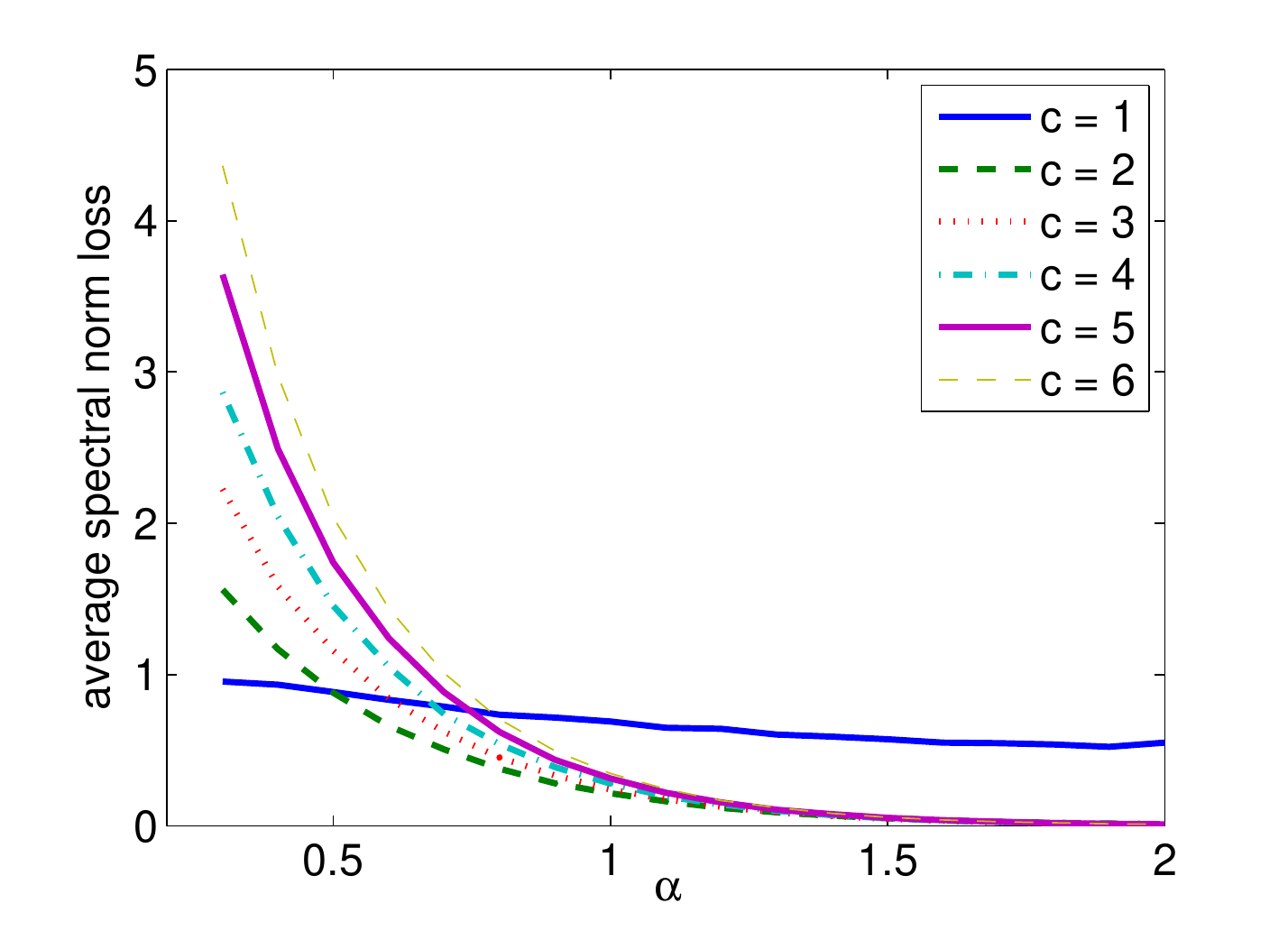} & \includegraphics[width =3.2in,height=2.0in]{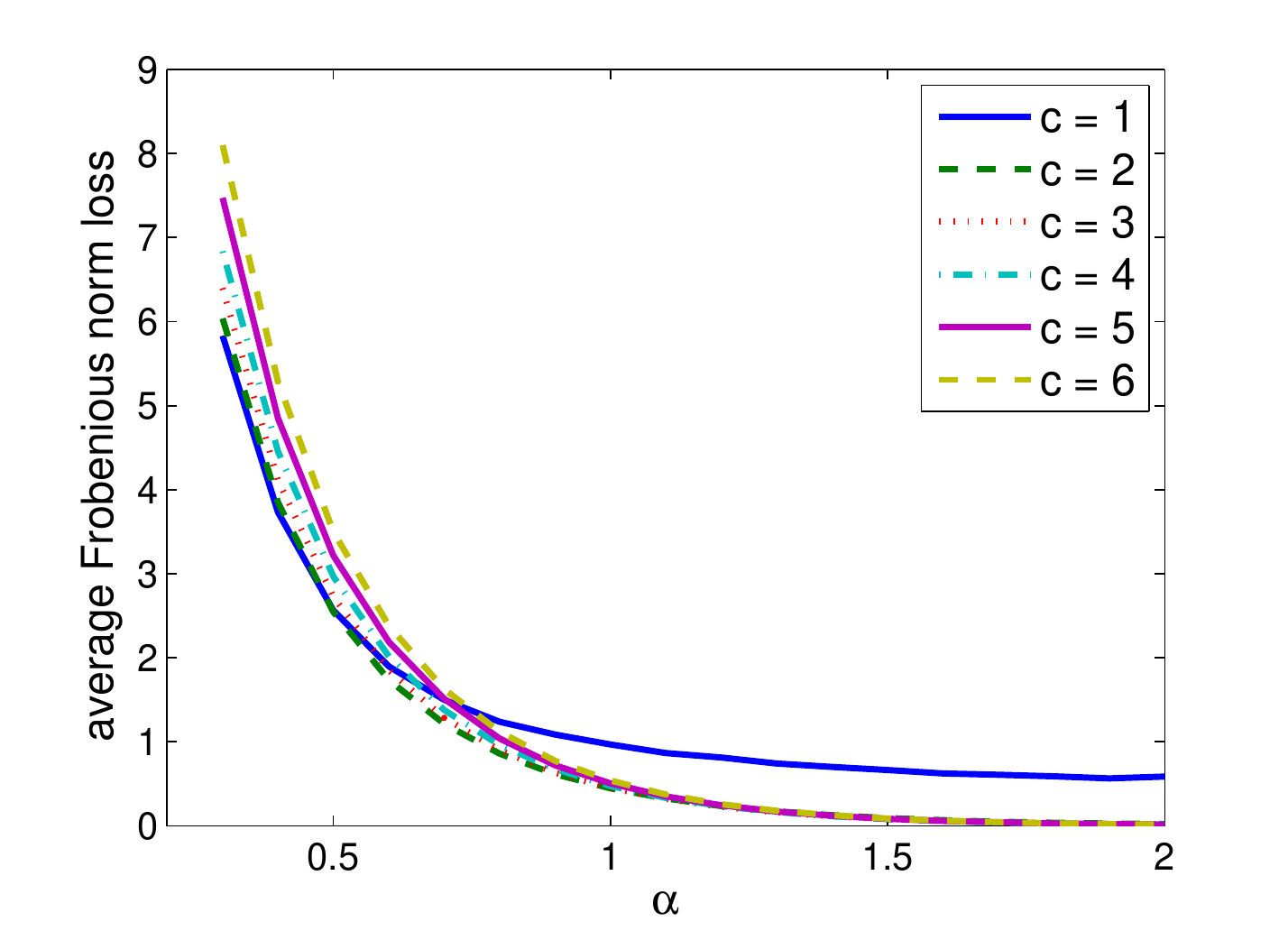}\\
\includegraphics[width =3.2in,height=2.0in]{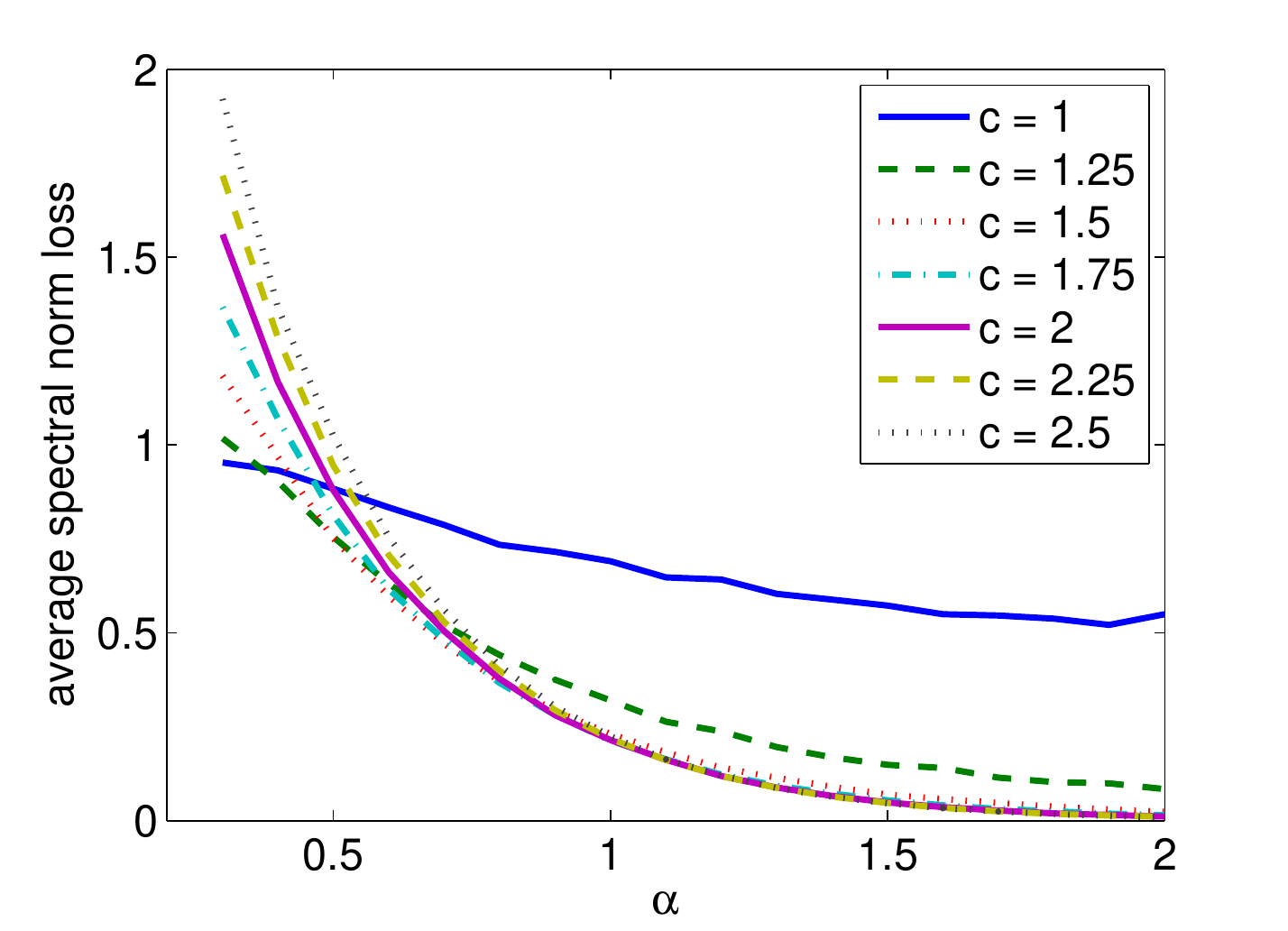} & \includegraphics[width =3.2in,height=2.0in]{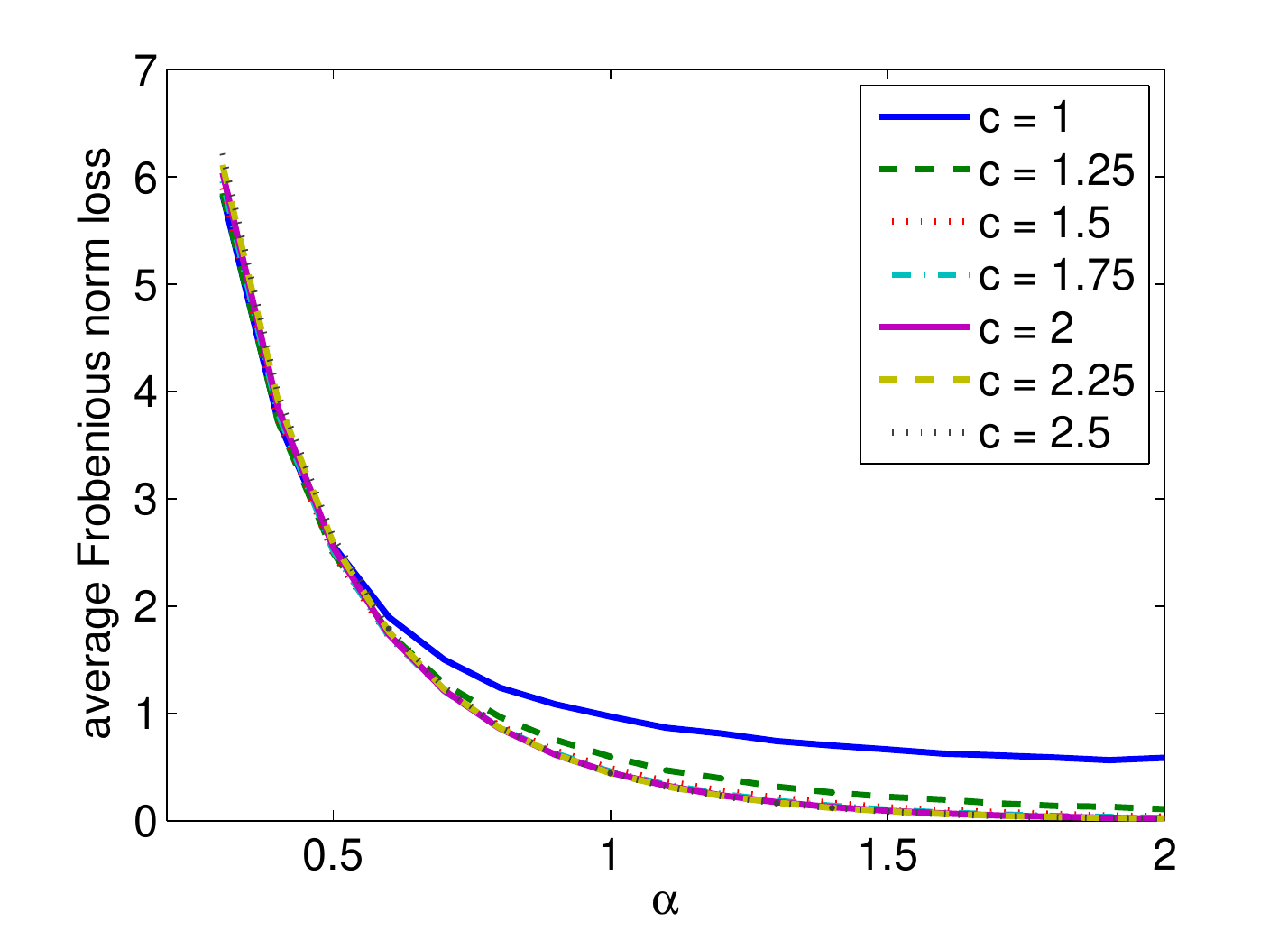}
\end{tabular}
\caption{Spectral norm  loss (left panel) and Frobenius norm loss  (right panel) as the thresholding constant $c$ varies. The singular values of $A$ are $\{j^{-\alpha}, j = 1, 2, ...\}$ with $\alpha$ varying from 0.3 to 2, $p_1 = p_2 =1000$, and $m_1 = m_2 = 50$.}\vspace{-.2in}
\label{fig:c_test}
\end{center}
\end{figure}

We now turn to the settings with the singular values being $\{j^{-\alpha}, \; j=1,2,..., \min(p_1,p_2)\}$ and various choices of $\alpha$, $p_1$ and $p_2$. Hence, no significant gap between adjacent singular values exists under these settings and we aim to demonstrate that our method continues to work well. We first consider $p_1=p_2 = 1000$, $m_1 = m_2 = 50$ and let $\alpha$ range from 0.3 to 2. Under this setting, we also study how the choice of thresholds affect the performance of our algorithm. For simplicity, we report results only for row thresholding as results for column thresholding are similar.  The average loss of $\hat A_{22}$ from Algorithm 2 with  $T_R \in \{c\sqrt{m_1/p_1}, c \in [1, 6]\}$ under both the spectral norm and Frobenius norm are given in Figure \ref{fig:c_test}. In general, the algorithm performs well provided that $\alpha$ is not too small and as expected, the average loss decreases with a higher decay rate in the singular values. This indicates that the existence of a significant gap between adjacent singular values is not necessary in practice, provided that the singular values decay sufficiently fast. When comparing the results across different choices of the threshold,  $c = 2$ as suggested in our theoretical analysis is indeed the optimal choice. Thus, in all subsequent numerical analysis, we fix $c = 2$.

\begin{figure}[htbp]
\begin{center}
\subfigure[Spectral norm loss; column thresholding]{\includegraphics[width =3.2in,height=2.0in]{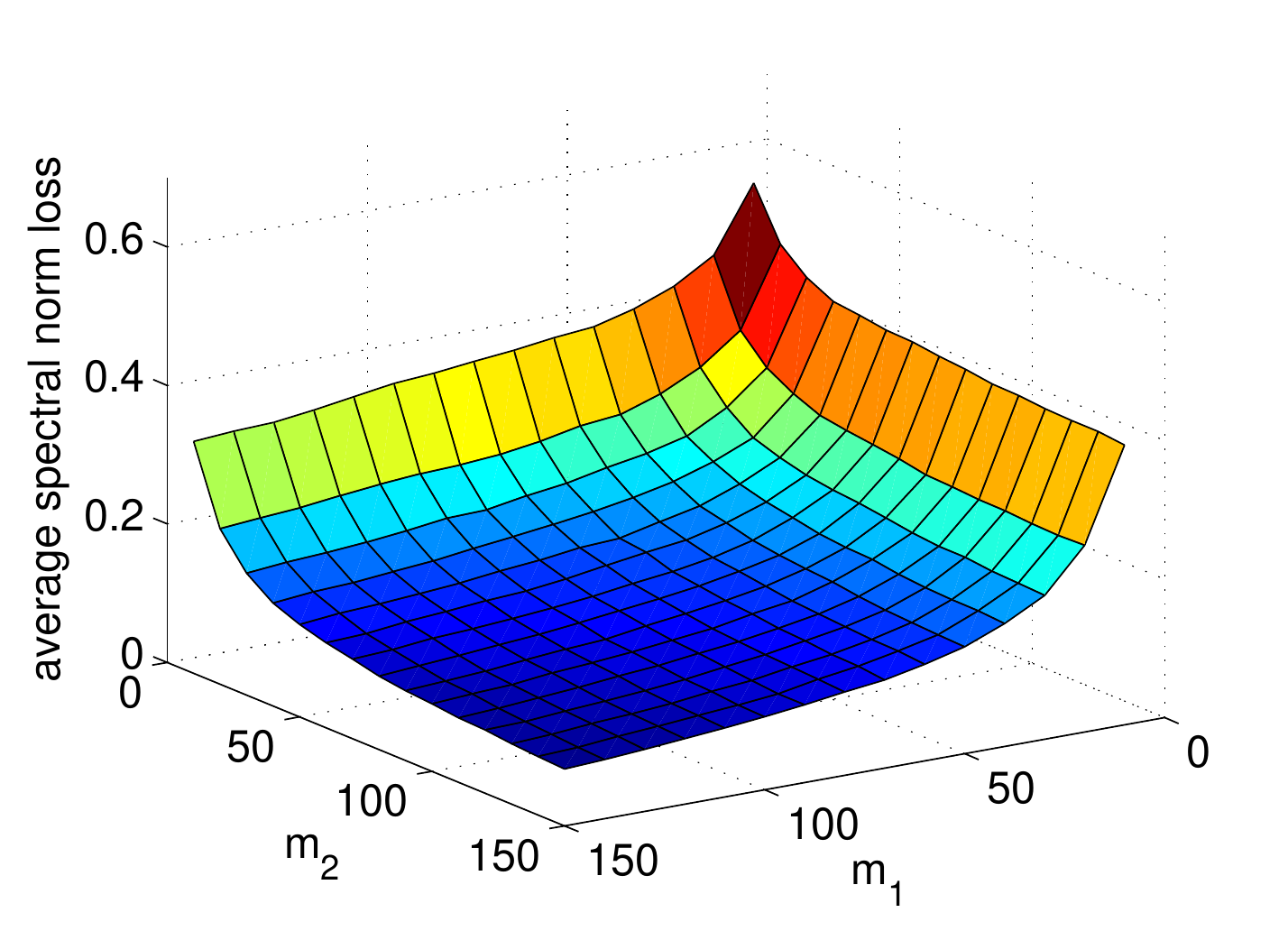}} 
\subfigure[Frobenius norm loss; column thresholding]{\includegraphics[width =3.2in,height=2.0in]{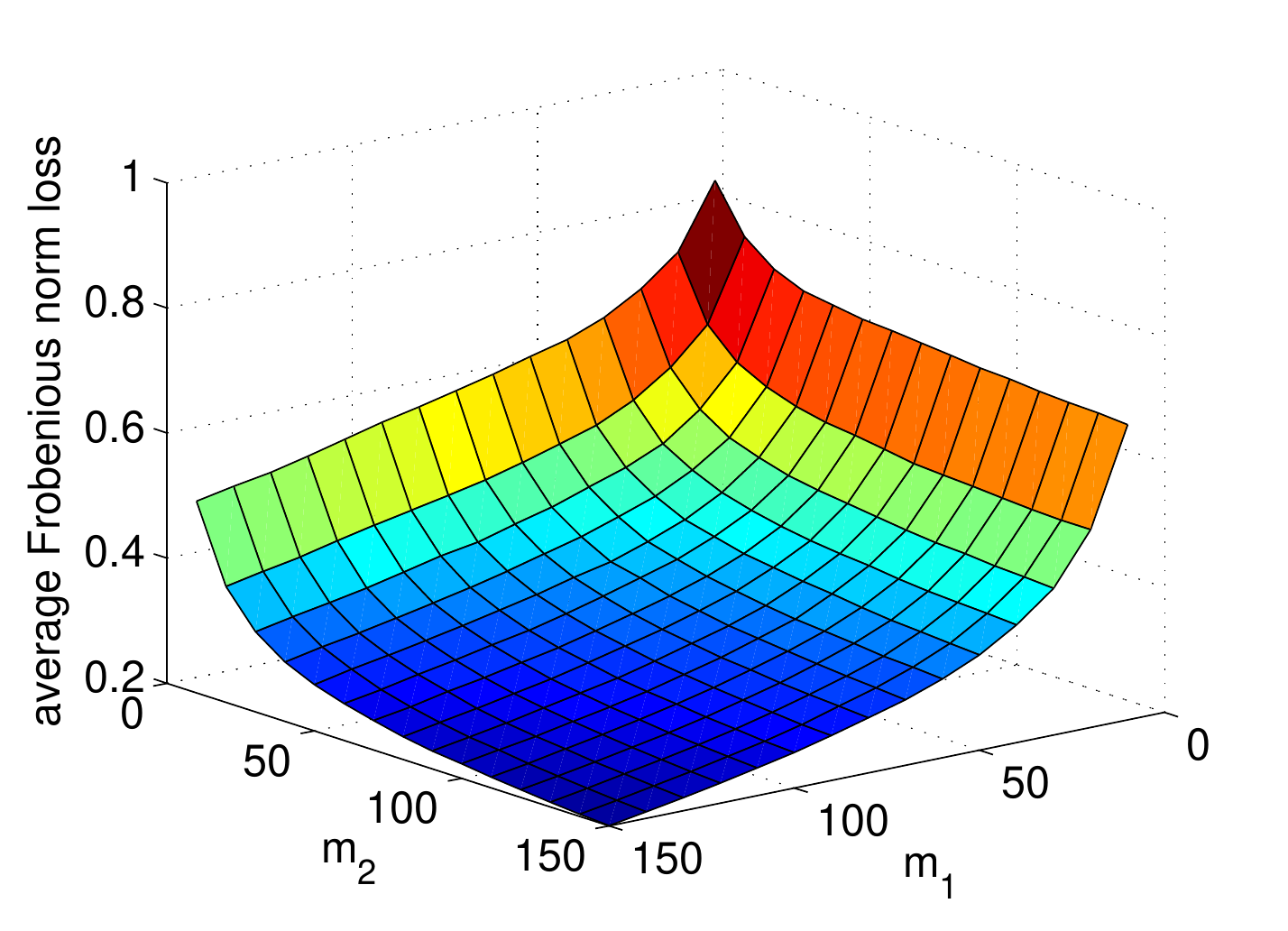}}\\
\subfigure[Spectral norm loss; row thresholding]{\includegraphics[width =3.2in,height=2.0in]{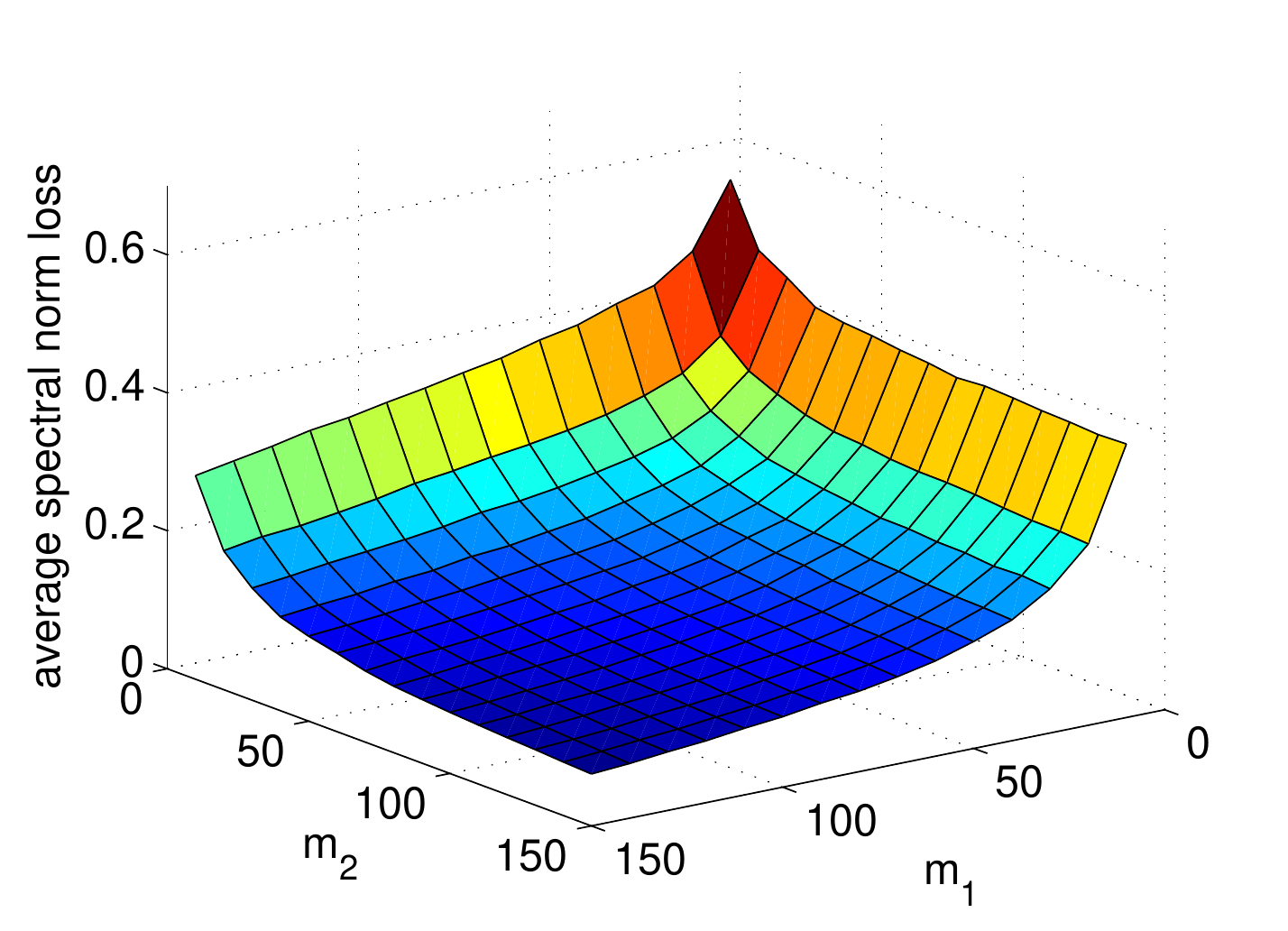}} 
\subfigure[Frobenius norm loss; row thresholding]{\includegraphics[width =3.2in,height=2.0in]{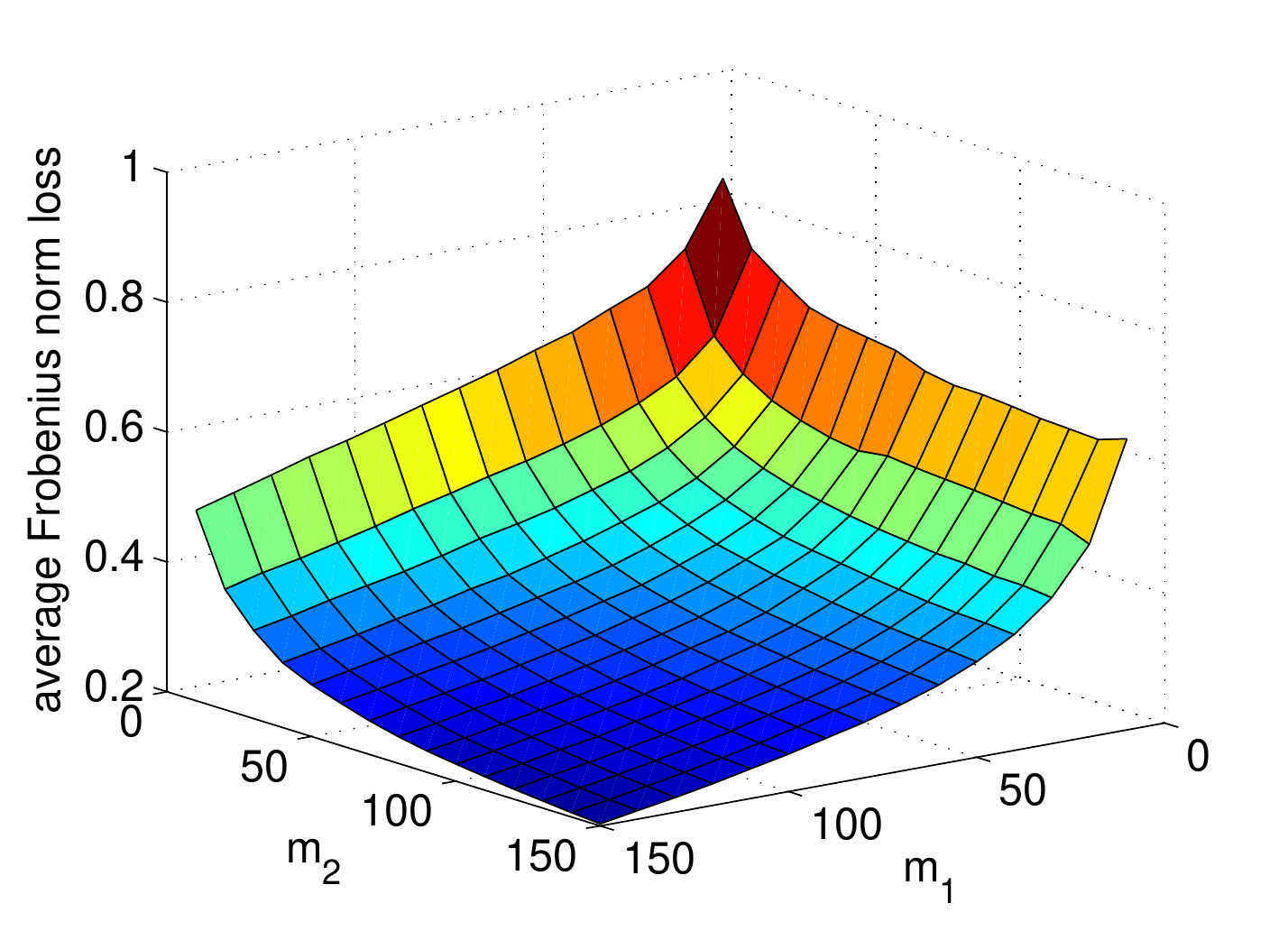}}
\caption{Spectral and Frobenius norm losses with column/row thresholding. The singular values of $A$ are $\{j^{-1}, j =1,2,...\}$, $p_1 = 300$, $p_2 =3000$, and $m_1$, $m_2=10, ...,150$.}\vspace{-.2in}
\label{fig:col_row_thresh}
\end{center}
\end{figure}

To investigate the impact of row versus column thresholding, we let the singular value decay rate be $\alpha = 1$, $p_1 = 300, p_2 = 3000$, and $m_1$ and $m_2$ varying from 10 to 150. The original matrix $A$ is generated the same way as before. We apply row and column thresholding with $T_R = 2\sqrt{p_1/m_1}$ and $T_C = 2\sqrt{p_2/m_2}$. It can be seen from Figure \ref{fig:col_row_thresh} that when the observed rows and columns are selected randomly, the results are not sensitive to the choice between row and column thresholding.

We next turn to the comparison between our proposed SMC algorithm and the penalized NNM method which recovers $A$ by (\ref{eq:PNNM}). 
The solution to  (\ref{eq:PNNM}) can be solved by the spectral regularization algorithm by \cite{mazumder} or the accelerated proximal gradient algorithm by \cite{Toh},  where these two methods provide similar results.
We use 5-fold cross-validation to select the tuning parameter $t$. 
Details on the implementation can be found in the Supplement. 

\begin{figure}[htbp]
\begin{center}
\subfigure[Spectral norm loss]{\includegraphics[width = .46\linewidth,height=.35\textwidth]{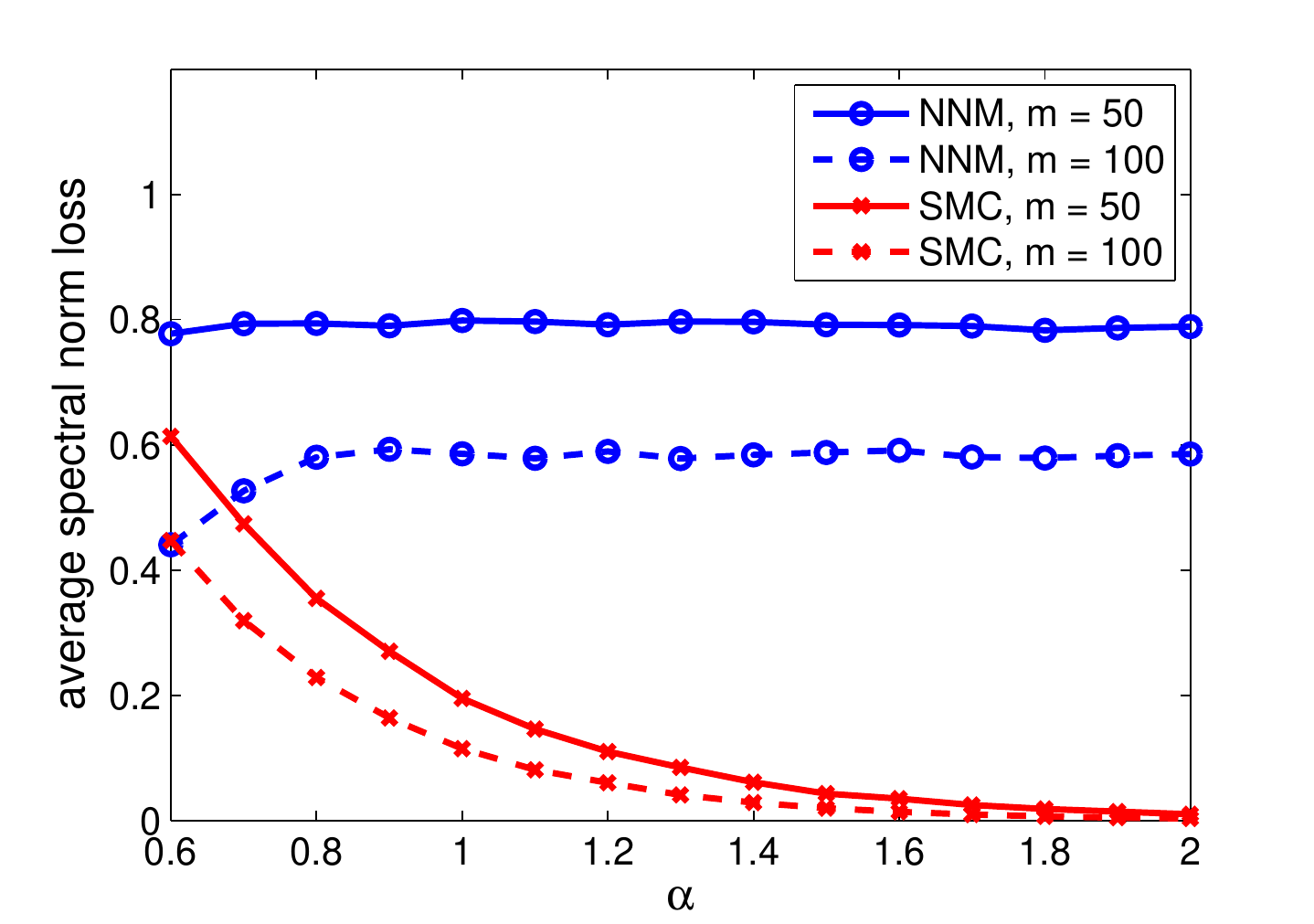}}
\subfigure[Frobenious norm loss]{\includegraphics[width = .46\linewidth,height=.35\textwidth]{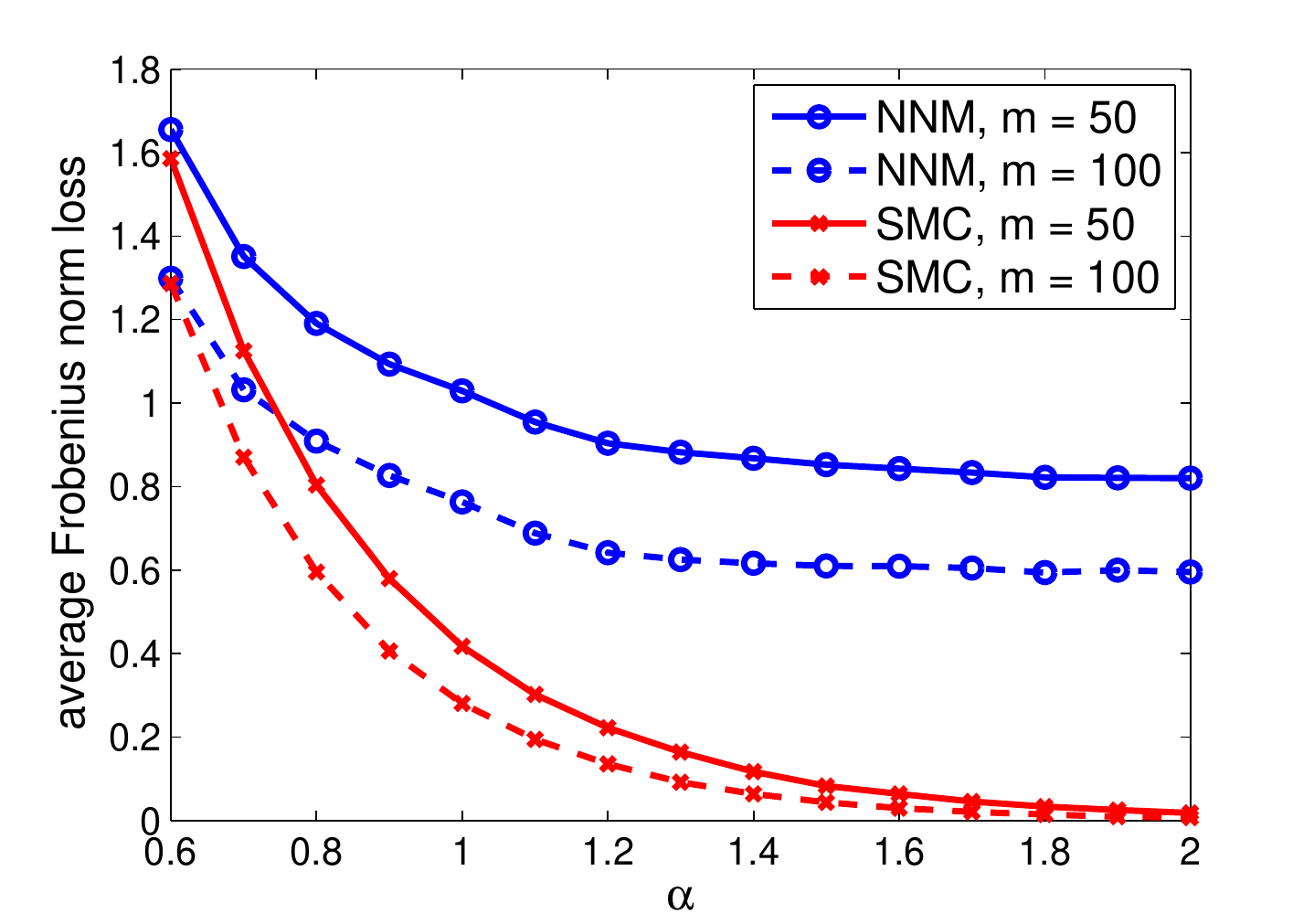}}
\caption{Comparison of  the proposed SMC method with the NNM method with 5-cross-validation for the settings with singular values of $A$ being $\{ j^{-\alpha}, j = 1, 2, ...\}$ for $\alpha$ ranging from 0.6 to 2, $p_1 = p_2 =500$, and $m_1 = m_2 = 50$ or $100$. }\vspace{-.2in}
\label{fig:compare_smc_nnm}
\end{center}
\end{figure}

We consider the setting where $p_1 = p_2 = 500$, $m_1 = m_2 = 50, 100$ and the singular value decay rate $\alpha$ ranges from 0.6 to 2. As shown in Figure \ref{fig:compare_smc_nnm}, 
the proposed SMC method substantially outperform the penalized NNM method with respect to both the spectral and Frobenius norm loss, especially as $\alpha$ increases.

Finally, we consider a simulation setting that mimics the ovarian cancer data application considered in the next section,  where $p_1=1148$, $p_2=1225$,  $m_1 = 230$, $m_2 = 426$ and the singular values of $A$ decay at a polynomial rate $\alpha$. Although the singular values of the full matrix are unknown, we estimate the decay rate based on the singular values of the  fully observed 552 rows of the matrix from the TCGA study, denoted by $\{\sigma_j, j = 1, ..., 522\}$.
A simple linear regression of $\{\log(\sigma_j), j = 1, ..., 522\}$ on $\{\log(j), j = 1, ..., 522\}$  estimates $\alpha$ as $0.8777$. 
In the simulation, we randomly generate $A\in \mathbb{R}^{p_1\times p_2}$ such that the singular values are fixed as $\{j^{-.8777}, j = 1, 2, \cdots\}$. For comparison, we also obtained results for $\alpha=1$ as well as those based on the penalized NNM method with 5-cross-validation. As shown in Table \ref{tb:relative_norm_loss}, the relative spectral  norm loss and relative Frobenius norm loss of the proposed method are reasonably small and substantially smaller than those from the penalized NNM method.

\begin{table}[htbp]
\begin{center}
\begin{tabular}{ccccc}
\hline
& \multicolumn{2}{c}{Relative spectral norm loss} & \multicolumn{2}{c}{Relative Frobenius norm loss} \\
\cmidrule(lr){2-3}\cmidrule(lr){4-5}
& SMC & NNM & SMC & NNM \\\hline
$\alpha = 0.8777$ & 0.1253 & 0.4614 
 & 0.2879 &  0.6122 
\\
$\alpha = 1 $ & 0.0732 & 0.4543 
& 0.1794 & 0.5671 
\\\hline
\end{tabular}
\end{center}
\caption{Relative spectral norm loss ($\|\hat A_{22}  - A_{22}\|/\|A_{22}\|$) and Frobenius norm loss ($\|\hat A_{22}  - A_{22}\|_F/\|A_{22}\|_F$) for $p_1=1148$, $p_2=1225$, $m_1=230$, $m_2 = 426$ and singular values of $A$ being $\{j^{-\alpha}: j=1, 2, \cdots\}$. 
}
\label{tb:relative_norm_loss}
\end{table}

\def\Gsc{\mathcal{G}}
\def\Ahat{\hat{A}}
\def\sPC{\mbox{\tiny PC}}
\def\PC{\mbox{PC}}
\def\bPC{\mbox{\bf PC}}
\def\smiRNA{\mbox{\tiny miRNA}}
\def\miRNAhat{\widehat{\mbox{miRNA}}}
\def\miRNA{\mbox{miRNA}}
\def\sTCGA{\mbox{\tiny TCGA}}
\def\sTOTH{\mbox{\tiny TOTH}}
\def\sDRES{\mbox{\tiny DRES}}
\def\sBONO{\mbox{\tiny BONO}}
\def\TCGAt{\mbox{TCGA}^{(t)} }
\def\TCGAv{\mbox{TCGA}^{(v)} }

\section{Application in Genomic Data Integration}
\label{application.sec}

In this section, we apply our proposed procedures to integrate multiple genomic studies of ovarian cancer (OC). 
OC is the fifth leading cause of cancer mortality among women, attributing to 14,000 deaths annually \citep{siegel2013cancer}.
OC is a relatively heterogeneous disease with 5-year survival rate varying substantially among different subgroups. The overall
5-year survival rate is near 90\% for stage I cancer. But the majority of the OC patients are diagnosed as stage III/IV diseases
and tend to develop resistance to chemotherapy, resulting a 5-year survival rate only about 30\% \citep{holschneider2000ovarian}.
On the other hand,  a small minority of advanced cancers are sensitive to chemotherapy and do not
replapse after treatment completion. Such a heterogeneity in disease progression is likely to be in part attributable to variations in
underlying biological characteristics of OC \citep{berchuck2005patterns}. This heterogeneity and the lack of successful treatment
strategies motivated multiple genomic studies of OC to identify molecular 
signatures that can distinguish OC subtypes, and in turn help to optimize and personalize treatment. For example, the Cancer Genome
Atlas (TCGA) comprehensively measured genomic and epigenetic abnormalities on high grade OC samples \citep{cancer2011integrated}. 
A gene expression risk score based on 193 genes, $\Gsc$, was trained on 230 training samples, denoted by $\TCGAt$, and shown as 
highly predictive of  OC survival when validated on the TCGA independent validation set of size 322, denoted by $\TCGAv$,  
as well as on several independent OC gene expression studies including those from  \cite{bonome2005expression} (BONO),  
\cite{dressman2007integrated}  (DRES) and \cite{tothill2008novel}  (TOTH). 

\begin{figure}[htbp]
	\begin{center}
		\caption{Imputation scheme for integrating multiple OC genomic studies.}
		\includegraphics[width = .95\linewidth]{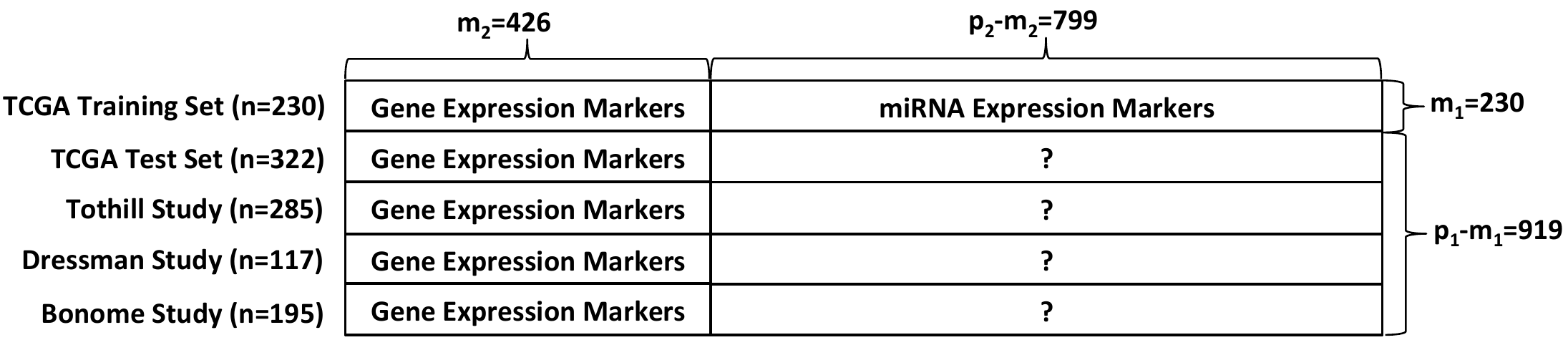} \vspace{-.2in}
		\label{fig-example}
	\end{center}
\end{figure}

The TCGA study also showed that 
clustering of miRNA levels  overlaps with gene-expression based clusters and is predictive of survival. 
It would be interesting to examine whether combining miRNA with $\Gsc$ could
improve survival prediction when compared to $\Gsc$ alone. One may use $\TCGAv$
to evaluate the added value of miRNA. However, $\TCGAv$ is of limited sample size. 
Furthermore, since miRNA was only measured for the TCGA study, its utility in prediction cannot
be directly validated using these independent studies. Here, we apply our proposed SMC method to impute 
the missing miRNA values and subsequently construct prediction rules based on both $\Gsc$ and the imputed
miRNA, denoted by $\miRNAhat$, for these independent validation sets. 
To facilitate the comparison with the analysis based on $\TCGAv$ alone where miRNA measurements
are observed, we only used the miRNA from $\TCGAt$ for imputation and reserved the miRNA data from
$\TCGAv$ for validation purposes. To improve the imputation, we also included additional 300 genes that were 
previously used in a prognostic gene expression signature for predicting ovarian cancer survival \citep{denkert2009prognostic}. 
This results in a total of $m_1=426$ unique gene expression variables available for imputation.
Detailed information on the data used for imputation is shown in Figure 
\ref{fig-example}. 
Prior to imputation, all gene expression and miRNA levels are log transformed and centered to have 
mean zero within each study to remove potential platform or batch effects. 
Since the observable rows (indexing subjects) can be viewed as random whereas the observable columns (indexing genes and miRNAs) 
are not random, we used row thresholding with threshold $T_R=2\sqrt{p_1/m_1}$  as suggested in the theoretical and simulation results.  
For comparison, we also imputed data using the penalized NNM method with tuning parameter $t$ selected via 5-fold cross-validation.

We first compared $\miRNAhat$ to the observed miRNA on $\TCGAv$. Our imputation yielded a rank 2 matrix for $\miRNAhat$ 
and the correlations between the two right  and left singular vectors $\miRNAhat$ to that of the observed miRNA
variables are .90, .71, .34, .14, substantially higher than that of those from the NNM method, with the corresponding values 
0.45, 0.06, 0.10, 0.05.
This suggests that the SMC imputation does a good job in recovering the leading projections of the miRNA measurements and outperforms the NNM method.

To evaluate the utility of $\miRNAhat$ for predicting OC survival, we used the $\TCGAt$ to select 117 miRNA markers that are marginally
associated with survival with a nominal $p$-value threshold of .05. We use the two leading 
principal components (PCs) of the 117 miRNA markers, $\miRNA^{\sPC} = (\miRNA^{\sPC}_1, \miRNA^{\sPC}_2)^T$, 
as predictors for the survival outcome in addition to $\Gsc$. The imputation enables us to integrate information
from 4 studies including $\TCGAt$, which could substantially improve efficiency and prediction performance. 
We first assessed the association between $\{\miRNA^{\sPC},\Gsc\}$ and OC survival by fitting
a stratified Cox model \citep{kalbfleisch2011statistical} to the integrated data
that combines $\TCGAv$ and the three additional studies via either the SMC or NNM methods. In addition, we fit
the Cox model to (i) $\TCGAv$ set alone with $\miRNA^{\sPC}$ obtained from the observed miRNA; and (ii)
each individual study separately with imputed $\miRNA^{\sPC}$. As shown in Table \ref{tab-example}(a), the  log hazard ratio (logHR) estimates for $\miRNA^{\sPC}$ from the integrated analysis, based on both SMC and NNM methods, are similar in magnitude to those obtained based on the observed miRNA values with $\TCGAv$.  However, the integrated analysis has substantially smaller standard error (SE) estimates due the increased sample sizes. The estimated logHRs are also reasonably consistent across studies when separate models were fit to individual studies. 

We also compared the prediction performance of the model based on 
$\Gsc$ alone to the model that includes both $\Gsc$ and  the imputed $\miRNA^{\sPC}$. 
Combining information from all 4 studies via standard meta analysis, the average improvement in C-statistic was $0.032$ (SE = $0.013$) for the SMC method and 
$0.001$ (SE = $0.009$)
for the NNM method, suggesting that the imputed $\miRNA^{\sPC}$ from the SMC method has much higher predictive value compared to
those obtained from the NNM method.

\begin{table}[ht]
	\centering
	\caption{Shown in (a) are the estimates of the log hazard ratio (logHR) along with their corresponding standard errors (SE) and $p$-values by fitting stratified Cox model integrating information from 4 independent studies with imputed miRNA based on the SMC method and the nuclear norm minimization (NNM); and Cox model to the TCGA test data with original observed miRNA (Ori.). Shown also are the estimates for each individual studies by fitting separate Cox models with imputed miRNA. }\label{tab-example}
	
	
	\vspace{.05in}
	
	\centerline{(a) Integrated Analysis with Imputed miRNA vs Single study with observed miRNA}\vspace{.05in}
	\begin{tabular}{r|rrr|rrr|rrr}  \hline
		& \multicolumn{3}{c|}{logHR} & \multicolumn{3}{c|}{SE} & \multicolumn{3}{c}{$p$-value} \\  
		& Ori. & SMC & NNM & Ori. & SMC & NNM & Ori. & SMC & NNM  \\ \hline
		$\Gsc$ & .067 & .143 & .168 & .041 & .034 & .028 & .104 & .000 & .000 \\ 
		$\miRNA^{\sPC}_1$  & -.012 & -.019 & -.013 & .009 & .006 & .012 & .218 & .001 & .283 \\ 
		$\miRNA^{\sPC}_2$  & .023 & .018 & -.005 & .014 & .009 & .014 & .092 & .039 & .725 \\ 
		\hline
	\end{tabular}\vspace{.3in}
	
	\centerline{(b) Estimates for Individual Studies with Imputed miRNA from the SMC method}\vspace{.05in}
	\begin{tabular}{r|rrrr|rrrr|rrrr}  \hline
		& \multicolumn{4}{c|}{logHR} & \multicolumn{4}{c|}{SE} & \multicolumn{4}{c}{$p$-value} \\
		& ${\sTCGA}$ & $\sTOTH$ & $\sDRES$ & $\sBONO$  & ${\sTCGA}$ & $\sTOTH$ & $\sDRES$ & $\sBONO$  & ${\sTCGA}$ & $\sTOTH$ & $\sDRES$ & $\sBONO$ \\  \hline
		$\Gsc$ & .051 & .377 & .174 & .311 & .048 & .069 & .132 & .117 & .286 & .000 & .187 & .008 \\ 
		$\miRNA^{\sPC}_1$& -.014 & -.021 & -.031 & -.010 & .011 & .012 & .014 & .014 & .207 & .082 & .030 & .484 \\ 
		$\miRNA^{\sPC}_2$ & .014 & .045 & -.021 & .036 & .016 & .018 & .022 & .019 & .391 & .009 & .336 & .054 \\ 
		\hline
	\end{tabular}\vspace{.3in}
	
	\centerline{(c) Estimates for Individual Studies with Imputed miRNA from the NNM method}\vspace{.05in}
	\begin{tabular}{r|rrrr|rrrr|rrrr}  \hline
		& \multicolumn{4}{c|}{logHR} & \multicolumn{4}{c|}{SE} & \multicolumn{4}{c}{$p$-value} \\
		& ${\sTCGA}$ & $\sTOTH$ & $\sDRES$ & $\sBONO$  & ${\sTCGA}$ & $\sTOTH$ & $\sDRES$ & $\sBONO$  & ${\sTCGA}$ & $\sTOTH$ & $\sDRES$ & $\sBONO$ \\  \hline
		$\Gsc$ & .082 & .405 & .361 & .258 & .037 & .066 & .114 & .088 & .028 & .000 & .002 & .003 \\ 
		$\miRNA^{\sPC}_1$ & -.045 & .016 & .055 & -.008 & .021 & .026 & .031 & .023 & .034 & .544 & .076 & .721 \\ 
		$\miRNA^{\sPC}_2$ & .008 & -.086 & -.043 & .019 & .026 & .027 & .034 & .029 & .758 & .002 & .201 & .496 \\ 
		\hline
	\end{tabular}
	
\end{table}

In summary, the results shown above suggest that our SMC procedure accurately recovers the leading PCs of the miRNA variables. In addition,
adding $\miRNA^{\sPC}$ obtained from imputation using the proposed SMC method could significantly improve the prediction performance, which confirms the value of our  method
for integrative genomic analysis. When comparing to the NNM method, the proposed SMC method produces summaries of miRNA that is more correlated with the truth and yields leading PCs that are more predictive of OC survival.

\clearpage

\section{Discussions}
\label{discussion.sec}

The present paper introduced a new framework of SMC where a subset of the rows and columns of an approximately low-rank matrix are observed. We proposed an SMC method for the recovery of the whole matrix with theoretical guarantees. The proposed procedure significantly outperforms the conventional NNM method for matrix completion, which does not take into account the special structure of the observations. As shown by our theoretical and numerical analyses,  the widely adopted NNM methods for matrix completion are not suitable for the SMC setting. These NNM methods perform particularly poorly when a small number of rows and columns are observed.

The key assumption in matrix completion is the matrix being approximately low rank. This is reasonable in the ovarian cancer application since as indicated in the results from the TCGA study (Cancer Genome Atlas Research Network, 2011), the patterns observed in the miRNA signature are highly correlated with the patterns observed in the gene expression signature. This suggests the high correlation among the selected gene expression  and miRNA variables.  Results from the imputation based on the approximate low rank assumption given in Section \ref{application.sec} are also encouraging with promising correlations with true signals and  good prediction performance from the imputed miRNA signatures. We expect that this imputation method will also work well in genotyping and sequencing applications, particularly for regions with reasonably high linkage disequilibrium. 

Another main assumption that is needed in the theoretical analysis is that there is a significant gap between the $r^{th}$ and $(r+1)^{th}$ singular values of $A$. This assumption may not be valid in real practice. In particular, the singular values of the ovarian dataset analyzed in Section \ref{application.sec} is decreasing smoothly without a significant gap. 
However, it has been shown in the simulation studies presented in Section \ref{simulation.sec} that, although there is no significant gap between any adjacent singular values of the matrix to be recovered, the proposed SMC method works well as long as the singular values  decay sufficiently fast. Theoretical analysis for the proposed SMC method under more general patterns of singular value decay warrants future research.

To implement the proposed Algorithm 2, major decisions include the choice of threshold values and choosing between column thresholding and row thresholding.  Based on both theoretical and numerical studies, optimal threshold values can be set as $T_C = 2\sqrt{p_2/m_2}$ for column thresholding and $T_R = 2 \sqrt{p_1/m_1}$ for row thresholding.  Simulation results in Section \ref{simulation.sec} show that when both rows and columns are randomly chosen, the results are very similar. In the real data applications, the choice between row thresholding and column thresholding depends on whether the rows or columns are more ``homogeneous", or closer to being randomly sampled. For example, in the ovarian cancer dataset analyzed in Section \ref{application.sec}, the rows correspond to the patients and the columns correspond to the gene expression levels and miRNA levels. Thus the rows are closer to random sample than the columns, consequently it is more natural to use the row thresholding in this case. 

We have shown both theoretically and numerically in Sections \ref{analysis.sec} and \ref{simulation.sec}  that Algorithm \ref{al:without_r} provides a good recovery of $A_{22}$. However, the naive implementation of this algorithm requires $\min(m_1, m_2)$ matrix inversions and multiplication operations in the for loop that calculates $\|D_{R, s}\|$ (or $\|D_{C, s}\|$), $s \in\{\hat r, \hat r + 1, \cdots, \min(m_1, m_2)\}$. Taking into account the relationship among $D_{R, s}$ (or $D_{C, s}$) for different $s$'s, it is possible to simultaneously calculate all $\|D_{R, s}\|$ (or $\|D_{C, s}\|$) and accelerate the computations. For reasons of space, we leave optimal implementation of Algorithm 2 as future work.

\subsection*{Acknowledgments}

We thank the Editor,  Associate Editor and referee for their detailed and constructive comments
which have helped to improve the presentation of the paper.



%
%
%

\renewcommand{\baselinestretch}{1.6}
\bibliographystyle{apa}

\newpage

\setcounter{page}{1}
\setcounter{section}{0}

\begin{center}
	{\LARGE Supplement to ``Structured Matrix Completion With}
	
	\bigskip	
	{\LARGE  Applications to Genomic Data Integration"	
		\footnote{Tianxi Cai is Professor of Biostatistics, Department of Biostatistics, Harvard School of Public Health, Harvard University, Boston, MA (E-mail: tcai@hsph.harvard.edu); T. Tony Cai is Dorothy Silberberg Professor of Statistics, Department of Statistics, The Wharton School, University of Pennsylvania, Philadelphia, PA (E-mail: tcai@wharton.upenn.edu); Anru Zhang is a Ph.D. student, Department of Statistics, The Wharton School, University of Pennsylvania, Philadelphia, PA (E-mail: anrzhang@wharton.upenn.edu). The research of Tianxi Cai was supported in part by  NIH Grants R01 GM079330 and U54 LM008748; the research of Tony Cai and Anru Zhang  was supported in part by NSF Grant DMS-1208982 and NIH Grant R01 CA127334. }}
	
	\bigskip\medskip
	{Tianxi Cai, ~~ T. Tony Cai ~ and~ Anru Zhang}
\end{center}

\begin{abstract}
	In this supplement we  provide additional simulation results and the proofs of the main theorems. Some key technical tools used in the proofs of the main results are also developed and proved.

\end{abstract}

\section{Additional Simulation Results}

\begin{figure}[htbp]
	\begin{center}
		\subfigure[Spectral norm loss]{\includegraphics[width = .45\linewidth]{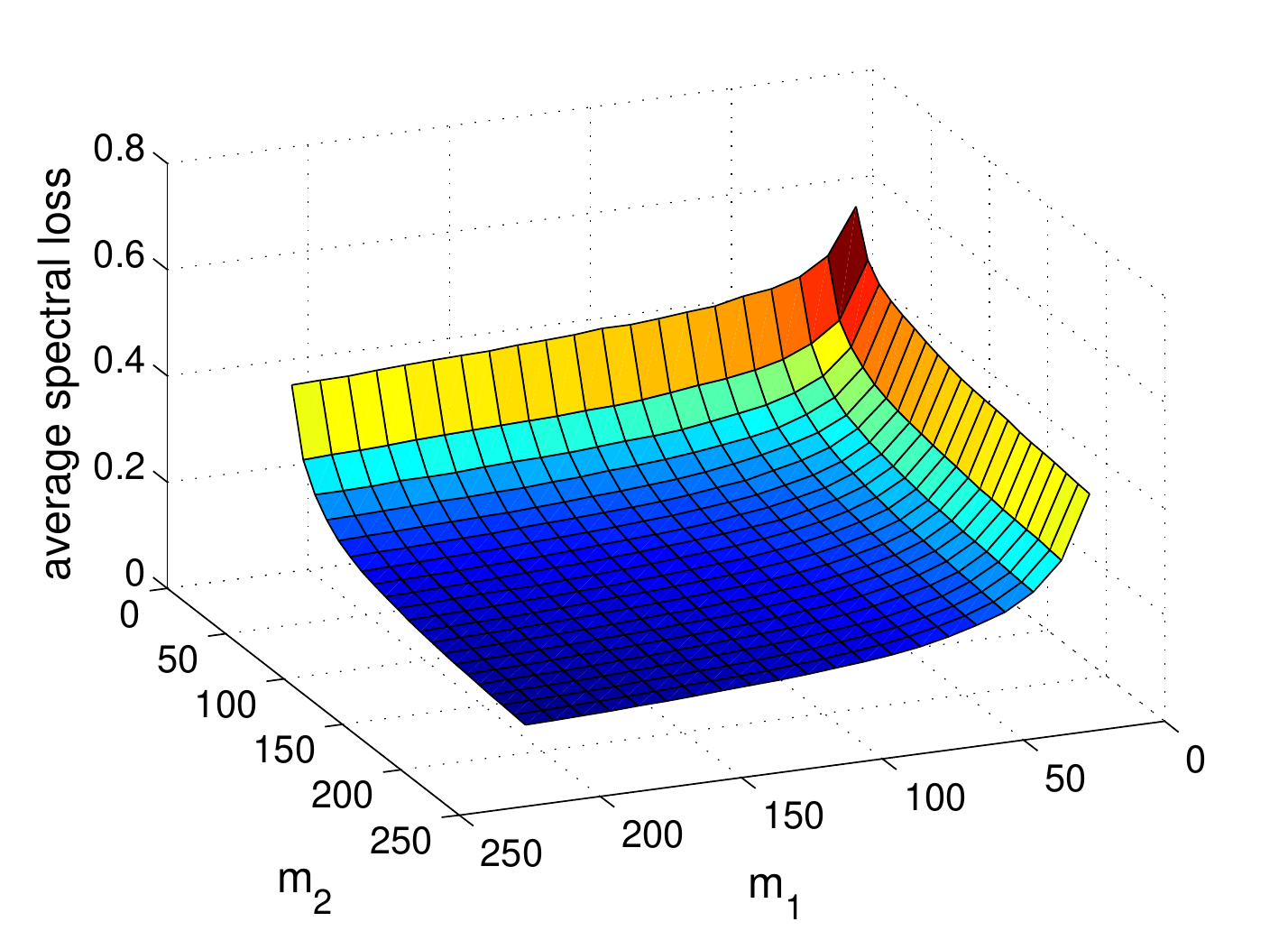}}
		\subfigure[Frobenious norm loss]{\includegraphics[width = .45\linewidth]{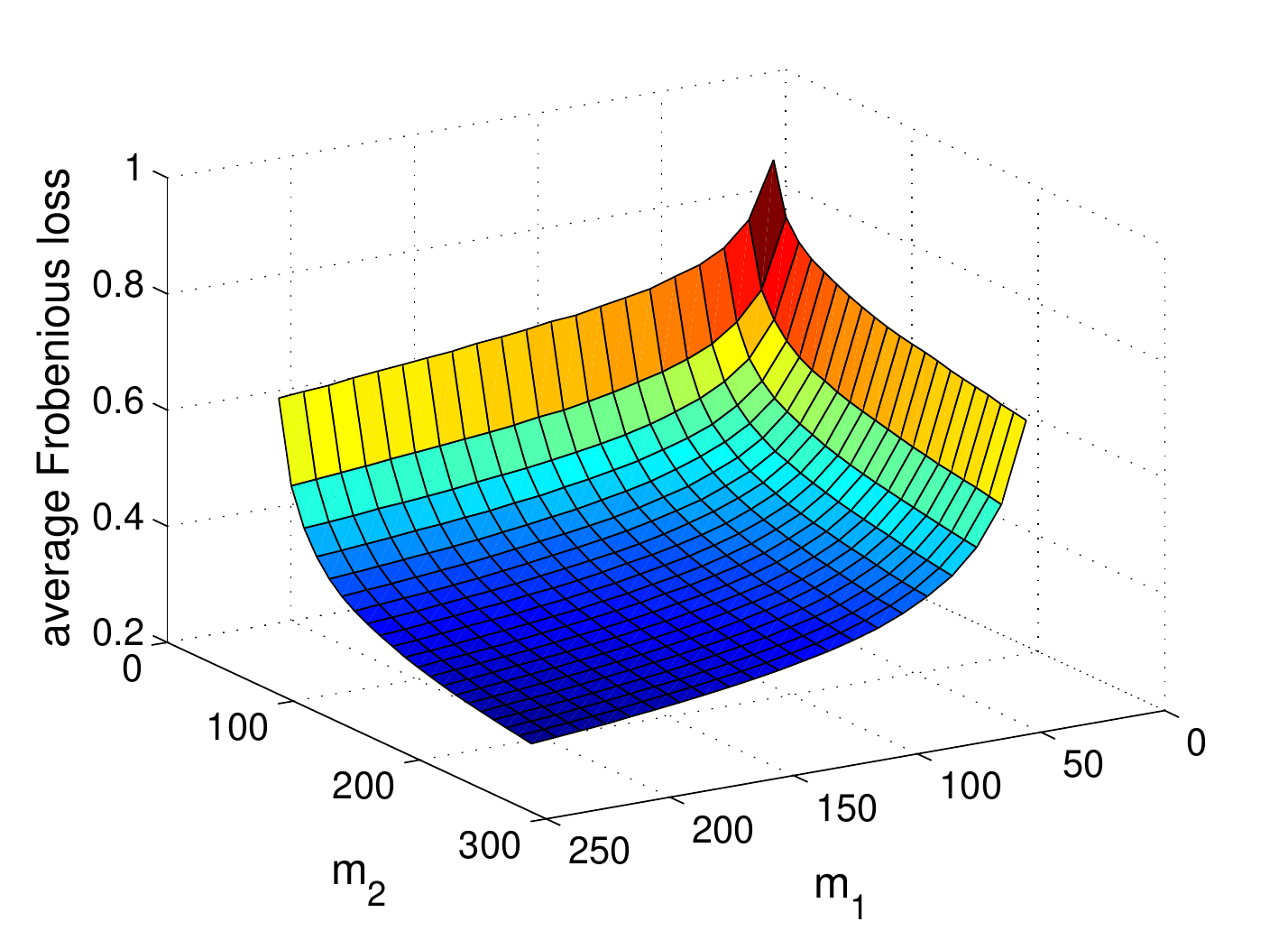}}
		\caption{Losses for the settings with singular values of $A$ being $\{j^{-1}, j = 1, 2, ...\}$, $p_1 = p_2 =1000$, $m_1, m_2 = 10, ..., 210$.}\vspace{-.2in}
	\end{center}
	\label{fig:m_varies}
\end{figure}

We consider the effect of the number of the observed rows and columns on the estimation accuracy. We let $p_1 = p_2 = 1000$, let the singular values of $A$ be $\{j^{-1}, j = 1, 2, ...\}$ and let $m_1$ and $m_2$ vary from $10$ to $210$. The singular spaces $U$ and $V$ are again generated randomly from the Haar measure. The estimation errors of $\hat{A}_{22}$ from Algorithm 2 with row thresholding and $T_R = 2\sqrt{p_1/m_1}$ over different choices of $m_1$ and $m_2$ are shown in Figure \ref{fig:m_varies}.
As expected, the average loss decreases as $m_1$ or $m_2$ grows. Another interesting fact is that the average loss is approximately symmetric with respect to $m_1$ and $m_2$. This implies that even with different numbers of observed rows and columns, Algorithm 2 has similar performance with row thresholding or column thresholding. 

\begin{figure}[htbp]
	\begin{center}
		\subfigure[Spectral norm loss]{\includegraphics[width = .46\linewidth,height=.35\textwidth]{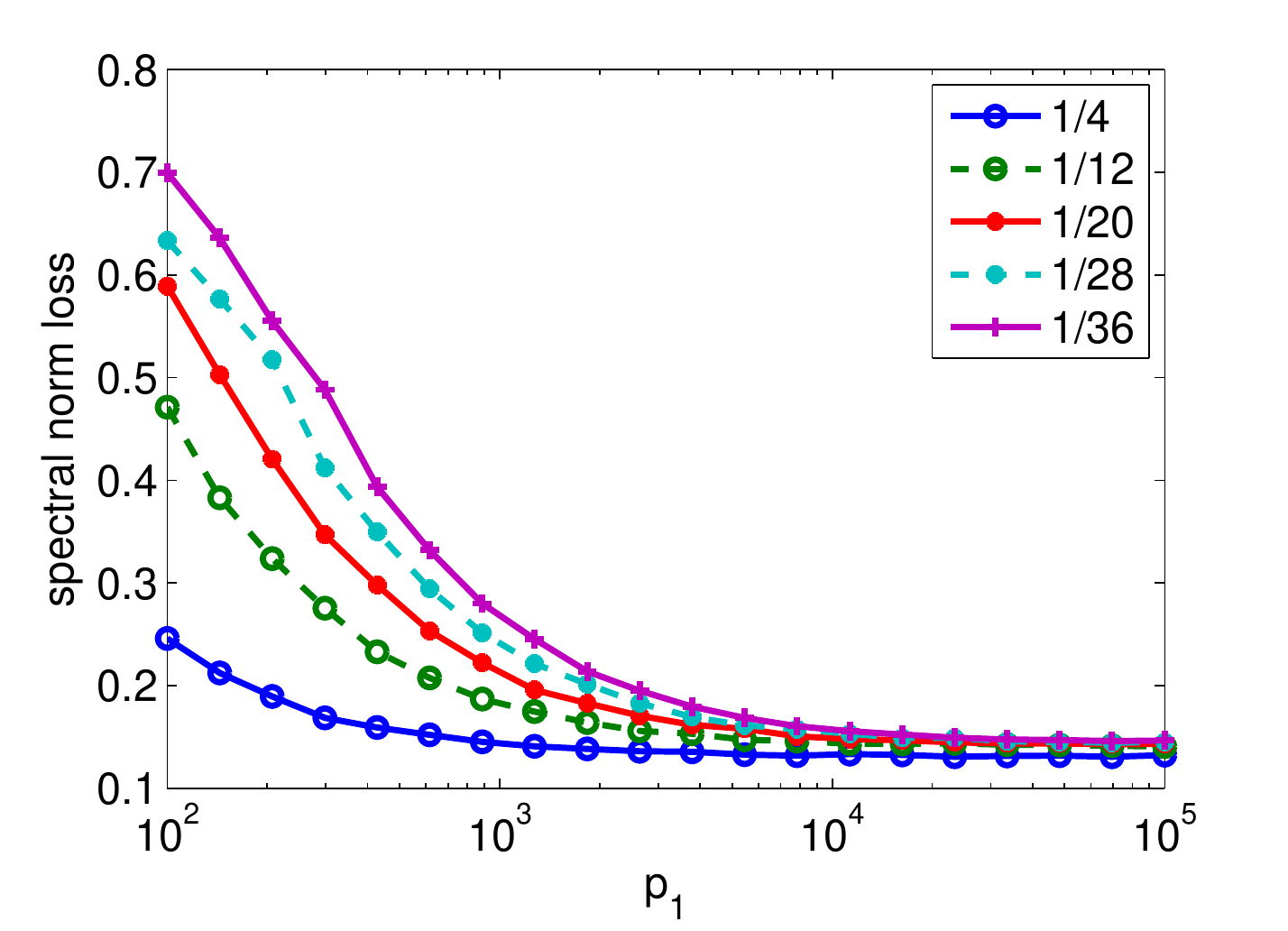}}
		\subfigure[Frobenious norm loss]{\includegraphics[width = .46\linewidth,height=.35\textwidth]{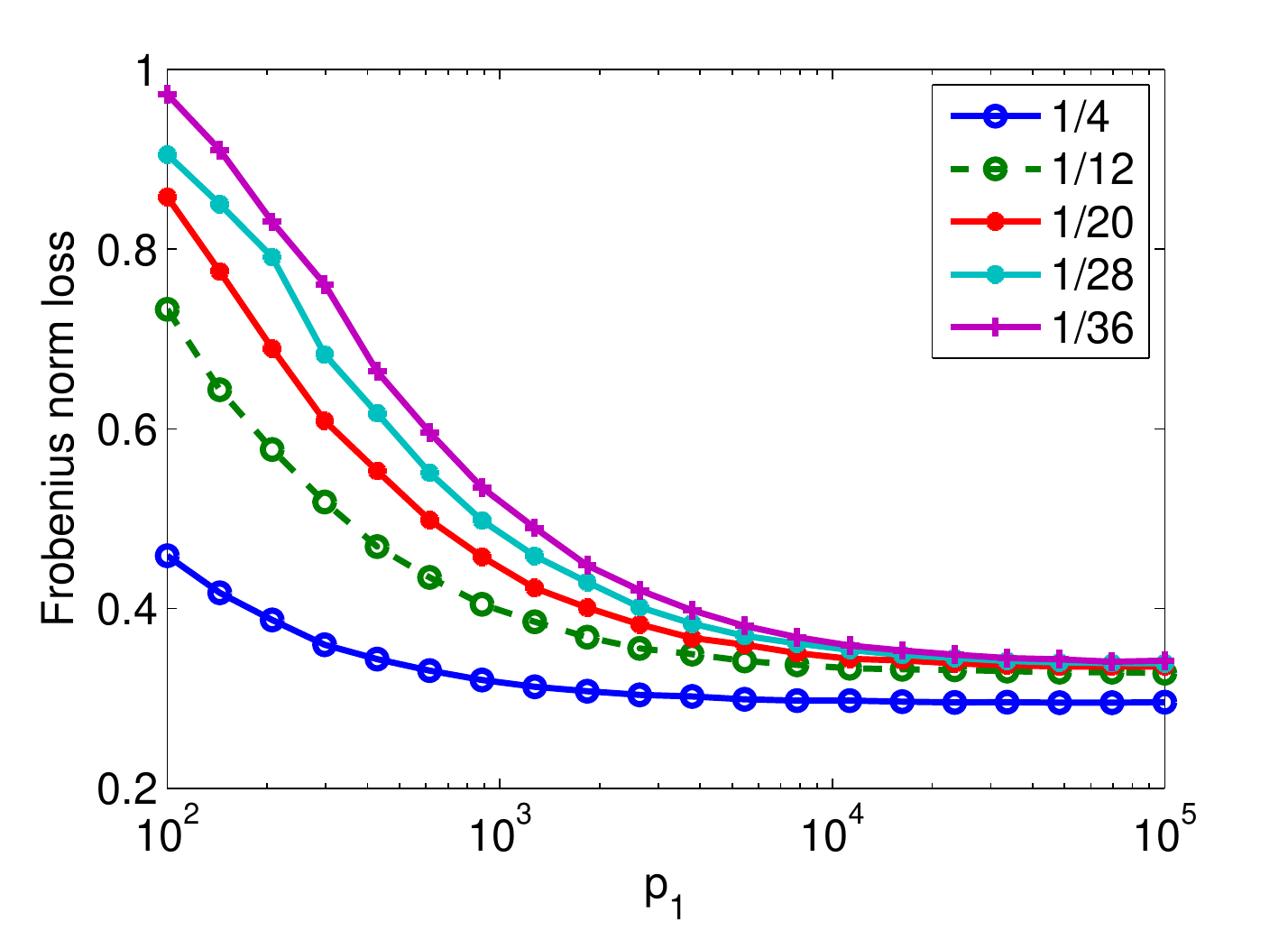}}
		\caption{Losses for settings with singular values of $A$ being $\{j^{-1}, j=1,2, 3...\}$, $p_2 =1000$, $m_2 = 50$, $m_1/p_1 = 1/4, 1/12, 1/20, 1/28, 1/36$, and $p_1 = 100, ...,100,000$.}\vspace{-.2in}
		\label{fig:m_ratio}
	\end{center}
\end{figure}

We are also interested in the performance of Algorithm 2 as $p_1$ and the ratio $m_1/p_1$ vary. To this end, we consider the  setting where $p_2 = 1000$,  $m_2 = 50$, and the singular values of $A$ are chosen as $\{j^{-1}, j = 1, 2, ...\}$.  The results are shown in Figure \ref{fig:m_ratio}. It can be seen that when $m_1/p_1$ increases, the recovery is generally more accurate; when $m_1/p_1$ is kept as a constant, the average loss does decrease but not converge to zero as $p_1$ increases. 

\section{Technical Tools}

We collect important technical tools in this section. The first lemma is about the inequalities of singular values in the perturbed matrix.
\begin{Lemma}\label{lm:X_Y_sv}
	Suppose $X\in \mathbb{R}^{p\times n}$, $Y\in \mathbb{R}^{p\times n}$, $rank(X) = a$, $rank(Y) = b$,
	\begin{enumerate}
		\item $\sigma_{a+b+1-r}(X+Y) \leq \min(\sigma_{a+1-r}(X), \sigma_{b+1-r}(Y))$ for $r\geq 1$;
		\item if we further have $X^{\intercal}Y = 0$, we must have $a+b\leq n$, $\sigma_r(X+Y) \geq \max(\sigma_r(X), \sigma_r(Y))$ for $r\geq 1$.
	\end{enumerate}
\end{Lemma}

\begin{Lemma}\label{lm:Schatten_q}
	Suppose $X\in \mathbb{R}^{p\times n}, Y\in\mathbb{R}^{n\times m}$ are two arbitrary matrices, denote $\|\cdot\|_q$, $\|\cdot\|$ as the Schatten-$q$ norm and spectral norm respectively, then we have
	\begin{equation}
	\|XY\|_q \leq \|X\|_q\cdot \|Y\|.
	\end{equation}
\end{Lemma}

The following two lemmas provide examples that illustrate NNM fails to recover $\hat A_{22}$.

\begin{Lemma}\label{lm:nuclear_norm_fail}
	Assume $A = B_1B_2^T$, where $B_1\in \mathbb{R}^{p_1\times r}$ and $B_2\in \mathbb{R}^{p_2\times r}$ are two i.i.d. standard Gaussian matrices. Let $A$ is divided into blocks as \eqref{eq:A_block}. Suppose
	\begin{equation}\label{ineq:condition_nuclear}
	r \leq \frac{1}{400} \min(p_1, p_2), \quad m_1 \leq \frac{1}{25}p_1, \quad m_2\leq \frac{1}{25}p_2,
	\end{equation}
	then the NNM \eqref{eq:nuclear_norm_minimization} fails to recover $A_{22}$ with probability at least $1 - 12\exp(-\min(p_1, p_2)/400)$.
\end{Lemma}

\begin{Lemma}\label{lm:nuclear_norm_1}
	Denote $1_{p}$ as the $p$-dimensional vector with all entries 1. Suppose $A = 1_{p_1}\cdot 1_{p_2}^{\intercal}$, and $A$ is divided into blocks as \eqref{eq:A_block}. Then the NNM \eqref{eq:nuclear_norm_minimization} yields
	$$\hat A_{22} = \min\left\{\sqrt{\frac{m_1m_2}{(p_1 - m_1)(p_2-m_2)}}, 1\right\}1_{p_1 - m_1} 1_{p_2 - m_2}^{\intercal}. $$
\end{Lemma}

The following result is on the norm of  a random submatrix of a given orthonormal matrix.
\begin{Lemma}\label{lm:U_Omega}
	Suppose $U\in\mathbb{R}^{p\times d}$ is a fixed matrix with orthonormal columns (hence $d\leq p$). Denote $W = \max_{1\leq i\leq p} \frac{p}{d}\cdot \sum_{j=1}^d u_{ij}^2$. Suppose we uniform randomly draw $n$ rows (with or without replacement) from $U$ and note the index as $\Omega$ and denote 
	$$U_{\Omega} = \begin{bmatrix}
	U_{\Omega(1)}\\
	\vdots\\
	U_{\Omega(n)}
	\end{bmatrix}. $$ When $ n \geq \frac{4Wd \left(\log d + c\right)}{(1-\alpha)^2}$ for some $0<\alpha<1$ and $c>1$, we have 
	$$\|\sigma_{\min}(U_{\Omega})\| \geq \sqrt{\frac{\alpha n}{p}} $$
	with probability $1 - 2e^{-c}$.
\end{Lemma}

The following results is about the spectral norm of the submatrix of a random orthonormal matrix.

\begin{Lemma}\label{lm:U_Haar}
	Suppose $U\in\mathbb{R}^{p\times d}$ ($d\leq p$) is with random orthonormal columns with Haar measure. For all $0<\alpha_1 < 1 < \alpha_2$, there exists constant $C, \delta>0$ depending only on $\alpha_1, \alpha_2$ such that when $p\geq n\geq \min\{Cd, p\}$, we have
	\begin{equation}\label{eq:lm_Haar}
	\sqrt{\frac{\alpha_1 n}{p}} \leq \sigma_{\min}(U_{[1:n, :]}) \leq \|U_{[1:n, :]}\| \leq \sqrt{\frac{\alpha_2 n}{p}}
	\end{equation}
	with probability at least $1 - \exp(-\delta n)$.
\end{Lemma}

\subsection*{Proof of  the Technical Lemmas}

{\noindent\bf Proof of Lemma \ref{lm:X_Y_sv}.} 
\begin{enumerate}
	\item First, by a well-known fact about best low-rank approximation,
	$$\sigma_{a+b+1-r} (X+Y) = \min_{M\in \mathbb{R}^{p\times n}, \rank(M) \leq a+b-r }\|X+Y - M\|. $$
	Hence,
	$$\sigma_{a+b+1-r}(X+Y) \leq \|X+Y - (X_{\max(a-r)} + Y)\| = \|X_{-\max(a-r)}\| = \sigma_{a+1-r}(X); $$
	similarly $\sigma_{a+b+1-r}(X+Y) \leq \sigma_{b+1-r}(Y)$. 
	
	\item When we further have $X^{\intercal}Y=0$, we know the column space of $X$ and $Y$ are orthogonal, then we have $\rank(X+Y) = \rank(X) + \rank(Y) = a+b$, which means $a+b \leq n$. Next, note that
	$$(X+Y)^{\intercal}(X+Y) = X^{\intercal}X+Y^{\intercal}Y + X^{\intercal}Y + Y^{\intercal}X = X^{\intercal}X + Y^{\intercal}Y,$$
	if we note $\lambda_r(\cdot)$ as the $r$-th largest eigenvalue of the matrix, then we have 
	\begin{equation*}
	\begin{split}
	\sigma^2_r(X+Y) = & \lambda_r((X+Y)^{\intercal}(X+Y)) = \lambda_r(X^{\intercal}X + Y^{\intercal}Y)\\
	\geq & \max(\lambda_r(X^{\intercal}X), \lambda_r(Y^{\intercal}Y)) = \max(\sigma_r^2(X), \sigma_r^2(Y)).
	\end{split}
	\end{equation*}
\end{enumerate}\quad $\square$

{\noindent\bf Proof of Lemma \ref{lm:Schatten_q}.} Since 
$$\|XY\|_q = \sqrt[q]{\sum_{i}\sigma^q_i(XY)}, \quad \|X\|_q = \sqrt[q]{\sum_i\sigma_i^q(X)}, $$
it suffices to show $\sigma_i(XY) \leq \sigma_i(X)\|Y\|$. To this end, we have
$$\sigma_i(X) = \min_{M\in \mathbb{R}^{p\times m}, \rank(M)\leq i-1} \|XY-M\| \leq \|XY - X_{\max(i-1)}Y\| = \|X_{-\max(i-1)}Y\| \leq \sigma_{i}(X)\|Y\|, $$
which finishes the proof of this lemma.\quad $\square$

\medskip
{\noindent\bf Proof of Lemma \ref{lm:nuclear_norm_fail}.}  Since $B_1$ and $B_2$ and their submatrices are all i.i.d. standard matrices, by the random matrix theory (Corollary 5.35 in \cite{Vershynin}), for $t>0$, we have with probability at least $1 - 12\exp(-t^2/2)$, the following inequalities hold,
\begin{equation}
\begin{split}
\lambda_r(A) \geq & \lambda_{\min} (B_1)\lambda_{\min}(B_2) \geq (\sqrt{p_1} - \sqrt{r} - t)(\sqrt{p_2} - \sqrt{r} - t)\\
\overset{\eqref{ineq:condition_nuclear}}{\geq} & \left(\frac{19}{20}\sqrt{p_1}-t\right)\left(\frac{19}{20}\sqrt{p_2} - t\right)
\end{split}\end{equation}
\begin{equation}
\|A_{1\bullet}\|  = \|B_{1, [1:m_1, :]} B_2^T\| \leq (\sqrt{m_1} + \sqrt{r} + t)(\sqrt{p_2} + \sqrt{r} + t) \overset{\eqref{ineq:condition_nuclear}}{\leq} \left(\frac{1}{4}\sqrt{p_1} + t\right)\left(\frac{21}{20}\sqrt{p_2} + t\right)
\end{equation}
and
\begin{equation}
\begin{split}
\|A_{21}\|  = & \|B_{1, [(m_1+1): p_1, :]} B_{2, [1:m_2, :]}^T\| \leq (\sqrt{p_1} + \sqrt{r} + t)(\sqrt{m_2} + \sqrt{r} + t) \\
\overset{\eqref{ineq:condition_nuclear}}{\leq} & \left(\frac{21}{20}\sqrt{p_1} + t\right)\left(\frac{1}{4}\sqrt{p_2} + t\right).
\end{split}
\end{equation}
Denote 
$$A_0 = \begin{bmatrix}
A_{11} & A_{12}\\
A_{21} & 0 \\
\end{bmatrix}$$
and set $t = \frac{1}{20}\min(\sqrt{p_1}, \sqrt{p_2})$. Since $\|A_0\|_\ast \leq \|A_{1\bullet}\|_\ast + \|A_{21}\|_\ast$, , we have
\begin{equation}
P\left(\|A\|_\ast \geq \frac{326}{400}\sqrt{p_1p_2} \right) \geq 1 - 12\exp(-\min(p_1, p_2)/400)
\end{equation}
and 
\begin{equation}
P\left(\|A_0\|_\ast \leq \frac{264}{400}\sqrt{p_1p_2} \right) \geq 1 - 12\exp(-\min(p_1, p_2)/ 400).
\end{equation}
Hence, with probability at least $1 - 12\exp(-\min(p_1, p_2)/400)$, $\|A_0\|_\ast < \|A\|_\ast$, which implies that  the NNM \eqref{eq:nuclear_norm_minimization} fails to recover $A_{22}$. \quad $\square$

\medskip
{\noindent\bf Proof of Lemma \ref{lm:nuclear_norm_1}.} For convenience, we denote $x\wedge y = \min(x, y)$ for any two real numbers $x, y$. First, we can extend the unit vectors $\frac{1}{\sqrt{m_1}} 1_{m_1}, \frac{1}{\sqrt{m_2}} 1_{m_2}$, $\frac{1}{\sqrt{p_1 - m_1}} 1_{p_1-m_1}$ and $\frac{1}{\sqrt{p_2 - m_2}} 1_{p_2 - m_2}$ into orthogonal matrices, which we denote as $U_{m_1}\in \mathbb{R}^{m_1 \times m_1}$, $U_{m_2} \in \mathbb{R}^{m_2 \times m_2}$, $U_{p_1-m_1}\in \mathbb{R}^{(p_1-m_1) \times (p_1-m_1)}$, $U_{p_2-m_2} \in \mathbb{R}^{(p_2 - m_2) \times (p_2-m_2)}$. Next, for all $A_{22}'\in \mathbb{R}^{(p_1-m_1)\times (p_2-m_2)}$, we must have
\begin{equation*}
\begin{split}
\left\|\begin{bmatrix}
A_{11} & A_{12}\\
A_{21} & A_{22}'
\end{bmatrix}\right\|_\ast = & \left\|\begin{bmatrix}
U_{m_1}^{\intercal} & 0\\
0 & U_{p_1 -m_1}^{\intercal}
\end{bmatrix}
\cdot\begin{bmatrix}
A_{11} & A_{12}\\
A_{21} & A'_{22}
\end{bmatrix}
\cdot\begin{bmatrix}
U_{m_2} & 0\\
0 & U_{p_2 - m_2}
\end{bmatrix}\right\|_\ast\\
\triangleq & \left\|\begin{bmatrix}
E_{11} & E_{12}\\
E_{21} & U_{p_1-m_1}^{\intercal}A'_{22}U_{p_2-m_2}
\end{bmatrix}\right\|_\ast,
\end{split}
\end{equation*}
where $E_{11}\in\mathbb{R}^{m_1\times m_2}, E_{12}\in \mathbb{R}^{m_1\times (p_2-m_2)}, E_{21}\in \mathbb{R}^{(p_1 - m_1)\times m_2}$ are with the first entry $\sqrt{m_1m_2}$, $\sqrt{m_1(p_2-m_2)}$ and $\sqrt{m_2(p_1-m_1)}$ respectively and other entries 0. Therefore, we can see
$$\left\|\begin{bmatrix}
E_{11} & E_{12}\\
E_{21} & U_{p_1-m_1}^{\intercal}A'_{22}U_{p_2-m_2}
\end{bmatrix}\right\|_\ast \geq \left\|\begin{bmatrix}
\sqrt{m_1m_2} & \sqrt{m_1(p_2 - m_2)}\\
\sqrt{m_2(p_1 - m_1)} & [U_{p_1-m_1}^{\intercal}A'_{22}U_{p_2-m_2}]_{[1,1]}
\end{bmatrix}\right\|_\ast $$
and the equality holds if and only if $U_{p_1-m_1}^{\intercal}A'_{22}U_{p_2-m_2}$ is zero except the first entry.

By some calculation, we can see the nuclear norm of 2-by-2 matrix
$$\left\|\begin{bmatrix}
\sqrt{m_1m_2} & \sqrt{m_1(p_2 - m_2)}\\
\sqrt{m_2(p_1 - m_1)} & x
\end{bmatrix}\right\|_\ast$$
achieves its minimum if and only if 
$$x = \sqrt{m_1m_2}\wedge \sqrt{(p_1 - m_1)(p_2- m_2)}.$$
Hence, $A_{22}'$ achieves the minimum of $\left\|\begin{bmatrix}
A_{11} & A_{12}\\
A_{21} & A_{22}'
\end{bmatrix}\right\|_\ast$ if and only if 
$$U_{p_1 - m_1}^{\intercal}A_{22}'U_{p_2 - m_2} = \begin{bmatrix}
\sqrt{m_1m_2}\wedge \sqrt{(p_1 - m_1)(p_2 - m_2)} & 0 & \cdots\\
0 & 0 & \\
\vdots & & \ddots
\end{bmatrix},$$
which means the minimizer $A_{22}' = \left(\sqrt{\frac{m_1m_2}{(p_1 - m_1)(p_2 - m_2)}}\wedge 1\right) \cdot 1_{p_1 - m_1}1_{p_2 - m_2}^{\intercal}$. \quad $\square$

\medskip
{\noindent\bf Proof of Lemma \ref{lm:U_Omega}.} The proof of this lemma relies on operator-Bernstein's inequality for sampling (Theorem 1 in \cite{Gross_sampling}). For two symmetric matrices $A$, $B$, we say $A \preceq B$ if $B - A$ is positive definite. By assumption, \{$U_{\Omega(j)\bullet}, j=1,\cdots, n$\} are uniformly random samples (with or without replacement) from $\{U_{i\bullet}, i=1,\cdots, n\}$. Suppose 
\begin{equation}
X_i = U_{i\bullet}^{\intercal}U_{i\bullet} - \frac{1}{p} I_d,\quad i=1,\cdots, p,
\end{equation}
then $X_i$ are symmetric matrices, $X_{\Omega(j)}, j=1,\cdots, n$ are uniformly random samples (with or without replacement) from $\{X_1, \cdots, X_p\}$. In addition, we have
$$EX_j = \frac{1}{p}\sum_{i=1}^p U_{i\bullet}^{\intercal}U_{i\bullet} - \frac{1}{p}I_d = \frac{1}{p}U^{\intercal}U - \frac{1}{p} I_d =  0$$
$$\|X_j\| \leq \max_{1\leq i\leq p} \left\| U_{i\bullet}^{\intercal}U_{i\bullet} - \frac{1}{p} I_d \right\| \leq \max_{1\leq i\leq p} \max\left\{\left\|U_{i\bullet}^{\intercal}U_{i\bullet}\right\|, \frac{1}{p}\left\|I_d\right\|\right\} \leq \frac{Wd}{p} $$
\begin{equation*}
\begin{split}
E X_j^2 = & \frac{1}{p}\sum_{i=1}^p \left(U_{i\bullet}^{\intercal}U_{i\bullet} - \frac{1}{p}I_d\right)^2 = \frac{1}{p} \sum_{i=1}^p \left(U_{i\bullet}^{\intercal}U_{i\bullet}U_{i\bullet}^{\intercal}U_{i\bullet} - \frac{2}{p}U_{i\bullet}^{\intercal}U_{i\bullet} + \frac{1}{p^2}I_d\right)\\
= & \frac{1}{p}\sum_{i=1}^p \|U_{i\bullet}\|_2^2\cdot U_{i\bullet}^{\intercal}U_{i\bullet} - \frac{1}{p^2}I_d\\
\preceq & \frac{1}{p}\cdot\frac{Wd}{p}\sum_{i=1}^p U_{i\bullet}^{\intercal}U_{i\bullet} - \frac{1}{p^2} I_d \preceq \frac{Wd-1}{p^2} I_d
\end{split}
\end{equation*}
For all $0<\alpha<1$, by Theorem 1 in \cite{Gross_sampling},
\begin{equation*}
\begin{split}
P\left(\|U_{\Omega}\| \leq \sqrt{\frac{\alpha n}{p}} \right) & = P\left(U_{\Omega}^{\intercal}U_{\Omega} \preceq \frac{\alpha n}{p} I_d\right)
=  P\left(\sum_{j=1}^nU_{\Omega(j)\bullet}^{\intercal}U_{\Omega(j)\bullet} \preceq \frac{\alpha n}{p} I_d\right)\\
& = P\left(\sum_{j=1}^n X_j \preceq -\frac{(1-\alpha)n}{p}I_d \right) \leq P\left(\left\|\sum_{j=1}^n X_j\right\| \geq \frac{(1-\alpha)n}{p} \right)\\
& \leq 2d\exp\left(-\min\left(\frac{\left((1-\alpha)n/p\right)^2}{4n(Wd-1)/p^2}, \frac{(1-\alpha)n/p}{2Wd/p} \right)\right) \\
& \leq 2d\exp\left(-\frac{n(1-\alpha)^2}{4Wd}\right) \leq 2\exp(-c).
\end{split}
\end{equation*}
The last inequality is due to the assumption that
$$ n \geq \frac{4Wd (\log d + c)}{(1-\alpha)^2}.$$ 
\quad $\square$

\medskip
{\noindent\bf Proof of Lemma \ref{lm:U_Haar}.} By the assumption on $n$, we have $n \geq p$ or $n\geq Cd$. When $n\geq p$, we know $n = p$ and $U_{[1:n,:]} = U$ is an orthogonal matrix, which means \eqref{eq:lm_Haar} is clearly true. Hence, we only need to prove the theorem under the assumption that $p\geq n$ is true. In this case, we must have $n\geq Cd$.

Since $U$ has random orthonormal columns with Haar measure, for any fixed vector $v\in \mathbb{R}^d$, $Uv$ is identitical distributed as 
$$\|x\|_2^{-1}\left(x_1, x_2, \cdots, x_p\right), \quad \text{where } x_1, \cdots, x_p\stackrel{iid}{\sim}N(0, 1)$$
Hence, $U_{[1:n, :]}v$ is identical distributed with $\|x\|_2^{-1}(x_1, \cdots, x_n)$ and 
\begin{equation}\label{eq:lm_Haar_identical}
\|U_{[1:n, :]}v\|_2 \text{ is identical distributed as } \sqrt{(\sum_{i=1}^nx_i^2)(\sum_{i=1}^px_i^2)^{-1}},
\end{equation}
which is the also the square root of Beta distribution. Denote
\begin{equation}
\alpha_1' = \frac{1 + \alpha_1}{2},\quad \alpha_2' = \frac{1+\alpha_2}{2}.
\end{equation}
By Lemma 1 in \cite{Laurent}, when $x_1, \cdots, x_p$ are i.i.d. standard normal, we have
$$1 - 2\sqrt{C'} \leq \frac{\sum_{i=1}^nx_i^2}{n} \leq 1 + 2\sqrt{C'} + 2C' $$
$$1 - 2\sqrt{\frac{C'n}{p}} \leq \frac{\sum_{i=1}^px_i^2}{p} \leq 1 + 2\sqrt{\frac{C'n}{p}} + \frac{2C'n}{p} $$
both hold with probability at least $1 - 4\exp(-C'n)$. Here we let $C'>0$ be small enough and only depending on $\alpha_1, \alpha_2$ such that
$$\alpha_1' \leq \frac{1 - 2\sqrt{C'}}{1 + 2\sqrt{C'} + 2C'} ,\quad \frac{1 + 2\sqrt{C'} + 2C'}{1 - 2\sqrt{C'}} \leq \alpha_2'. $$
Combining the previous inequalities and \eqref{eq:lm_Haar_identical}, we have for any fixed unit vector $v\in\mathbb{R}^d$, 
\begin{equation}\label{eq:alpha_1 alpha_2}
\frac{\alpha_1' n}{p} \leq \|U_{[1:n, :]}v\|_2^2 \leq \frac{\alpha_2' n}{p}
\end{equation}
with probability at least $1 - 4\exp(-C'n)$, where $C'$ only depends on $\alpha_1', \alpha_2'$. Next, based on Lemma 2.5 in \cite{epsilon_net}, we can construct an $\varepsilon$-net on the unit sphere of $\mathbb{R}^{d}$ as $B$, such that $|B|\leq (1+2/\varepsilon)^d$, where $\varepsilon > 0$ is to be determined later. Under the event that
$\{\forall v\in B,  \text{\eqref{eq:alpha_1 alpha_2} holds}\}$, we suppose 
$$\kappa_1 = \min_{\|v\|_2 = 1} \|U_{[1:n, :]}v\|_2^2,\quad \kappa_2 = \max_{\|v\|_2 = 1} \|U_{[1:n, :]}v\|_2^2.$$
For any $v$ in the unit sphere of $\mathbb{R}^{d}$, there must exists $v'\in B$ such that $\|v - v'\|_2 \leq \varepsilon$, which yields,
$$\|U_{[1:n, :]}v\|_2 \leq \|U_{[1:n, :]}v'\|_2 + \|U_{[1:n, :]}(v - v')\|_2\leq \sqrt{\alpha_2'n/p} + \kappa_2\varepsilon $$
$$\|U_{[1:n, :]}v\|_2 \geq \|U_{[1:n, :]}v'\|_2 - \|U_{[1:n, :]}(v - v')\|_2\geq \sqrt{\alpha_1'n/p} - \varepsilon \kappa_2 $$
These implies that $\kappa_2 \leq \sqrt{\alpha_2'n/p}/(1-\varepsilon)$, $\kappa_1 \geq \sqrt{\alpha_1'n/p} - \varepsilon \kappa_2 \geq \sqrt{\alpha_1'n/p} - \sqrt{\alpha_2'n/p}\cdot \varepsilon/(1-\varepsilon)$. Hence, we can take $\varepsilon$ depending on $\alpha_1, \alpha_2$ such that $\kappa_2 \leq \sqrt{\alpha_2n/p}$, $\kappa_1 \geq \sqrt{\alpha_1n/p}$, which implies \eqref{eq:lm_Haar}.

Finally we estimate the probability that the event  $\{\forall v\in B, \text{\eqref{eq:alpha_1 alpha_2} holds}\}$ happens. We choose $C\geq 4d\log (1+2/\varepsilon)/C'$ that only depends on $\alpha_1$ and $\alpha_2$. If $n\geq Cd$, 
$$C'n/2 \geq d\log (1+2/\varepsilon) + \log 4. $$
so
$$1 - (1+2/\varepsilon)^d\cdot 4\exp(-C'n) = 1 - \exp(d\log(1+2/\varepsilon) + \log 4-C' n) \geq 1 -\exp(-nC'/2) $$
Finally, we finish the proof of the lemma by setting $\delta = C'/2$.\quad $\square$

\section{Proofs of the Results in the Main Paper}

We prove Proposition \ref{th:noiseless}, Theorems \ref{th:hat A} and \ref{th:main_without_r}, Lemma \ref{lm:hat_r geq r}, Lemma \ref{lm:inequalities_Theorem_2}, Theorem \ref{th:lower_bound}, Corollary \ref{cr:random_column_row} and Corollary \ref{cr:randomUV} in this section.

\subsection*{Proof of Proposition \ref{th:noiseless}}  

Since $A_{1\bullet}$ is of rank $r$, which is the same as $A$,
all rows of $A$ must be linear combinations of the rows of $A_{1\bullet}$. This implies all rows of $A_{\bullet1}$ 
is a linear combination of $A_{11}$. Since rank($A_{\bullet1}$)$=r$, we must have $\text{rank}(A_{11})\geq r$. Besides, $\text{rank}(A_{11})\leq \text{rank}(A) = r$ since $A_{11}$ is a submatrix of $A$. So $\text{rank}(A_{11}) = r$.
Simiarly, rows of $A_{\bullet 1}$ is the linear combination of $A_{11}$, so we have 
$$A_{21} = A_{21}P_{A_{11}} = A_{21}A_{11}^{\intercal}(A_{11}A_{11}^{\intercal})^\dagger A_{11} = A_{21}V\Sigma U^{\intercal}(U\Sigma^2U^{\intercal})^\dagger A_{11} = \left(A_{21}V\Sigma^{-1}U^{\intercal}\right)A_{11}, $$
namely rows of $A_{21}$ is a linear combination of $A_{11}$. By the argument before, we know $A_{22}$ can be represented as the same linear combination of $A_{12}$ as $A_{21}$ by $A_{11}$, so we have
$A_{22} = \left(A_{21}V\Sigma^{-1}U^{\intercal}\right)A_{12} = A_{21}V\Sigma^{-1}U^{\intercal}A_{12} = A_{21} A_{11}^\dagger A_{12},$
which concludes the proof. \quad $\square$

\subsection*{Proof of Theorem \ref{th:hat A}} 

Suppose $M\in \mathbb{R}^{m_1\times r}, N\in\mathbb{R}^{m_2 \times r}$ are column orthonormalized matrices of $U_{11}$ and $V_{11}$. $\hat M\in \mathbb{R}^{m_1\times r}$ and $\hat N\in \mathbb{R}^{m_2\times r}$  are the first $r$ left singular vectors of $A_{1\bullet}$ and $A_{\bullet1}$, respectively. Also, recall that we use $P_{U} = U(U^{\intercal}U)^{\dagger}U^{\intercal}$ to represent the projection onto the column space of $U$.
\begin{enumerate}
	\item We first give the lower bound for $\sigma_{\min}(\hat M^{\intercal} M)$, $\sigma_{\min}(\hat N^{\intercal}N)$ by the unilateral perturbation bound result in \cite{Perturbation}. Since,
	$$P_{U_{11}}A_{1\bullet} = P_{U_{11}} U_{1\bullet}\Sigma V^{\intercal} = [U_{11}\Sigma_1, P_{U_{11}}U_{12}\Sigma_2]V^{\intercal}, \quad
	P_{U_{11}^\bot}A_{1\bullet} = P_{U_{11}^\bot}U_{1\bullet}\Sigma V^{\intercal} = [0, P_{U_{11}^\bot}U_{12}\Sigma_2]V^{\intercal}, $$
	by $V$ is an orthogonal matrix, we can see 
	$$\sigma_{r}(P_{U_{11}}A_{1\bullet}) = \sigma_{r}([U_{11}\Sigma_1\quad P_{U_{11}}U_{12}\Sigma_2]) \geq \sigma_{r}(U_{11}\Sigma_1) \geq \sigma_r(A)\sigma_{\min}(U_{11}),$$ 
	$$\|P_{U^\bot_{11}}A_{1\bullet}\| = \|P_{U^\bot_{11}}U_{12}\Sigma_2\| \leq\|P_{U_{11}^\bot}U_{12}\|\|\Sigma_2\| \leq \sigma_{r+1}(A).$$
	So $\sigma_{r}(P_{U_{11}}A_{1\bullet}) \geq \|P_{U_{11}^\bot}A_{1\bullet}\|$. Besides, $\rank(P_{U_{11}}A_{1\bullet}) \leq r$. Apply the unilateral perturbation bound result in \cite{Perturbation} by setting $X = P_{U_{11}}A_{1\bullet}$, $Y = P_{U_{11}^\bot}A_{1\bullet}$, we have
	\begin{equation}\label{eq:hat M M}
	\sigma_{\min}^2(\hat M^{\intercal} M) \leq 1- \left(\frac{\|Y \cdot P_{X^{\intercal}}\|\cdot \sigma_{r+1}(A)}{\sigma_r^2(A)\sigma_{\min}^2(U_{11}) - \sigma_{r+1}^2(A)}\right)^2.
	\end{equation}
	Moreover, 
	$A_{1\bullet} = [U_{11}~U_{12}]  \mbox{diag}(\Sigma_1,\Sigma_2)
	V^{\intercal} = [U_{11}\Sigma_1~U_{12}\Sigma_2]V^{\intercal}, $
	and hence,
	\begin{equation*}
	\begin{split}
	& \|YP_{X^{\intercal}}\| = \left\|P_{U_{11}^\bot}A_{1\bullet} \cdot P_{(P_{U_{11}} A_{1\bullet})^{\intercal}}\right\| = \left\|[0 \quad P_{U_{11}^\bot}U_{12}\Sigma_2]V^{\intercal} \cdot P_{V\cdot[U_{11}\Sigma_1\quad P_{U_{11}} U_{12}\Sigma_2]^{\intercal}}\right\|\\
	= &  \left\|[0 \quad P_{U_{11}^\bot}U_{12}\Sigma_2] \cdot P_{[U_{11}\Sigma_1\quad P_{U_{11}} U_{12}\Sigma_2]^{\intercal}}\right\|
	=  \sup_{x\in \mathbb{R}^{p_2}, \|x\|_2 = 1} [0 \quad P_{U_{11}^\bot}U_{12}\Sigma_2] \cdot P_{[U_{11}\Sigma_1\quad P_{U_{11}} U_{12}\Sigma_2]^{\intercal}} x.
	\end{split}
	\end{equation*}
	When $\|x\|_2=1$, let $y$ denote the projection of $x$ onto the column space of $[U_{11} \Sigma_1~~ P_{U_{11}}U_{12}\Sigma_2]^{\intercal}$. Then $\|y\|_2\leq 1$ and $y$ is in the column space of $[U_{11} \Sigma_1~~ P_{U_{11}}U_{12}\Sigma_2]^{\intercal}$. Hence, 
	$$\frac{\|y_{[1:m_1]}\|_2} {\|y_{[(m_1+1): p_1]}\|_2} \geq \frac{\sigma_{\min}(U_{11}\Sigma_1)}{ \|P_{U_{11}}U_{12}\Sigma_2\|} \geq \frac{\sigma_{\min}(U_{11})\sigma_r(A)} {\sigma_{r+1}(A)} \ \mbox{and} \
	\|y_{[(m_1+1):p_1]} \|_2^2 + \|y_{[1:m_1]}\|_2^2 \leq 1, $$
	which implies $\|y_{[(m_1+1): p_1]}\|_2^2 \leq {\sigma_{r+1}^2(A)}/{\sigma^2_{\min}(U_{11})\sigma^2_r(A)+\sigma^2_{r+1}(A)}$.
	Hence for all $x\in \mathbb{R}^{p_2}$ such that $\|x\|_2 = 1$,
	\begin{equation*}
	\begin{split}
	\left\|[0 \quad P_{U^\bot_{11} }U_{12}\Sigma_2] \cdot P_{[U_{11}\Sigma_1\quad P_{U_{11}} U_{12}\Sigma_2]^{\intercal}} x\right\| \leq & \|P_{U_{11}^\bot} U_{12}\Sigma_2\|\cdot \|y_{[m_1 + 1: p_1]}\|_2\\
	\leq & \sigma_{r+1}(A)\frac{\sigma_{r+1}(A)}{\sqrt{\sigma_{r+1}^2(A) + \sigma_{\min}^2(U_{11})\sigma_r^2(A)}}.
	\end{split}
	\end{equation*}
	This yields
	$\|YP_{X^{\intercal}}\|=\|P_{U_{11}^\bot}A_{1\bullet} \cdot P_{(P_{U_{11}} A_{1\bullet})}\| \leq \sigma^2_{r+1}(A) /\sqrt{\sigma_{r+1}^2(A) + \sigma_{\min}^2(U_{11})\sigma_r^2(A)}.$
	Combining \eqref{eq:hat M M}, we have
	\begin{equation}\label{eq:hatMM_result}
	\begin{split}
	\sigma_{\min}^2(\hat M^{\intercal}M) \geq & 1 - \left(\frac{\sigma^3_{r+1}(A)}{\sqrt{\sigma_{r+1}^2(A) + \sigma_{\min}^2(U_{11})\sigma_r^2(A)}\left(\sigma^2_r(A)\sigma_{\min}^2(U_{11}) - \sigma_{r+1}^2(A)\right)}\right)^2.
	\end{split}
	\end{equation}
	Since $\sigma_{\min}(U_{11})\sigma_r(A)\geq 2\sigma_{r+1}(A)$, we have 
	$$ \sigma^2_{\min}(\hat M^{\intercal} M) \geq  1 - \left(\frac{1}{\sqrt{5}\cdot 3}\right)^2 \geq \frac{44}{45}.$$
	Similarly, we also have $\sigma_{\min}^2(\hat N^{\intercal}N) \geq \frac{44}{45}$.
	\item 
	Following by \eqref{eq:hat A},
	\begin{equation*}
	\begin{split}
	& \hat A_{22} =   U_{2\bullet}\Sigma V_{1\bullet}^{\intercal} \hat N\left(\hat M^{\intercal}(U_{1\bullet} \Sigma V^{\intercal}_{1\bullet}) \hat N\right)^{-1} \hat M^{\intercal} U_{1 \bullet} \Sigma V^{\intercal}_{2\bullet}\\
	= & \left(U_{2 1}\Sigma_1 V_{11}^{\intercal}\hat N + U_{2 2}\Sigma_2 V_{12}^{\intercal}\hat N\right)\left(\hat M^{\intercal} U_{11}\Sigma_1 V_{11}^{\intercal} \hat N + \hat M^{\intercal} U_{12}\Sigma_2V_{12}^{\intercal}\hat N\right)^{-1}
	\left(\hat M^{\intercal} U_{11}\Sigma_1 V_{2 1}^{\intercal} + \hat M^{\intercal} U_{12}\Sigma_2 V^{\intercal}_{2 2}\right).
	\end{split} 
	\end{equation*}
	Let ``L", ``M", ``R" stand for ``Left", ``Middle" and ``Right", 
	\begin{equation}\label{eq:B_L}
	B_L =  U_{2 1}\Sigma_1 V_{11}^{\intercal}\hat N, \quad E_L =  U_{2 2}\Sigma_2 V_{12}^{\intercal}\hat N;
	\end{equation}
	\begin{equation}\label{eq:B_M}
	B_M = \hat M^{\intercal} U_{11}\Sigma_1 V_{11}^{\intercal} \hat N, \quad E_M = \hat M^{\intercal} U_{12}\Sigma_{2}V_{12}^{\intercal} \hat N; 
	\end{equation}
	\begin{equation}\label{eq:B_R}
	B_R = \hat M^{\intercal} U_{11}\Sigma_1 V_{2 1}^{\intercal},\quad E_R = \hat M^{\intercal} U_{12}\Sigma_2 V^{\intercal}_{2 2}. 
	\end{equation}
	By Lemma \ref{lm:Schatten_q} in the Supplement, we can see the following properties of these matrices,
	\begin{equation}\label{ineq:E}
	\|E_L\|\leq \sigma_{r+1}(A), \quad \|E_M\|\leq \sigma_{r+1}(A),\quad \|E_R\|\leq \sigma_{r+1}(A), 
	\end{equation}
	\begin{equation}\label{ineq:E_q}
	\|E_L\|_q \leq \|\Sigma_2\|_q, \quad \|E_M\|_q\leq \|\Sigma_2\|_q, \quad \|E_R\|_q\leq \|\Sigma_2\|_q,
	\end{equation}
	\begin{align}
	& \sigma_{\min}(B_M) =  \sigma_{\min}\left(\hat M^{\intercal}(P_{M}U_{11}) \Sigma_1 (V_{11}^\intercal P_{N}) \hat N\right) = \sigma_{\min} \left((\hat M^{\intercal}M) (M^\intercal U_{11}) \Sigma_1 (V_{11}^{\intercal} N) (N^{\intercal}\hat N) \right) \nonumber \\
	\geq & \sigma_{\min}(\Sigma_1)\sigma_{\min}(U_{11})\sigma_{\min}(V_{11})\sigma_{\min}(\hat M^{\intercal} M)\sigma_{\min}(\hat N^{\intercal} N)\geq \frac{44}{45}\sigma_r(A)\sigma_{\min}(U_{11})\sigma_{\min}(V_{11}), \label{ineq:sigma_min B_M}
	\end{align}
	\begin{equation}
	\|B_M^{-1}\| = \sigma_{\min}^{-1}(B_M) \leq \frac{45}{44\sigma_r(A)\sigma_{\min}(U_{11})\sigma_{\min}(V_{11})},
	\end{equation}
	\begin{equation}\label{eq:A22_expansion}
	\hat A_{22} = (B_L+E_L)(B_M+E_M)^{-1}(B_R+E_R),\quad B_L B_M^{-1} B_R = U_{2 1}\Sigma_1 V_{2 1}^{\intercal},
	\end{equation}
	\begin{equation}\label{ineq:B_LB_M-1}
	\begin{split}
	\| B_L B_M^{-1}\| = & \|U_{21}\Sigma_1(V_{11}^{\intercal}\hat N) (V_{11}^{\intercal}\hat N)^{-1}\Sigma^{-1} (\hat M^\intercal U_{11})^{-1}\|= \|U_{2 1}(\hat M^{\intercal}U_{11})^{-1}\|\\
	\leq & \|(\hat M^\intercal MM^\intercal U_{11})^{-1}\| \leq \frac{1}{\sigma_{\min}(M^\intercal U_{11})\sigma_{\min}(\hat M^{\intercal} M)} \leq \frac{\sqrt{45/44}}{\sigma_{\min}(U_{11})},
	\end{split}
	\end{equation}
	\begin{equation}\label{ineq:B_M-1B_R}
	\| B_M^{-1} B_R\|  =  \|(V_{11}\hat N)^{-1}V_{2 1}^{\intercal} \| \leq \frac{\sqrt{45/44}}{\sigma_{\min}(V_{11})}.
	\end{equation}
	By \eqref{ineq:E}, \eqref{ineq:sigma_min B_M} and the assumption \eqref{eq:assumption}, we can see $\sigma_{\min}(B_M)>\|E_M\|$, so
	$$\hat A_{22} \overset{\eqref{eq:A22_expansion}}{=} (B_L+E_L)(B_M^{-1} - B_M^{-1}E_MB_M^{-1} + B_M^{-1}E_MB_M^{-1}E_MB_M^{-1} -\cdots) (B_R+E_R);$$
	\begin{align*}
	& \|\hat A_{22} - B_LB_M^{-1}B_R\|_q
	\leq   \big\|B_L B_M^{-1}E_M\sum_{i=0}^{\infty}(-B_M^{-1}E_M)^{i}B_M^{-1} B_R\big\|_q + \big\|E_L\sum_{i=0}^\infty (-B_M^{-1}E_M)^i B_M^{-1} B_R\big\|_q
	\\
	& \hspace{1.4in} + \big\|B_L  B_M^{-1}\sum_{i=0}^\infty(-E_M B_M^{-1})^{i}  E_R\big\|_q + \big\|E_L B_M^{-1}\sum_{i=0}^{\infty}(-E_M B_M^{-1})^i  E_R\big\|_q\\
	\leq & \|B_L B_M^{-1}\|\|E_M\|_q \sum_{i=0}^{\infty} \|E_M\|^{i}\|B_M^{-1}\|^{i} \|B_M^{-1}B_R\| + \|E_L\|_q \sum_{i=0}^\infty \|B_M^{-1}\|^i\|E_M\|^i \|B_M^{-1}B_R\|\\
	& + \|B_LB_M^{-1}\|\sum_{i=0}^\infty \|E_M\|^{i}\|B_M^{-1}\|^i\|E_R\|_q + \|E_L\|\sum_{i=0}^\infty\|B_M^{-1}\|^{i+1}\|E_M\|^{i}\|E_R\|_q\\
	\overset{\eqref{ineq:E}\eqref{ineq:E_q}}{\leq} & \frac{\|B_LB_M^{-1}\|\|B_M^{-1}B_R\| + \|B_M^{-1}B_R\| + \|B_LB_M^{-1}\| + \|B_M^{-1}\|\sigma_{r+1}(A) }{1-\sigma_{r+1}(A)\|B_M^{-1}\|}\|\Sigma_2\|_q\\
	\overset{\eqref{ineq:B_LB_M-1}\eqref{ineq:B_M-1B_R}}{\leq} & \frac{1}{1 - \sigma_{r+1}(A)\|B_M^{-1}\|}\left(\frac{45/44}{\sigma_{\min}(U_{11})\sigma_{\min}(V_{11})} + \frac{\sqrt{45/44}}{\sigma_{\min}(U_{11})} + \frac{\sqrt{45/44}}{\sigma_{\min}(V_{11})} + \frac{45}{88}\right)\|\Sigma_2\|_q\\
	\leq & \frac{\|A_{-\max(r)}\|_q}{1 - \frac{45\sigma_{r+1}(A)}{44\sigma_r(A)\sigma_{\min}(U_{11})\sigma_{\min}(V_{11})}}\left(\frac{45/44}{\sigma_{\min}(U_{11})\sigma_{\min}(V_{11})} + \frac{\sqrt{45/44}}{\sigma_{\min}(U_{11})} + \frac{\sqrt{45/44}}{\sigma_{\min}(V_{11})} + \frac{45}{88}\right)\\
	\leq & \frac{88}{43}\|A_{-\max(r)}\|_q\left(\frac{45/44}{\sigma_{\min}(U_{11})\sigma_{\min}(V_{11})} + \frac{\sqrt{45/44}}{\sigma_{\min}(U_{11})} + \frac{\sqrt{45/44}}{\sigma_{\min}(V_{11})} + \frac{45}{88}\right).
	\end{align*}
	Finally, since $A_{22} = U_{21}\Sigma_1 V_{21}^{\intercal} + U_{22}\Sigma_2V_{22}^{\intercal} \overset{\eqref{eq:A22_expansion}}{=} B_LB_M^{-1}B_R + U_{22}\Sigma_2V_{22}^{\intercal}$, we have
	\begin{equation*}
	\begin{split}
	\|\hat A_{22} - A_{22}\|_q \leq & \|\hat A_{22} - B_LB_M^{-1}B_R\|_q + \|U_{2 2}\Sigma_2V_{2 2}^{\intercal}\|_q \\
	\leq & 3\|A_{-\max(r)}\|_q \left(1+\frac{1}{\sigma_{\min}(U_{11})}\right)\left(1+\frac{1}{\sigma_{\min}(V_{11})}\right).   \quad \quad \square
	\end{split}
	\end{equation*}
\end{enumerate}

\subsection*{Proof of Theorem \ref{th:main_without_r}}

We only present proof for row thresholding as the column thresholding is essentially the same by working with $A^T$. 
Suppose $M, N$ are orthonormal basis of column vectors of $U_{11}, V_{11}$. We denote $U^{(1)}_{[:, 1:r]} = \hat M$, $V^{(2)}_{[:, 1:r]} = \hat N$, which are exactly the same as the $\hat M$ and $\hat N$ in Algorithm 1. Similarly to the proof of Theorem \ref{th:hat A}, we have \eqref{eq:hatMM_result}. Due to the assumption that $\sigma_r(A)\sigma_{\min}(U_{11})\sigma_{\min}(V_{11}) \geq 4\sigma_{r+1}(A)$, \eqref{eq:hatMM_result} yields
\begin{equation}\label{eq:hatMM_hatNN}
\sigma^2_{\min}(\hat M^{\intercal} M)\geq 3824/3825, \quad \sigma^2_{\min}(\hat N^{\intercal} N)\geq 3824/3825.
\end{equation}
As shown in the Supplementary material, we have
\begin{Lemma}\label{lm:hat_r geq r}
	Under the assumption of Theorem \ref{th:main_without_r}, we have $\hat r \geq r$. 
\end{Lemma}
We next show \eqref{eq:th_without_r} with the condition that $\hat r \geq r$ in steps.
\begin{enumerate}
	\item Note that $A_{11} = U_{11}\Sigma_{1}V_{11}^{\intercal} + U_{12}\Sigma_2 V_{12}^{\intercal}$, we consider the decompositions of $Z$ and let $$Z_{11} = U^{(2)\intercal}U_{11}\Sigma_1V_{11}^{\intercal}V^{(1)} + U^{(2)\intercal}U_{12}\Sigma_2V_{12}^{\intercal}V^{(1)}, $$
	\begin{equation}\label{eq:Z_11_decompose}
	Z_{11, [1:\hat r, 1:\hat r]} = U_{[:, 1:\hat r]}^{(2)\intercal}U_{11}\Sigma_1V_{11}^{\intercal}V^{(1)}_{[:, 1:\hat r]} + U^{(2)\intercal}_{[:, 1:\hat r]}U_{12}\Sigma_2V_{12}^{\intercal}V^{(1)}_{[:, 1:\hat r]} \triangleq B_{M,\hat r} + E_{M, \hat r}, 
	\end{equation}
	\begin{equation}\label{eq:Z_21_decompose}
	Z_{21, [:, 1:\hat r]} = U_{21}\Sigma_1V_{11}^{\intercal}V^{(1)}_{[:, 1:\hat r]} + U_{22}\Sigma_2V_{12}^{\intercal}V^{(1)}_{[:, 1:\hat r]} \triangleq B_{L,\hat r} + E_{L, \hat r}, \end{equation}
	\begin{equation}\label{eq:Z_12_decompose}
	Z_{12, [1:\hat r, :]} = U_{[:, 1:\hat r]}^{(2)\intercal}U_{11}\Sigma_1V_{21}^{\intercal} + U^{(2)\intercal}_{[:, 1:\hat r]}U_{12}\Sigma_2V_{22}^{\intercal} \triangleq B_{R,\hat r} + E_{R, \hat r}. 
	\end{equation}
	Note that the square matrix $U^{(2)\intercal}_{[:,1:r]}M \in \mathbb{R}^{r \times r}$ is a submatrix of $U^{(2)\intercal}_{[:,1:\hat r]}M \in \mathbb{R}^{\hat r \times r}$, we know
	\begin{equation}
	\sigma_{\min}(U_{[:, 1:\hat r]}^{(2)\intercal}M)\geq \sigma_{\min}(U_{[:, 1:r]}^{(2)\intercal}M) = \sigma_{\min}(\hat M M) \overset{\eqref{eq:hatMM_hatNN}}{\geq} \sqrt{\frac{3824}{3825}}.
	\end{equation}
	Similarly, $\sigma_{\min}(V_{[:, 1:\hat r]}^{(1)\intercal}N) \geq \sqrt{\frac{3824}{3825}}$.
	By $M, N$ are the orthonormal basis of column vectors of $U_{11}, V_{11}$, we have $P_{M} = MM^{\intercal}$, $P_N = NN^{\intercal}$, and
	\begin{equation}\label{ineq:UU_11}
	\begin{split}
	\sigma_{\min}(U_{[:, 1:\hat r]}^{(2)\intercal} U_{11}) 
	\geq & \sigma_{\min}(U_{[:, 1:\hat r]}^{(2)\intercal}M)\sigma_{\min}(M^{\intercal} U_{11}) \geq \sqrt{\frac{3824}{3825}}\sigma_{\min}(U_{11});
	\end{split}
	\end{equation}
	similarly, we also have
	\begin{equation}\label{ineq:VV_11}
	\sigma_{\min}(V_{[:, 1:\hat r]}^{(1)\intercal} V_{11})\geq \sqrt{\frac{3824}{3825}}\sigma_{\min}(V_{11}).
	\end{equation}
	\eqref{ineq:UU_11} and \eqref{ineq:VV_11} immediately yield
	\begin{equation}\label{ineq:sigma_r(B_M,hat r)}
	\sigma_{r}(B_{M, \hat r})\geq \frac{3824}{3825}\sigma_{\min}(U_{11})\sigma_{\min}(\Sigma_1)\sigma_{\min}(V_{11}) = \frac{3824}{3825}\sigma_{r}(A)\sigma_{\min}(U_{11})\sigma_{\min}(V_{11}).
	\end{equation}
	Besides, we also have
	\begin{equation}\label{ineq:E_M, hat r_bound}
	\|E_{M, \hat r}\| \overset{\eqref{eq:Z_11_decompose}}{\leq} \|\Sigma_2\| = \sigma_{r+1}(A)
	\end{equation}
	\item Next, we consider the SVD of $Z_{11, [1:\hat r, 1:\hat r]}$
	\begin{equation}\label{eq:Z_11_svd}
	Z_{11, [1:\hat r, 1:\hat r]} = J\Lambda K^{\intercal},\quad J,\Lambda, K\in\mathbb{R}^{\hat r\times \hat r}.
	\end{equation}
	For convenience, we denote  $ \Lambda_1 = \Lambda_{[1:r, 1:r]},  \Lambda_2 = \Lambda_{[(r+1):\hat r, (r+1):\hat r]},$
	\begin{equation}\label{eq:J_1J_2K_1_K_2}
	J_1 = J_{[:, 1:r]},\quad J_2 = J_{[:, (r+1): \hat r]},\quad K_1 = K_{[:, 1:r]},\quad K_2 = K_{[:, (r+1): \hat r]},
	\end{equation}
	Suppose $M_Z\in \mathbb{R}^{\hat r\times r}$ is an orthonormal basis of the column space of $B_{M, \hat r}$; $N_Z\in \mathbb{R}^{\hat r\times r}$ is an orthonormal basis of the column space of $B_{M, \hat r}^{\intercal}$. Denote ${\rm span}(\cdot)$ as the linear span of the column space of the matrix. We want to show ${\rm span}(M_Z)$ is close to ${\rm span}(J_1)$; while ${\rm span}(N_Z)$ is close to ${\rm span}(K_1)$. So in the rest of this step, we try to establish bounds for $\sigma_{\min}(J_1^{\intercal}M_Z)$ and $\sigma_{\min}(K_1^{\intercal}N_Z)$. Actually,
	$$Z_{11,[1:\hat r, 1:\hat r]} = B_{M, \hat r} + E_{M, \hat r} =\left( B_{M, \hat r} + P_{M_Z}E_{M, \hat r}\right)  + P_{M_Z^\perp}E_{M, \hat r}.$$
	Now we set $X = (B_{M, \hat r} + P_{M_Z}E_{M, \hat r})$, $Y = P_{M_Z^\perp}E_{M, \hat r}$, then we have
	\begin{equation*}
	\begin{split}
	\sigma_r(X) \geq & \sigma_r(B_{M, \hat r}) - \|P_{M_Z}E_{M, \hat r}\| \overset{\eqref{ineq:sigma_r(B_M,hat r)}}{\geq} \frac{3824}{3825}\sigma_r(A)\sigma_{\min}(U_{11})\sigma_{\min}(V_{11}) - \sigma_{r+1}(A),\\
	\overset{\eqref{ineq:assumption_theorem2}}{\geq} & \sigma_{r+1}(A) \overset{\eqref{ineq:E_M, hat r_bound}}{\geq} \|E_{M, \hat r}\| \geq \|Y\|.
	\end{split}
	\end{equation*}
	Besides, by the definition of $B_{M, \hat r}$ and $M_Z$ we know $\rank(X)\leq r$. Also based on the definition of $Y$, we know $P_{X}Y = 0$. Now the unilateral perturbation bound in \cite{Perturbation} yields
	\begin{equation}
	\sigma_{\min}^2 (M_Z^{\intercal}J_1) \geq 1 - \left(\frac{\sigma_r(X)\cdot \|Y\|}{\sigma^2_r(X) - \|Y\|^2 }\right)^2.
	\end{equation}
	The right hand side of the inequality above is an increasing function of $\sigma_r(X)$. Since $\sigma_r(X) \geq \frac{3824}{3825}\sigma_r(A)\sigma_{\min}(U_{11})\sigma_{\min}(V_{11}) - \sigma_{r+1}(A) \geq (3-\frac{4}{3825}) \sigma_{r+1}(A) \geq (3-\frac{4}{3825})\|Y\|$, 
	\begin{equation}\label{ineq:J_1^TM_Z}
	\sigma_{\min}^2(J_1^{\intercal}M_Z) \geq 1 - \left(\frac{3 - 4/3825}{(3 - 4/3825)^2 - 1}\right)^2 \geq 0.859.
	\end{equation}
	Similarly, we also have
	\begin{equation}\label{ineq:K_1^TN_Z}
	\sigma_{\min}^2(K_1^{\intercal}N_Z) \geq 0.859.
	\end{equation}
	
	\item We next derive useful expressions of $A_{22}$ and $\hat A_{22}$.
	First we introduce the following quantities,
	\begin{equation}\label{eq:JK_begin}
	J_1^{\intercal}Z_{11, [1:\hat r, 1:\hat r]}K_1 \overset{\eqref{eq:Z_11_decompose}}{=} J_1^{\intercal}B_{M,\hat r}K_1 + J_1^{\intercal}E_{M, \hat r} K_1 \triangleq B_{M1} + E_{M1},
	\end{equation}
	\begin{equation}
	J_2^{\intercal}Z_{11, [1:\hat r, 1:\hat r]}K_2 \overset{\eqref{eq:Z_11_decompose}}{=} J_2^{\intercal}B_{M,\hat r}K_2 + J_2^{\intercal}E_{M, \hat r} K_2 \triangleq B_{M2} + E_{M2},
	\end{equation}
	\begin{equation}
	Z_{21, [:, 1:\hat r]}K_1 \overset{\eqref{eq:Z_21_decompose}}{=} B_{L,\hat r}K_1 + E_{L, \hat r} K_1 \triangleq B_{L1} + E_{L1},
	\end{equation}
	\begin{equation}
	Z_{21, [:, 1:\hat r]}K_2 \overset{\eqref{eq:Z_21_decompose}}{=} B_{L,\hat r}K_2 + E_{L, \hat r} K_2 \triangleq B_{L2} + E_{L2},
	\end{equation}
	\begin{equation}
	J_1^{\intercal}Z_{12, [1:\hat r, :]} \overset{\eqref{eq:Z_12_decompose}}{=} J_1^{\intercal}B_{R,\hat r} + J_1^{\intercal}E_{R, \hat r} \triangleq B_{R1} + E_{R1},
	\end{equation}
	\begin{equation}\label{eq:JK_last}
	J_2^{\intercal}Z_{11, [1:\hat r, :]} \overset{\eqref{eq:Z_12_decompose}}{=} J_2^{\intercal}B_{R,\hat r} + J_2^{\intercal}E_{R, \hat r} \triangleq B_{R2} + E_{R2}.
	\end{equation}

	Since
	\begin{equation}\label{eq:B_L1B_M1-1BR1}
	\begin{split}
	& B_{L1}B_{M1}^{-1}B_{R1} = B_{L, \hat r}K_1 \left(J_1^{\intercal} B_{M, \hat r} K_1\right)^{-1}J_1^{\intercal}B_{R, \hat r}\\
	= & U_{21}\Sigma_1 V_{11}^{\intercal}V_{[:, 1:\hat r]}^{(1)} K_1 \left(J_1^{\intercal} U^{(2)\intercal}_{[:, 1:\hat r]} U_{11} \Sigma_1V_{11}^{\intercal}V_{[:,1:\hat r]}^{(1)} K_1\right)^{-1} J_1^{\intercal} U_{[:, 1:\hat r]}^{(2)\intercal}U_{11}\Sigma_1V_{21}^{\intercal} =  U_{21}\Sigma_1V_{21}^{\intercal},
	\end{split}
	\end{equation}
	we can characterize $A_{22}, \hat A_{22}$ by these new notations as
	\begin{equation}\label{eq:A_22_proof}
	A_{22} = U_{21}\Sigma_1V_{21}^{\intercal} + U_{22}\Sigma_2V_{22}^{\intercal} \overset{\eqref{eq:B_L1B_M1-1BR1}}{=} B_{L1}B_{M1}^{-1}B_{R1} + U_{22}\Sigma_2V_{22}^{\intercal},
	\end{equation}
	\begin{align}
	\hat A_{22} = & Z_{21, [:, 1:\hat r]}Z_{11, [1:\hat r, 1:\hat r]}^{-1}Z_{12,[1:\hat r, :]}
	\overset{\eqref{eq:Z_11_svd}}{=} Z_{21, [:, 1:\hat r]}K \left(J^{\intercal}Z_{11, [1:\hat r, 1:\hat r]}K\right)^{-1}J^{\intercal}Z_{12,[1:\hat r, :]} \nonumber \\
	= & \left(Z_{21,[1:\hat r]} K_1 + Z_{21,[1:\hat r]} K_2\right) \left(J_1^{\intercal}Z_{11,[1:\hat r, 1:\hat r]}K_1 + J_2^{\intercal}Z_{11,[1:\hat r, 1:\hat r]}K_2\right)^{-1}\left(J_1^{\intercal}Z_{12, [1:\hat r]} + J_2^{\intercal} Z_{12,[1:\hat r]}\right) \nonumber \\
	\overset{\eqref{eq:JK_begin}-\eqref{eq:JK_last}}{=} & \sum_{k=1}^2 (B_{Lk} +E_{Lk})(B_{Mk} +E_{Mk})^{-1}(B_{Rk} +E_{Rk}) \label{eq:hatA_22_proof}
	\end{align}
	\item We now establish a number of bounds for the terms on the right hand side of \eqref{eq:JK_begin}-\eqref{eq:JK_last}. 
	\begin{Lemma}\label{lm:inequalities_Theorem_2} Based on the assumptions above, we have
		\begin{equation}\label{ineq:sigma_min_B_M1}
		\sigma_{\min}(B_{M1}) \geq 3.43\sigma_{r+1}(A);
		\end{equation}
		\begin{equation}\label{ineq:B_L1B_M1^-1}
		\|B_{L1}B_{M1}^{-1}\| \leq \frac{\sqrt{3825/3824}}{\sqrt{0.859}\sigma_{\min}(U_{11})}, \quad \|B_{M1}^{-1}B_{R1}\| \leq \frac{\sqrt{3825/3824}}{\sqrt{0.859}\sigma_{\min}(V_{11})},
		\end{equation}
		\begin{equation}\label{ineq:E_Mt}
		\|E_{Mt}\|_q \leq \|A_{-\max(r)}\|_q, \, \|E_{Lt}\|_q \leq \|A_{-\max(r)}\|_q,\, \|E_{Rt}\|_q \leq \|A_{-\max(r)}\|_q, \quad t = 1, 2,
		\end{equation}
		\begin{equation}\label{ineq:B_L2+E_L2B_M2E_M2}
		\|(B_{L2}+E_{L2} )(B_{M2} + E_{M2})^{-1}\| \leq T_R +  \frac{1}{1 - 1/3.43}\left(\frac{\sqrt{3825/3824}}{\sqrt{0.859}\sigma_{\min}(U_{11})} + \frac{1}{3.43}\right), 
		\end{equation}
		\begin{equation}\label{ineq:B_R2}
		\|B_{R2}\|_q \leq \frac{2\sqrt{3825/3824}}{\sqrt{0.859\sigma_{\min}(V_{11})}}\|A_{-\max(r)}\|_q.
		\end{equation}
	\end{Lemma}
	The proof of Lemma \ref{lm:inequalities_Theorem_2} is given in the Supplement.

	\item We finally give the upper bound of $\|\hat A_{22} - A_{22}\|_q$. By \eqref{eq:A_22_proof} and \eqref{eq:hatA_22_proof}, we can split the loss as,
	\begin{equation}\label{ineq:split_hatA22-A22}
	\begin{split}
	\hat A_{22} - A_{22} = & \left(\left(B_{L1}+E_{L1}\right)\left(B_{M1}+E_{M1}\right)^{-1}\left(B_{R1}+E_{R1}\right) - B_{L1}B_{M1}^{-1}B_{R1}\right)\\
	& + \left(B_{L2}+E_{L2}\right)\left(B_{M2}+E_{M2}\right)^{-1}\left(B_{R2}+E_{R2}\right)  - U_{22}\Sigma_2 V_{22}^{\intercal}.
	\end{split}
	\end{equation}
	We will analyze them separately. First, $\|U_{22}\Sigma_2V_{22}^{\intercal}\|_q \leq \|A_{-\max(r)}\|_q$; second, 
	\begin{align}
	& \|(B_{L2}+E_{L2})(B_{M2}+E_{M2})^{-1}(B_{R2} + E_{M2}) \|_q \nonumber \\
	\leq & \|(B_{L2}+E_{L2})(B_{M2}+E_{M2})^{-1}\| \cdot \left(\|B_{R2}\|_q + \|E_{M2}\|_q\right)\nonumber \\
	\overset{\eqref{ineq:B_L2+E_L2B_M2E_M2}\eqref{ineq:B_R2}}{\leq} & \left(T_R + \frac{3.43}{2.43}\left(\frac{\sqrt{3825/3824}}{\sqrt{0.859}\sigma_{\min}(U_{11})} + \frac{1}{3.43}\right)\right)\left(\frac{2\sqrt{3825/3824}}{\sqrt{0.859}\sigma_{\min}(V_{11})} + 1\right)\|A_{-\max(r)}\|_q \nonumber \\
	\leq & \left(T_R + \frac{1.524}{\sigma_{\min}(U_{11})} + 0.412\right)\left(\frac{2.16}{\sigma_{\min}(V_{11})} + 1\right)\|A_{-\max(r)}\|_q. \label{ineq:second_term_proof}
	\end{align}
	The analysis of $\left(\left(B_{L1}+E_{L1}\right)\left(B_{M1}+E_{M1}\right)^{-1}\left(B_{R1}+E_{R1}\right) - B_{L1}B_{M1}^{-1}B_{R1}\right)$ is similar to the proof of Theorem \ref{th:hat A}. We have
	\begin{align}
	& \left\|(B_{L1}+E_{L1})(B_{M1}+E_{M1})^{-1}(B_{R1}+E_{R1}) - B_{L1}B_{M1}^{-1}B_{R1}\right\|_q\nonumber \\
	\leq & \left\|B_{L1}(B_{M1}^{-1}E_{M1}\sum_{i=0}^{\infty}(-B_{M1}^{-1}E_{M1})^{i}B_{M1}^{-1})B_{R1}\right\|_q + \left\|E_{L1}\left(\sum_{i=0}^\infty (-B_{M1}^{-1}E_{M1})^i B_{M1}^{-1}\right)B_{R1}\right\|_q\nonumber \\
	& + \left\|B_{L1}\left( B_{M1}^{-1}\sum_{i=0}^\infty(-E_{M1} B_{M1}^{-1})^{i}\right)  E_{R1}\right\|_q + \left\|E_{L1}\left(B_{M1}^{-1}\sum_{i=0}^{\infty}(-E_{M1} B_{M1}^{-1})^i\right) E_{R1}\right\|_q\nonumber \\
	\leq & \|B_{L1} B_{M1}^{-1}\|\|E_{M1}\|_q \sum_{i=0}^{\infty} \|E_{M1}\|^{i}\|B_{M1}^{-1}\|^{i} \|B_{M1}^{-1}B_{R1}\| + \|E_{L1}\|_q \sum_{i=0}^\infty \|B_{M1}^{-1}\|^i\|E_{M1}\|^i \|B_{M1}^{-1}B_{R1}\|\nonumber \\
	& + \|B_{L1}B_{M1}^{-1}\|\sum_{i=0}^\infty \|E_{M1}\|^{i}\|B_{M1}^{-1}\|^i\|E_{R1}\|_q + \|E_{L1}\|\sum_{i=0}^\infty\|B_{M1}^{-1}\|^{i+1}\|E_{M1}\|^{i}\|E_{R1}\|_q\nonumber \\
	\overset{\eqref{ineq:E_Mt}}{\leq} & \frac{\|\Sigma_2\|_q}{1-\sigma_{r+1}(A)\|B_{M1}^{-1}\|}
	\left(\|B_{L1}B_{M1}^{-1}\|\|B_{M1}^{-1}B_{R1}\| + \|B_{M1}^{-1}B_{R1}\| + \|B_{L1}B_{M1}^{-1}\| + \|B_{M1}^{-1}\|\sigma_{r+1}(A) \right)\nonumber \\
	\overset{\eqref{ineq:B_L1B_M1^-1}\eqref{ineq:sigma_min_B_M1}}{\leq} & \left(\frac{1.65}{\sigma_{\min}(U_{11})\sigma_{\min}(V_{11})} + \frac{1.53}{\sigma_{\min}(V_{11})} + \frac{1.53}{\sigma_{\min}(V_{11})} + 0.42\right)\|A_{-\max(r)}\|_q. \label{ineq:third_term_proof}
	\end{align}
	From \eqref{ineq:second_term_proof}, \eqref{ineq:third_term_proof}, \eqref{ineq:split_hatA22-A22}, and the fact that $\sigma_{\min}(U_{11})\leq 1$ and $T_R \geq \frac{1.36}{\sigma_{\min}(U_{11})} + 0.35$, 
	\begin{equation}
	\begin{split}
	\|\hat A_{22} - A_{22}\|_q \leq & \left(2.16T_R + \left(\frac{4.95}{\sigma_{\min}(U_{11})} + 2.42\right)\right)\left(\frac{1}{\sigma_{\min}(V_{11})} + 1\right) \|A_{-\max(r)}\|_q\\
	\leq & \left(2.16T_R + 4.31\left(\frac{1.36}{\sigma_{\min}(U_{11})} + 0.35\right)\right)\left(\frac{1}{\sigma_{\min}(V_{11})} + 1\right)\|A_{-\max(r)}\|_q\\
	\leq & 6.5T_R\left(\frac{1}{\sigma_{\min}(V_{11})} + 1\right) \|A_{-\max(r)}\|_q.
	\end{split}
	\end{equation}
	This concludes the proof.\quad $\square$
\end{enumerate}

\subsection*{Proof of Lemma \ref{lm:hat_r geq r}.}
In order to prove this lemma, we just need to prove that the for-loop in Algorithm 2 will break for some $s \geq r$. This can be shown by proving the break condition \begin{equation}
\|D_{R,s}\| = \|Z_{21, [1:s]} Z_{11, [1:s, 1:s]}^{-1}\| \leq T_R, 
\end{equation}
hold for $s = r$.

We adopt the definitions in \eqref{eq:B_L}, \eqref{eq:B_M}, \eqref{eq:B_R}, then we have
\begin{equation*}
\begin{split}
Z_{11, [1:r, 1:r]} & = U^{(2)\intercal}_{[:, 1:r]}A_{11}V^{(1)}_{[:, 1:r]} = \hat M^{\intercal} A_{11} \hat N\\
& = \hat M^{\intercal}U_{11}\Sigma_1 V_{11}^{\intercal} \hat N + \hat M^{\intercal} U_{12}\Sigma_2 V_{12}^{\intercal} \hat N \\
& = B_M + E_M,
\end{split}
\end{equation*}
$$Z_{21, [:, 1:r]} = A_{21}V^{(1)}_{[:, 1:r]} = \left(U_{21}\Sigma_1V_{11}^{\intercal} + U_{22}\Sigma_2 V_{12}^{\intercal}\right)\hat N = B_L + E_L. $$
Hence,
\begin{equation*}
\begin{split}
\left\|Z_{21, [:, 1:r]}Z_{11, [1:r, 1:r]}^{-1}\right\|  = & \|(B_L+E_L)(B_M+E_M)^{-1}\|\\
\leq & \left\|B_LB_M^{-1}\sum_{i=0}^{\infty}(-E_M B_M^{-1})^i \right\| + \left\|E_L B_M^{-1}\sum_{i=0}^\infty(-E_M B_M^{-1})^i\right\|\\
\leq & \left(\|B_LB_M^{-1}\| + \|E_L\|\|B_M^{-1}\|\right)\frac{1}{1-\|E_M B_M^{-1}\|}\\
\overset{\eqref{ineq:sigma_min B_M},\eqref{ineq:B_L1B_M1^-1}}{\leq} & \left(\frac{\sqrt{45/44}}{\sigma_{\min}(U_{11})} + \frac{45\sigma_{r+1}(A)}{44\sigma_r(A)\sigma_{\min}(U_{11})\sigma_{\min}(V_{11})}\right) \frac{1}{1 -\frac{45\sigma_{r+1}(A)}{44\sigma_r(A)\sigma_{\min}(U_{11})\sigma_{\min}(V_{11})} }\\
\leq & \frac{1.36}{\sigma_{\min}(U_{11})} + 0.35 \leq T_R,
\end{split}
\end{equation*}
which finished the proof of the lemma. \quad $\square$

\subsection*{Proof of Lemma \ref{lm:inequalities_Theorem_2}.}
First, since $M_Z \in \mathbb{R}^{\hat r \times r}$ and $N_Z \in \mathbb{R}^{\hat r\times r}$ are an orthonormal basis of $B_{M, \hat r}$ and $B_{M,  \hat r}^{\intercal}$, we have $P_{M_Z} = M_ZM_Z^{\intercal}$ and $P_{N_Z} = N_ZN_Z^{\intercal}$ and
\begin{equation}
\begin{split}
\sigma_{\min}(B_{M1}) = & \sigma_{\min}(J_1^{\intercal}B_{M, \hat r}K_1) = \sigma_{\min}(J_1^{\intercal}M_Z M_Z^{\intercal} B_{M, \hat r}N_Z N_Z^{\intercal} K_1)\\
\geq & \sigma_{\min}(J_1^{\intercal}M_Z)\sigma_{\min}(M_Z^{\intercal}B_{M, \hat r}N_Z)\sigma_{\min}(N^{\intercal}_ZK_1)\\
\overset{\eqref{ineq:J_1^TM_Z}\eqref{ineq:K_1^TN_Z}}{\geq} & 0.859 \sigma_r(B_{M, \hat r}) \overset{\eqref{ineq:sigma_r(B_M,hat r)}}{\geq} \frac{0.859\cdot 3824}{3825}\sigma_r(A)\sigma_{\min}(U_{11})\sigma_{\min}(V_{11})\overset{\eqref{ineq:assumption_theorem2}}\geq 3.43 \sigma_{r+1}(A).
\end{split}
\end{equation}
which gives \eqref{ineq:sigma_min_B_M1}.
\begin{equation}\label{ineq:B_L1B_M1^-1_process}
\begin{split}
& \|B_{L1}B_{M1}^{-1}\| = \left\|B_{L, \hat r} K_1 \left(J_1^{\intercal}B_{M, \hat r} K_1\right)^{-1}\right\|\\
= & \left\| U_{21}\Sigma_1V_{11}^{\intercal}V_{[:, 1:\hat r]}^{(1)}K_1 \left(J_1^{\intercal}U_{[:, 1:\hat r]}^{(2)\intercal}U_{11}\Sigma_1 V_{11}^{\intercal}V_{[:, 1:\hat r]}^{(1)}K_1\right)^{-1} \right\|
=  \left\|U_{21}\left(J_1^{\intercal}U_{[:, 1:\hat r]}^{(2)\intercal}U_{11}\right)^{-1}\right\|\\
\leq & \frac{1}{\sigma_{\min}(J_1^{\intercal}U_{[:, 1:\hat r]}^{(2)\intercal}U_{11})} =  \frac{1}{\sigma_{\min}(J_1^{\intercal} P_{M_Z} (U^{(2)\intercal}_{[:, 1:\hat r]}U_{11}))} = \frac{1}{\sigma_{\min}((J_1^{\intercal} M_Z) (M_Z^{\intercal} U^{(2)\intercal}_{[:, 1:\hat r]}U_{11}))}\\
\leq & \frac{1}{\sigma_{\min}(J_1^{\intercal}M_Z)} \cdot \frac{1}{\sigma_{\min}(U_{[:,1:\hat r]}^{(2)\intercal} U_{11})}
\overset{\eqref{ineq:UU_11}\eqref{ineq:J_1^TM_Z}}{\leq} \frac{\sqrt{3825/3824}}{\sqrt{0.859}\sigma_{\min}(U_{11})},
\end{split}
\end{equation}
which gives the first part of \eqref{ineq:B_L1B_M1^-1}. Here we used the fact that $\Sigma_1 V_{11}^{\intercal}V_{[:, 1:\hat r]}^{(1)}K_1$ is a square matrix; $M_Z$ is the orthonormal basis of the column space of $Z_{11,[1:\hat r, 1:\hat r]} = U_{[:, 1:\hat r]}^{(2)\intercal}U_{11}\Sigma_1 V_{11}^{\intercal}V_{[:, 1:\hat r]}^{(1)}$. Similarly we have the later part of \eqref{ineq:B_L1B_M1^-1},
\begin{equation}\label{ineq:B_M1B_R1}
\|B_{M1}^{-1}B_{R1}\| \leq \frac{\sqrt{3825/3824}}{\sqrt{0.859}\sigma_{\min}(V_{11})}.
\end{equation}
Based on the definitions, we have the bound for all `` $E$" terms in \eqref{eq:JK_begin}-\eqref{eq:JK_last}, i.e. \eqref{ineq:E_Mt}.
Now we move on to \eqref{ineq:B_L2+E_L2B_M2E_M2}. By the SVD of $Z_{11, [1:\hat r, 1:\hat r]}$ \eqref{eq:Z_11_svd} and the partition \eqref{eq:J_1J_2K_1_K_2}, we know 
$$\left([J_1 ~ J_2]^{\intercal}Z_{11, [1:\hat r, 1:\hat r]} [K_1 ~ K_2]\right)^{-1} = \begin{bmatrix}
\Lambda_1 & 0 \\
0 & \Lambda_2
\end{bmatrix}^{-1} = \begin{bmatrix}
\left(J_1^{\intercal}Z_{11, [1:\hat r, 1:\hat r]} K_1\right)^{-1} & 0 \\
0 & \left(J_2^{\intercal}Z_{11, [1:\hat r, 1:\hat r]} K_2\right)^{-1}
\end{bmatrix}. $$
Hence, we have 
\begin{equation}
\begin{split}
& \left\|(B_{L2} + E_{L2})(B_{M2} + E_{M2})^{-1}\right\| = \left\|Z_{21,[:, 1:\hat r]} K_2 \left(J_2^{\intercal}Z_{11, [1:\hat r, 1:\hat r]} K_2\right)^{-1}\right\|\\
= & \left\|Z_{21,[:, 1:\hat r]} [K_1 ~ K_2] \left([J_1 ~ J_2]^{\intercal}Z_{11, [1:\hat r, 1:\hat r]} [K_1 ~ K_2]\right)^{-1} - Z_{21,[1:\hat r]} K_1 \left(J_1^{\intercal}Z_{11, [1:\hat r, 1:\hat r]} K_1\right)^{-1}\right\|\\
\leq & \left\| Z_{21,[:, 1:\hat r]} \left(Z_{11, [1:\hat r, 1:\hat r]}\right)^{-1} \right\| + \left\|(B_{L1} + E_{L1})(B_{M1} + E_{M1})^{-1}\right\|\\
\leq & T_R + \left\|B_{L1}\cdot B_{M1}^{-1}\sum_{i=0}^\infty (-E_{M1}B_{M1}^{-1})^i\right\| + \left\|E_{L1}\cdot B_{M1}^{-1}\sum_{i=0}^\infty (-E_{M1}B_{M1}^{-1})^i\right\| \\
\leq & T_R + \left(\|B_{L1}B_{M1}^{-1}\| + \|E_{L1}\|\|B_{M1}^{-1}\|\right) \frac{1}{1 - \|E_{M1}\|\|B_{M1}^{-1}\|}\\
\overset{\eqref{ineq:sigma_min_B_M1}\eqref{ineq:B_L1B_M1^-1}\eqref{ineq:E_Mt}}{\leq} & T_R + \left(\frac{\sqrt{3825/3824}}{\sqrt{0.859}\sigma_{\min}(U_{11})} + \frac{1}{3.43}\right)\cdot \frac{1}{1 - 1/3.43},
\end{split}
\end{equation}
which proves \eqref{ineq:B_L2+E_L2B_M2E_M2}. Since $Z_{11, [1:\hat r, 1:\hat r]} = B_{M,\hat r} + E_{M, \hat r}$ and by definition, $\rank(B_{M, \hat r})\leq r$, by Lemma \ref{lm:X_Y_sv}, we know
\begin{equation}
\sigma_{r+i} (Z_{11, [1:\hat r, 1:\hat r]}) \leq \sigma_i(E_{M, \hat r}), \quad \forall i\geq 1.
\end{equation} 
Then
\begin{equation}\label{ineq:B_M2}
\begin{split}
\|B_{M2}\|_q \leq & \|B_{M2} + E_{M2}\|_q + \|E_{M2}\|_q \leq \|J_2^{\intercal}Z_{11, [1:\hat r, 1:\hat r]} K_2\|_q + \|E_{M2}\|_q\\
= & \sqrt[q]{\sum_{i=r+1}^{\hat r} \sigma_i^q(Z_{11, [1:\hat r, 1:\hat r]})} + \|E_{M2}\|_q \leq \sqrt[q]{\sum_{i=1}^{\hat r - r} \sigma_i^q(E_{M, \hat r})} + \|E_{M2}\|_q\\
\leq & \|E_{M, \hat r}\|_q + \|E_{M2}\|_q \overset{\eqref{ineq:E_Mt}}{\leq} 2\|A_{-\max(r)}\|_q.
\end{split}
\end{equation}
Same to the process of \eqref{ineq:B_L1B_M1^-1_process}, we know 
\begin{equation}\label{ineq:V_11V_[]K_1}
\frac{1}{\sigma_{\min}(V_{11}^{\intercal}V_{[:, 1:\hat r]}^{(1)}K_1)} \leq \frac{\sqrt{3825/3824}}{\sqrt{0.859}\sigma_{\min}(V_{11})}.
\end{equation}
Also, $\|V_{21}^{\intercal}\|\leq 1$. Hence,
\begin{equation}
\begin{split}
\|B_{R2}\|_q \overset{\eqref{eq:JK_last}}{=} & \|J_2^{\intercal} B_{R, \hat r}\|_q = \|J_2^{\intercal} U_{[:, 1:\hat r]}^{(2)\intercal}U_{11}\Sigma_1V_{21}^{\intercal}\|_q\\
= & \|J_2^{\intercal} U_{[:, 1:\hat r]}^{(2)\intercal}U_{11}\Sigma_1(V_{11}^{\intercal}V_{[:, 1:\hat r]}^{(1)} K_1) (V_{11}^{\intercal}V_{[:, 1:\hat r]}^{(1)} K_1)^{-1}V_{21}^{\intercal}\|_q\\
\leq & \|B_{M2}\|_q \cdot \|(V_{11}^{\intercal}V_{[:, 1:\hat r]}^{(1)} K_1)^{-1}\| \cdot \|V_{21}^{\intercal}\|\\
\overset{\eqref{ineq:B_M2}\eqref{ineq:V_11V_[]K_1}}{\leq} & \frac{2\sqrt{3825/3824}}{\sqrt{0.859}\sigma_{\min}(V_{11})}\|A_{-\max(r)}\|_q.
\end{split}
\end{equation}
which proves \eqref{ineq:B_R2}. \quad $\square$

\subsection*{Proof of Theorem \ref{th:lower_bound}.} The idea of proof is to construct two matrices $A^{(1)}, A^{(2)}$ both in $\mathcal{F}_c(M_1, M_2)$ such that they have the identical first $m_1$ rows and $m_2$ columns, but differ much in the remaining block. Suppose $a,b,c>0$ are fixed numbers, $\varepsilon$ is a small real number. We first consider the following 2-by-2 matrix 
\begin{equation}\label{eq:B_lower_bound}
B(\varepsilon) = \begin{bmatrix}
a & c\\
b & \frac{bc}{a} + \varepsilon
\end{bmatrix}.
\end{equation}
Suppose the larger and smaller singular value of $B(\varepsilon)$ are $\lambda_{\max}(\varepsilon)$ and $\lambda_{\min}(\varepsilon)$, then we have 
\begin{equation}
\lambda_{\max}(\varepsilon) \to \|B(0)\| =  \frac{\sqrt{(a^2 + b^2)(a^2+c^2)}}{a}
\end{equation}
as $\varepsilon \to 0$; since $\lambda_{\max}(\varepsilon)\cdot \lambda_{\min}(\varepsilon) = |\text{det}(B)| = a|\varepsilon|$, we also have
\begin{equation}\label{eq:lambda_min_varepsilon}
\lambda_{\min}(\varepsilon)/|\varepsilon| \to \frac{a^2}{\sqrt{(a^2+b^2)(a^2+c^2)}}
\end{equation}
as $\varepsilon \to 0$. If $B(\varepsilon)$ defined in \eqref{eq:B_lower_bound} has SVD
\begin{equation}\label{eq:svd_B}
B(\varepsilon) = \begin{bmatrix}
u_{11} & u_{12}\\
u_{21} & u_{21}
\end{bmatrix} \cdot \begin{bmatrix}
\lambda_{\max}(\varepsilon) & 0\\
0 & \lambda_{\min}(\varepsilon)
\end{bmatrix}\cdot
\begin{bmatrix}
v_{11} & v_{12}\\
v_{21} & v_{21}
\end{bmatrix}^{\intercal}
\end{equation}
then we also have
\begin{equation}\label{eq:u v_lower_bound}
u_{11} \to \frac{a}{\sqrt{a^2+b^2}}\quad , u_{21} \to \frac{b}{\sqrt{a^2+b^2}},\quad v_{11}\to \frac{a}{\sqrt{a^2+c^2}},\quad v_{21} \to \frac{c}{\sqrt{a^2+c^2}}.
\end{equation}
as $\varepsilon \to 0$.

Now we set $a = 1$, $b = \sqrt{1 - M_1^2}/M_1 - \eta$, $c = \sqrt{1 - M_2^2}/M_2 - \eta$, $d = bc/a$, where $\eta$ is some small positive number to be specify later. We construct $A_{11}, A_{12}, A_{21}, A^{(1)}_{22}$ and $A^{(2)}_{22}$ such that,
\begin{equation}
A_{11} = 
\begin{bmatrix}
aI_r & 0\\
0 & 0 \\
\end{bmatrix}_{m_1\times m_2}, \quad
A_{12} = 
\begin{bmatrix}
cI_r & 0 \\
0 & 0\\
\end{bmatrix}_{m_1\times (p_2 -m_2)}, \quad
A_{21} = \begin{bmatrix}
bI_r & 0 \\
0 & 0\\
\end{bmatrix}_{(p_1 - m_1)\times m_2};
\end{equation}
\begin{equation}
A_{22}^{(1)} = 
\begin{bmatrix}
(d+\varepsilon)I_r & 0 \\
0 & 0\\
\end{bmatrix}_{(p_1-m_1) \times (p_2-m_2)},\quad
A_{22}^{(2)} = 
\begin{bmatrix}
(d-\varepsilon)I_r & 0 \\
0 & 0\\
\end{bmatrix}_{(p_1-m_1) \times (p_2-m_2)}.
\end{equation}
Here we use $I_r$ to note the identity matrix of dimension $r$. Then we construct $A^{(1)}$ and $A^{(2)}$ as
\begin{equation}
A^{(1)} = \begin{bmatrix}
A_{11} & A_{12}\\
A_{21} & A_{22}^{(1)}
\end{bmatrix},
\quad 
A^{(2)} = \begin{bmatrix}
A_{11} & A_{12}\\
A_{21} & A_{22}^{(2)}
\end{bmatrix},
\end{equation}
where $A^{(1)}$ and $A^{(2)}$ are with identical first $m_1$ rows and $m_2$ columns. Since the SVD of $B(\varepsilon)$ is given as \eqref{eq:svd_B}, the SVD of $A^{(1)}$ can be written as 
$$A^{(1)} = \begin{bmatrix}
U_{11}^{(1)} & U_{12}^{(1)}\\
U_{21}^{(1)} & U_{22}^{(1)}
\end{bmatrix} \cdot \begin{bmatrix}
\Sigma_1^{(1)} & 0\\
0 & \Sigma_2^{(1)}
\end{bmatrix}\cdot \begin{bmatrix}
V_{11}^{(1)} & V_{12}^{(1)}\\
V_{21}^{(1)} & V_{22}^{(1)}
\end{bmatrix}^{\intercal},$$ 
where 
$$U_{11} = \begin{bmatrix}
u_{11} I_r \\
0 
\end{bmatrix}_{m_1\times r}, \quad U_{12}=\begin{bmatrix}
u_{12} I_r \\
0 
\end{bmatrix}_{m_1\times r}, \quad U_{21} = \begin{bmatrix}
u_{21} I_r \\
0 
\end{bmatrix}_{(p_1 - m_1)\times r}, \quad U_{22} = \begin{bmatrix}
u_{22}I_r\\
0 
\end{bmatrix}_{(p_1 - m_1)\times r};$$ 
$$V_{11} = \begin{bmatrix}
v_{11}I_r \\
0 
\end{bmatrix}_{m_2\times r},\quad V_{12} = \begin{bmatrix}
v_{12}I_{r} \\
0 
\end{bmatrix}_{m_2\times r}, \quad V_{21} = \begin{bmatrix}
v_{21}I_r \\
0 
\end{bmatrix}_{(p_2 - m_2)\times r}, \quad V_{22} = \begin{bmatrix}
v_{22} I_r \\
0 
\end{bmatrix}_{(p_2 - m_2)\times r};$$ 
$$\Sigma_1 = \lambda_{\max}(\varepsilon) I_r,\quad \Sigma_2 = \lambda_{\min}(\varepsilon)I_r.$$
Hence, 
$$\sigma_{\min}(U_{11}) = u_{11} = \frac{a}{\sqrt{a^2+b^2}}\to \frac{1}{1 + \left(\frac{\sqrt{1-M_1^2}}{M_1} - \eta\right)^2} > M_1,\quad \text{as } \varepsilon\to 0$$
$$\sigma_{\min}(V_{11}) = v_{11} = \frac{a}{\sqrt{a^2+c^2}}\to \frac{1}{1 + \left(\frac{\sqrt{1-M_2^2}}{M_2} - \eta\right)^2} > M_2,\quad \text{as } \varepsilon\to 0.$$
Also, $\|\Sigma_2^{(1)}\| \to 0$ as $\varepsilon \to 0$. So we have $A^{(1)} \in \mathcal{F}_r(M_1, M_2)$ when $\varepsilon$ is small enough. Similarly $A^{(2)}\in \mathcal{F}_r(M_1, M_2)$ when $\varepsilon$ is small enough. Now we also have $\|A_{-\max(r)}^{(1)}\|_q = \left(q \lambda_{\min}(\varepsilon)^q\right)^{1/q} = q^{1/q}\lambda_{\min}(\varepsilon)$, $\|A_{-\max(r)}^{(2)}\|_q = \left(q \lambda_{\min}(-\varepsilon)^q\right)^{1/q} = q^{1/q}\lambda_{\min}(-\varepsilon)$. $\|A_{22}^{(1)} - A_{22}^{(2)}\|_q = (q(2|\varepsilon|)^q)^{1/q} = 2|\varepsilon|q^{1/q}$.

Finally for any estimate $\hat A_{22}$, we must have
\begin{equation}\label{ineq:lower_bound_big_formula}
\begin{split}
& \max\left\{\frac{\|\hat A_{22} - A_{22}^{(1)}\|_q}{\|A^{(1)}_{-\max(r)}\|_q}, \frac{\|\hat A_{22} - A_{22}^{(2)}\|_q}{\|A^{(2)}_{-\max(r)}\|_q}\right\} \geq \frac{\frac{1}{2}\left\|\left(\hat A_{22} - A_{22}^{(1)}\right) -\left(\hat A_{22}- A_{22}^{(2)}\right)\right\|_q}{\min\left\{\|A^{(1)}_{-\max(r)}\|_q, \|A^{(2)}_{-\max(r)}\|_q \right\}}\\
\geq & \frac{2|\varepsilon|}{2\min\left\{\lambda_{\min}(\varepsilon), \lambda_{\min}(-\varepsilon)\right\}} \overset{\eqref{eq:lambda_min_varepsilon}}{\to} \frac{\sqrt{(a^2+b^2)(a^2+c^2)}}{a^2}\\
= & \sqrt{\left(1 + (\frac{\sqrt{1-M_1^2}}{M_1} - \eta)^2\right)\left(1 + (\frac{\sqrt{1-M_2^2}}{M_2} - \eta)^2\right)}
\end{split}
\end{equation} 
as $\varepsilon \to 0$. Since $A^{(1)}, A^{(2)}\in \mathcal{F}_r(M_1, M_2)$ and are with identical first $m_1$ rows and $m_2$ columns, we must have
$$\inf_{\hat A_{22}}\sup_{A\in \mathcal{F}_{r}(M_1, M_2)} \frac{\|\hat A_{22} - A_{22}\|_q}{\|A_{-\max(r)}\|_q} \geq \sqrt{\left(1 + (\frac{\sqrt{1-M_1^2}}{M_1} - \eta)^2\right)\left(1 + (\frac{\sqrt{1-M_2^2}}{M_2} - \eta)^2\right)}. $$
Let $\eta \to 0$, since $M_1, M_2<1$, we have
\begin{equation}
\inf_{\hat A_{22}}\sup_{A\in \mathcal{F}_{r}(M_1, M_2)} \frac{\|\hat A_{22} - A_{22}\|_q}{\|A_{-\max(r)}\|_q} \geq \frac{1}{M_1M_2} \geq \frac{1}{4}\left(\frac{1}{M_1}+1\right)\left(\frac{1}{M_2} + 1\right),
\end{equation}
which finished the proof of theorem.
\quad $\square$

\subsection*{Proof of Corollary \ref{cr:random_column_row}.}
We first prove the second part of the corollary. We set $\alpha = (136/165)^2$. Since $U_{[:, 1:r]}\in\mathbb{R}^{p_1\times r}$ is with orthonormal columns, by Lemma \ref{lm:U_Omega} and 
$$m_1\geq 12.5W_r^{(1)}r(\log r + c)\geq  \frac{4}{(1-\alpha)^2}\cdot W^{(1)}_rr(\log r + c),$$ 
we have
\begin{equation}\label{eq:sigma_min(U)}
\sigma_{\min}(U_{11}) = \sigma_{\min}(U_{[\Omega_1, 1:r]}) \geq \sqrt{\frac{\alpha m_1}{p_1}}
\end{equation}
with probability at least $ 1 - 2\exp(-c)$. When \eqref{eq:sigma_min(U)} holds, by the condition, we know
\begin{equation*}
\begin{split}
\sigma_{r+1}(A) & \leq \sigma_r(A)\sigma_{\min}(V_{11})\frac{1}{5}\sqrt{\frac{m_1}{p_1}} \leq\sigma_r(A)\sigma_{\min}(V_{11})\frac{1}{5\sqrt{\alpha}}\cdot \sigma_{\min}(U_{11})
\leq \frac{1}{4}\sigma_r(A)\sigma_{\min}(V_{11})\sigma_{\min}(U_{11}).
\end{split}
\end{equation*}
When $T_R \geq 2\sqrt{p_1/m_1}$, we have 
$$\frac{1.36}{\sigma_{\min}(U_{11})} + 0.35 \leq 1.36\sqrt{\frac{p_1}{\alpha m_1}} + 0.35 \leq 2\sqrt{\frac{p_1}{m_1}} \leq T_R$$
Hence we can apply Theorem \ref{th:main_without_r}, for $1\leq q\leq \infty$ we must have
\begin{equation}
\left\|\hat A_{22} - A_{22}\right\|_q \leq 6.5T_R \left\|A_{-\max(r)}\right\|_q \left(\frac{1}{\sigma_{\min}(V_{11})} + 1\right),
\end{equation}
which finishes the proof of the second part of Corollary \ref{cr:random_column_row}. Besides, the proof for the third part is the same as the second part after we take the transpose of the matrix.

For the first part, the proof is also similar. Again we set $\alpha = (136/165)^2$. Then we have
$$m_1 \geq \frac{4}{(1-\alpha)^2}W_r^{(1)} r(\log r+c), \quad m_2 \geq \frac{4}{(1-\alpha)^2}W_r^{(2)} r(\log r+c),$$ 
so
\begin{equation}\label{ineq:sigma_min(U, V)}
\sigma_{\min}(U_{11}) = \sigma_{\min} (U_{[\Omega_1, 1:r]}) \geq \sqrt{\frac{\alpha m_1}{p_1}}, \quad \sigma_{\min}(V_{11}) = \sigma_{\min} (V_{[\Omega_2, 1:r]}) \geq \sqrt{\frac{\alpha m_2}{p_2}}
\end{equation}
with probability at least $1 - 4\exp(-c)$.  When \eqref{ineq:sigma_min(U, V)} holds, we have
$$\sigma_{r+1}(A)\leq \sigma_r(A)\frac{1}{6}\sqrt{\frac{m_1m_2}{p_1p_2}} \leq \sigma_r(A)\frac{1}{6\alpha}\sigma_{\min}(U_{11})\sigma_{\min}(V_{11}) \leq \frac{1}{4}\sigma_r(A)\sigma_{\min}(V_{11})\sigma_{\min}(U_{11}). $$
When $T_R = 2\sqrt{p_1/m_1}$ or $T_C  = 2\sqrt{p_2/m_2}$, similarly to the first part we have
$$\frac{1.36}{\sigma_{\min}(U_{11})} + 0.35 \leq T_R,\quad \text{  or  }\quad \frac{1.36}{\sigma_{\min}(V_{11})}+0.35 \leq T_C. $$
Hence we can apply Theorem \ref{th:main_without_r} and get
\begin{equation*}
\begin{split}
\left\|\hat A_{22} - A_{22}\right\|_q \leq & 6.5 T_R \|A_{-\max(r)}\|_q \left(\frac{1}{\sigma_{\min}(V_{11})} + 1\right)\leq 6.5 \cdot 2\sqrt{\frac{p_1}{m_1}} \cdot \left(\sqrt{\frac{p_2}{\alpha m_2}} + 1\right)\|A_{-\max(r)}\|_q \\
\leq & 29\|A_{-\max(r)}\|_q\sqrt{\frac{p_1p_2}{m_1m_2}}. 
\end{split}
\end{equation*}
\quad $\square$

\subsection*{Proof of Corollary \ref{cr:randomUV}.} Suppose $0< \alpha_1<1$, since $U_{[:, 1:r]}\in\mathbb{R}$ is with random orthonormal columns of Haar measure, we can apply Lemma \ref{lm:U_Haar} and find some $c >0$ and $\delta>0$ such that when $p_1\geq m_1\geq cr$,
\begin{equation}\label{eq:sigma_min_U_11_random_U}
\sigma_{\min}(U_{11}) = \sigma_{\min}(U_{[1:m_1, 1:r]}) \geq \frac{136}{165}\sqrt{\frac{m_1}{p_1}}
\end{equation}
with probability at least $1 - \exp(-\delta m_1)$. When \eqref{eq:sigma_min_U_11_random_U} happen, we have
$$\sigma_{r+1}(A) \leq \sigma_r(A)\sigma_{\min}(V_{11})\frac{1}{5}\sqrt{\frac{m_1}{p_1}} \leq \sigma_{r}(A)\sigma_{\min}(V_{11})\sigma_{\min}(U_{11}), $$
$$\frac{1.36}{\sigma_{\min}(U_{11})} + 0.35 \leq 1.36\cdot\frac{165}{136}\sqrt{\frac{p_1}{m_1}} + 0.35  \leq 2\sqrt{\frac{p_1}{m_1}}. $$
Hence we can apply Theorem \ref{th:main_without_r}, for $1\leq q\leq \infty$, we have
\begin{equation}
\left\|\hat A_{22} - A_{22}\right\|_q \leq 6.5T_R \left\|A_{-\max(r)}\right\|_q \left(\frac{1}{\sigma_{\min}(V_{11})} + 1\right),
\end{equation}
which finishes the proof of the corollary. \quad $\square$

\subsection{Description of Cross-Validation}

In this section, we describe the cross-validation used in penalized nuclear norm minimization \eqref{eq:PNNM} in the numerical comparison in Sections \ref{simulation.sec} and \ref{application.sec}.

First, we construct a grid $T$ of non-negative numbers based on a pre-selected positive integer $N$. Denote 
$$t_{\max}^{PN} = \left\|\begin{bmatrix}
A_{11} & A_{12}\\
A_{21} & 0
\end{bmatrix}\right\|,$$ 
i.e. the largest singular value of the observed blocks. For penalized nuclear norm minimization, we let $T = \left\{t_{\max}^{PN}, t^{PN}_{\max}\cdot10^{-3(1/N)}, \cdots, t^{PN}_{\max}\cdot10^{-3(N/N)} \right\}$.

Next, for a given positive integer $K$, we randomly divide the integer set $\{1,\cdots, m_1\}$ into two groups of size $m^{(1)} \approx \frac{(K-1)n}{K}$, $m^{(2)}\approx\frac{n}{K}$ for $H$ times. For $h=1,\cdots, H$, we denote by $J_1^h$ and $J_2^h \subseteq\{1, 2, \cdots, m_1\}$ the index sets of the two groups for the $h$-th split. Then the penalized nuclear norm minimization estimator \eqref{eq:PNNM} is applied to the first group of data: $A_{11}, A_{21}, (A_{12})_{[J_1^h, :]}$, i.e. the data of the observation set $\Omega = \{(i,j): 1\leq j \leq m_2, \text{ or } i\in J_1^h, m_2+1\leq j\leq p_2\}$, with each value of the tuning parameter $t \in T$ and denote the result by $\hat A^{PN}_h(t)$. Note that we did not use the observed block $A_{[J_2^h, (m_2+1):p_2]}$ in calculating $\hat A^{PN}_h(t)$. Instead, $A_{[J_2^h, (m_2+1):p_2]}$ is used to evaluate the performance of the tunning parameter $t\in T$. Set
\begin{equation}\label{CV}
\hat{R}(t) =\frac{1}{H} \sum_{h=1}^H \left\| \left[\hat A^{PN}_h(t)\right]_{[J_2^h, (m_2+1):p_2]} - A_{[J_2^h, (m_2+1):p_2]} \right\|_F^2.
\end{equation}
Finally, the tuning parameter is chosen as
$$t_\ast = \argmin_{t\in T} \hat R(t) $$
and the final estimator $\hat{A}^{PN}$ is calculated using this choice of the tuning parameter  $t_\ast$.

In all the numerical studies with penalized nuclear norm minimization in Sections \ref{simulation.sec} and \ref{application.sec}, we use 5-cross-validation (i.e., $K=5$), $N=10$ to select the tuning parameter.

\end{document}